\documentclass[useAMS]{mn2e}
\usepackage{psfig}
\newcommand{\ltsima} {$\; \buildrel < \over \sim \;$}
\newcommand{\gtsima} {$\; \buildrel > \over \sim \;$}
\newcommand{\lta} {\lower.5ex\hbox{\ltsima}}
\newcommand{\gta} {\lower.5ex\hbox{\gtsima}}

\newcommand{\atlas}{ATLAS$^{\mathrm{3D}}$}

\title [Intrinsic shapes of early-type galaxies]
{The \atlas\ project - XXIV. The intrinsic shape
  distribution of early-type galaxies}

\author[A. Weijmans et al.]
  {Anne-Marie Weijmans$^{1,2}$\thanks{E-mail:
      amw23@st-andrews.ac.uk}\thanks{Dunlap Fellow},
   P. T. de Zeeuw$^{3,4}$,  Eric Emsellem$^{3,5}$,  Davor
   Krajnovi\'c$^{6}$, \and
   Pierre-Yves Lablanche$^{5}$, Katherine
   Alatalo$^{7,8}$, Leo Blitz$^7$, Maxime Bois$^9$, \and Fr\'ed\'eric
   Bournaud$^{10}$, Martin Bureau$^{11}$, Michele Cappellari$^{11}$, Alison
   F. Crocker$^{12}$, \and Roger
   L. Davies$^{11}$, Timothy A. Davis$^{3}$,  Pierre-Alain Duc$^{10}$, Sadegh
   Khochfar$^{13}$, \and Harald Kuntschner$^{3}$, Richard
   M. McDermid$^{14,15,16}$, Raffaella
   Morganti$^{17,18}$, \and Thorsten
   Naab$^{19}$, Tom Oosterloo$^{17,18}$, Marc Sarzi$^{20}$, Nicholas
   Scott$^{21}$, Paolo Serra$^{17,22}$,\and Gijs Verdoes Kleijn$^{18}$
   \&  Lisa M. Young$^{23}$\\
  $^{1}$School of Physics and Astronomy, University of St Andrews,
  North Haugh, St Andrews KY16 9SS, UK \\
  $^2$Dunlap Institute for Astronomy \& Astrophysics, University of
  Toronto, 50 St. George Street, Toronto, ON M5S 3H4, Canada \\
   $^3$European Southern Observatory, Karl-Schwarzschild-Str. 2, 85748
   Garching, Germany\\
 $^4$Sterrewacht Leiden, Leiden University, Postbus 9513, 2300 RA
 Leiden, the Netherlands\\
$^5$Universit\'e Lyon 1, Observatoire de Lyon, Centre de Recherche
Astrophysique de Lyon \\ $^{}$and Ecole Normale Sup\'erieure de Lyon, 9 avenue
Charles Andr\'e, F-69230 Saint-Genis Laval, France\\
$^{6}$Leibniz-Institut f\"ur Astrophysik Potsdam (AIP), An der
Sternwarte 16, D-14482 Potsdam, Germany \\
$^7$Department of Astronomy, Campbell Hall, University of California,
Berkeley, CA 94720, USA\\
$^8$  Infrared Processing and Analysis Center, California Institute of
Technology, Pasadena, California 91125, USA  \\
$^9$Observatoire de Paris, LERMA and CNRS, 61 Av. de l'Observatoire,
F-75014 Paris, France\\\
$^{10}$Laboratoire AIM Paris-Saclay, CEA/IRFU/SAp – CNRS – Universit\'e
Paris Diderot, 91191 Gif-sur-Yvette Cedex, France\\
$^{11}$Sub-department of Astrophysics, Department of Physics, University of Oxford, Denys Wilkinson Building, Keble Road, Oxford OX1 3RH\\
$^{12}$Ritter Astrophysical Observatory, University of Toledo, Toledo, OH 43606, USA\\ 
$^{13}$Institute for Astronomy, University of Edinburgh, Royal Observatory, Edinburgh EH9 3HJ, UK\\
$^{14}$Gemini Observatory, Northern Operations Centre, 670 N. A`ohoku
 Place, Hilo, HI 96720, USA\\
$^{15}$Department of Physics and Astronomy and MQ Research Centre in Astronomy, Astrophysics and Astrophotonics, Macquarie University, NSW 2109, Australia\\
$^{16}$Australian Astronomical Observatory, PO Box 296, NSW 1710, Australia\\
$^{17}$Netherlands Institute for Radio Astronomy (ASTRON), Postbus 2, 7990 AA Dwingeloo, The Netherlands\\
$^{18}$Kapteyn Astronomical Institute, University of Groningen,
Postbus 800, 9700 AV Groningen, The Netherlands\\
$^{19}$Max-Planck-Institut f\"ur Astrophysik,
Karl-Schwarzschild-Str. 1, 85741 Garching, Germany\\
$^{20}$Centre for Astrophysics Research, University of Hertfordshire, Hatfield, Herts AL1 9AB, UK\\
$^{21}$Centre for Astrophysics and Supercomputing, Swinburne
University of Technology, Hawthorn, Victoria 3122, Australia\\
$^{22}$CSIRO Astronomy and Space Science, Australia Telescope National
Facility, PO Box 76, Epping, NSW 1710, Australia \\
 $^{23}$Physics Department, New Mexico Institute of Mining and
 Technology, Socorro, NM 87801, USA\\
\vspace{-1.8cm}  
 }

\pagerange{\pageref{firstpage}--\pageref{lastpage}} \pubyear{2014}

\begin{document}

\maketitle
\label{firstpage}

\begin{abstract} 
 We use the \atlas sample to perform a study of the intrinsic shapes
  of early-type galaxies, taking advantage of the available combined photometric
  and kinematic data. Based on our ellipticity
  measurements from the Sloan Digital Sky Survey Data Release 7, and
  additional imaging from the Isaac Newton Telescope, we first invert
  the shape distribution of fast and slow rotators under the
  assumption of axisymmetry. The so-obtained intrinsic shape
  distribution for the fast rotators can be described with a Gaussian
  with a mean flattening of $q=0.25$ and standard deviation $\sigma_q
  = 0.14$, and an additional tail towards
  rounder shapes. The slow rotators are much rounder, and are well
  described with a Gaussian with mean $q = 0.63$ and $\sigma_q
  =0.09$. We then checked that our results were consistent when
  applying a different and independent method to obtain intrinsic shape
  distributions, by fitting the observed ellipticity distributions
  directly using Gaussian parametrisations for the intrinsic axis ratios. Although both fast
  and slow rotators are identified as early-type galaxies in
  morphological studies, and in many previous shape studies are therefore
  grouped together, their shape distributions are significantly
  different, hinting at different formation scenarios. The intrinsic
  shape distribution of the fast rotators shows similarities with the
  spiral galaxy population. Including the observed kinematic
  misalignment in our intrinsic shape study shows that the fast
  rotators are predominantly axisymmetric, with only very little room
  for triaxiality. For the slow rotators though there are very
    strong indications that they are (mildly) triaxial.   
\end{abstract}



\section{Introduction}
\label{sec:introduction}

Shape is a very basic property of a galaxy, yet it contains strong
constraints for its formation history, with different merger,
accretion and assembly scenarios resulting in different shapes. Still,
intrinsic shapes of individual galaxies are not readily obtained:
detailed photometry and kinematical information is needed to construct
a dynamical model of a galaxy, and constrain its shape (e.g. Statler
1994\nocite{1994ApJ...425..500S}; Statler, Lambright \& Bak
2001\nocite{2001ApJ...549..871S}; van den Bosch \& van de Ven
2009\nocite{2009MNRAS.398.1117V}). Therefore, many studies to obtain
intrinsic shapes of galaxies have focused on large samples, using
statistical methods to obtain the underlying intrinsic shape \emph{distribution} of a particular galaxy population (e.g., Hubble
1926\nocite{1926ApJ....64..321H}; Sandage, Freeman \& Stokes
1970\nocite{1970ApJ...160..831S}; Lambas, Maddox \& Loveday
1992\nocite{1992MNRAS.258..404L}; Tremblay \& Merrit
1996\nocite{1996AJ....111.2243T}; Ryden
2004\nocite{2004ApJ...601..214R}; Vincent \& Ryden
2005\nocite{2005ApJ...623..137V}; Ryden
2006\nocite{2006ApJ...641..773R}; Kimm \& Yi
2007\nocite{2007ApJ...670.1048K}; Padilla \& Strauss
2008\nocite{2008MNRAS.388.1321P}; M\'endez-Abreu et al.
2010\nocite{2010A&A...521A..71M}; Yuma, Ohta \& Yabe
2012\nocite{2012ApJ...761...19Y}). These studies rely on measurements
of the observed ellipticities $\epsilon = 1 - b/a$, with $b/a$ the
observed axis ratio of the galaxy image, and, in principle, do not
require kinematic information (although as we mention later inclusion
of kinematic misalignment provides additional constraints on the shape
distribution, e.g. Binney 1985\nocite{1985MNRAS.212..767B}; Franx,
Illingworth \& de Zeeuw 1991\nocite{1991ApJ...383..112F}). Especially
the Sloan Digital Sky Survey (SDSS) has been a major provider for
imaging used in shape studies: recent results based on this survey
include the non-circularity of discs in spiral galaxies (Ryden
2004\nocite{2004ApJ...601..214R}; Padilla \& Strauss
2008\nocite{2008MNRAS.388.1321P}) and the presence of triaxial and
prolate galaxies in the early-type galaxy population (Vincent \& Ryden
2005\nocite{2005ApJ...623..137V}; Kimm \& Yi
2007\nocite{2007ApJ...670.1048K}).

The selection of the galaxy populations in these previous studies has
been predominantly based on morphology, colour and structural
parameters such as S\'ersic index (S\'ersic 1968\nocite{1968adga.book.....S}).
With the advent of integral-field spectroscopic studies we have an
additional parameter to base our sample selection on: kinematic
structure. In this paper we exploit this opportunity to make a
stricter selection by using the \atlas\ sample: a volume-limited
survey of 260 nearby early-type galaxies (Cappellari et al. 2011a,
hereafter Paper I\nocite{2011MNRAS.413..813C}), that includes
integral-field spectroscopy obtained by the SAURON
spectrograph (Bacon et al. 2001\nocite{2001MNRAS.326...23B}). We are
now able to make a distinction between two classes of early-type
galaxies, fast and slow rotators, based on their extended kinematic
properties, and as such obtain a cleaner galaxy population sample.

In Section 2 we describe the properties of the \atlas\ sample and the
dataset that we use for the shape inversion in this paper, while in
Section 3 we explain our methods and show our results for an
  axisymmetric shape inversion. Section 4
contains a discussion and interpretation of our results, and we
  further investigate the assumption of axisymmetry, by including
  kinematic misalignment angles in our shape analysis. We
summarize our work in Section 5, and provide additional
  formularium for our shape distributions in the appendices.


\section{Observations}
\label{sec:obs}

\subsection{Sample}

The \atlas\ sample was selected from a volume-limited parent sample of
871 galaxies in the nearby Universe. This parent sample consists of
all galaxies within a distance of 42 Mpc, down to a total luminosity
of -21.5 $M_K$, based on the 2MASS extended source catalog (Jarrett et
al. 2000\nocite{2000AJ....119.2498J}). The sample had to be
observable with the William Herschel Telescope (WHT) from La Palma, Spain, so
that only galaxies with sky declination $| \delta -29^{\circ} | <
35^{\circ} $ were included. Finally, the dusty region near the Galaxy
equatorial plane $|b| < 15^{\circ} $was excluded, with $b$ the
galactic latitude. From this parent sample, early-type galaxies were
morphologically selected based on visual inspection of multi-colour
images from SDSS DR7 (Abazajian et
al. 2009\nocite{2009ApJS..182..543A}) or $B$-band DSS2-blue
images\footnote{available on-line at http://archive.eso.org/dss}, resulting in a sample of 260 galaxies. The main selection
criteria here were the apparent lack of spiral arms in face-on, and dust lanes in
edge-on systems, indicating that our selected galaxies are
indeed early-types. For more details on the selection and properties
of the \atlas\ sample, we refer the reader to Paper I. Important for our work here is
to keep in mind that our sample is complete and has integral-field
kinematics available for all galaxies (Paper I; Krajnovi\'c et
al. 2011, hereafter paper II; Emsellem et al. 2011, hereafter
Paper III\nocite{2011MNRAS.414..888E}), allowing us to
perform a shape inversion on fast and slow rotators separately.

\begin{figure}
\begin{center} 

\psfig{figure=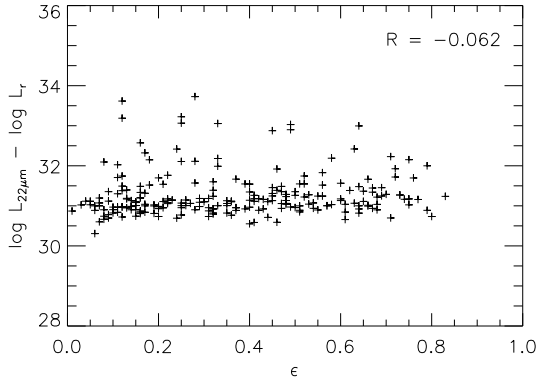,width=8cm} 

\end{center}
\caption{Ratio between mid-infrared and optical flux (expressed
  as a difference in magnitudes) as a function of ellipticity for 231
  galaxies in our sample. There is no correlation between these two
  quantities, indicating that our sample is not contaminated with
  late-type galaxies of preferred orientations, see text for
  details. The linear Pearson correlation coefficient $R$
  is printed in the top right corner.}
\label{fig:randomangles}
\end{figure}

\subsection{Investigation of selection bias}

For a statistical shape analysis as described in this paper to work,
we must have a galaxy sample that is randomly oriented on the sky,
such that our assumption of random viewing angles is a valid one. Our
selection criteria for removing late-type galaxies from our sample
(i.e., presence of spiral arms and/or dust lanes) may however differ
in reliability for different viewing angles and could in principle
introduce a bias in our sample. For instance, if our method of
detecting dust lanes in edge-on galaxies is not effective enough to
identify all edge-on late-type galaxies present, then our sample would
be contaminated with an extra population of flat spirals.  We also
note that although edge-on galaxies with large scale dust lanes were
excluded from the \atlas sample, galaxies with small, central dust
features were not, as these are not related to galaxy-wide spiral
arms. To investigate whether a bias is present, we extract band W4
22$\mu$m from the archive of the Wide-field Infrared Survey Explorer
(WISE; Wright et al. 2010\nocite{2010AJ....140.1868W}) for 231
galaxies in our sample. These fluxes are presented in Davis et
al. (2014)\nocite{2014arXiv1403.4850D} and are measured within
elliptical apertures, see the
on-line\footnote{http://wise2.ipac.caltech.edu/docs/release/allsky/}
WISE documentation for more details. If indeed our sample suffers from
harbouring edge-on spiral galaxies, which are dustier than early-type
galaxies, then we expect the ratio between the dust-tracing
mid-infrared and the star-tracing optical fluxes to change as a
function of ellipticity. Figure~\ref{fig:randomangles} shows that this
is not the case: there is no correlation between mid-infrared to
optical flux ratio and ellipticity for the galaxies in our
sample. This is confirmed by the linear Pearson correlation
coefficient $R$, which is small (-0.062).  The mid-infrared fluxes
correlate with the optical fluxes as expected (Temi et
al. 2009\nocite{ 2009ApJ...707..890T}, Davis et al. 2014\nocite{2014arXiv1403.4850D}), which is
a necessary condition for our test to work. We therefore conclude that
the \atlas sample of early-type galaxies is indeed randomly
distributed on the sphere of viewing angles.

\subsection{Observed shape and misalignment distribution}

The ellipticities of the galaxies in our sample were measured and
presented in Paper
II\nocite{2011MNRAS.414.2923K}, and we refer the reader there for
details. Briefly, for 212 galaxies in our sample SDSS DR7 $r$-band imaging is available
(Abazajian et al. 2009\nocite{2009ApJS..182..543A}) and for 46
galaxies not covered by this survey we obtained comparable $r$-band
imaging with the Wide Field Camera on the Isaac Newton Telescope (INT)
on La Palma. These observations and their data reduction are
presented in Scott et al. (2013, Paper XXI)\nocite{2013MNRAS.432.1894S}. For
the two remaining galaxies we used 2MASS K-band observations
instead. 

Since we are interested in the global shapes of the galaxies, and to
avoid  our analysis being dominated by e.g. central bars, we measured the
ellipticities using the moment of inertia of the surface
  brightness distribution on the sky subtracted images,
with bright stars and neighbouring galaxies masked (see Paper II
  for a detailed description of this method). This way, all
components in the galaxy contribute to a global ellipticity
measurement, which would not be the case if we measured ellipticity
only at a fixed radius. However, this method of measuring ellipticity
does introduce a bias towards the shapes at larger radius. Only pixels above
a certain threshold, 3 times the rms of the sky, were included in the
measurements. For galaxies that were dominated by bars, we lowered the
threshold to 0.5 or 1 times the sky rms, to better probe the underlying
stellar disc. This resulted in ellipticity measurements representative
of the galaxy out to typically 2.5 to 3 effective radii. We compared
these global ellipticity values to radial profiles, determined by
fitting ellipses along isophotes with \textsc{kinemetry} (Krajnovi\'c et
al. 2006\nocite{2006MNRAS.366..787K}), and found that these values
agree well: the standard deviation of the differences between the two
measurements was 0.03, see Paper II.

\begin{figure}
\begin{tabular}{c}
\psfig{figure=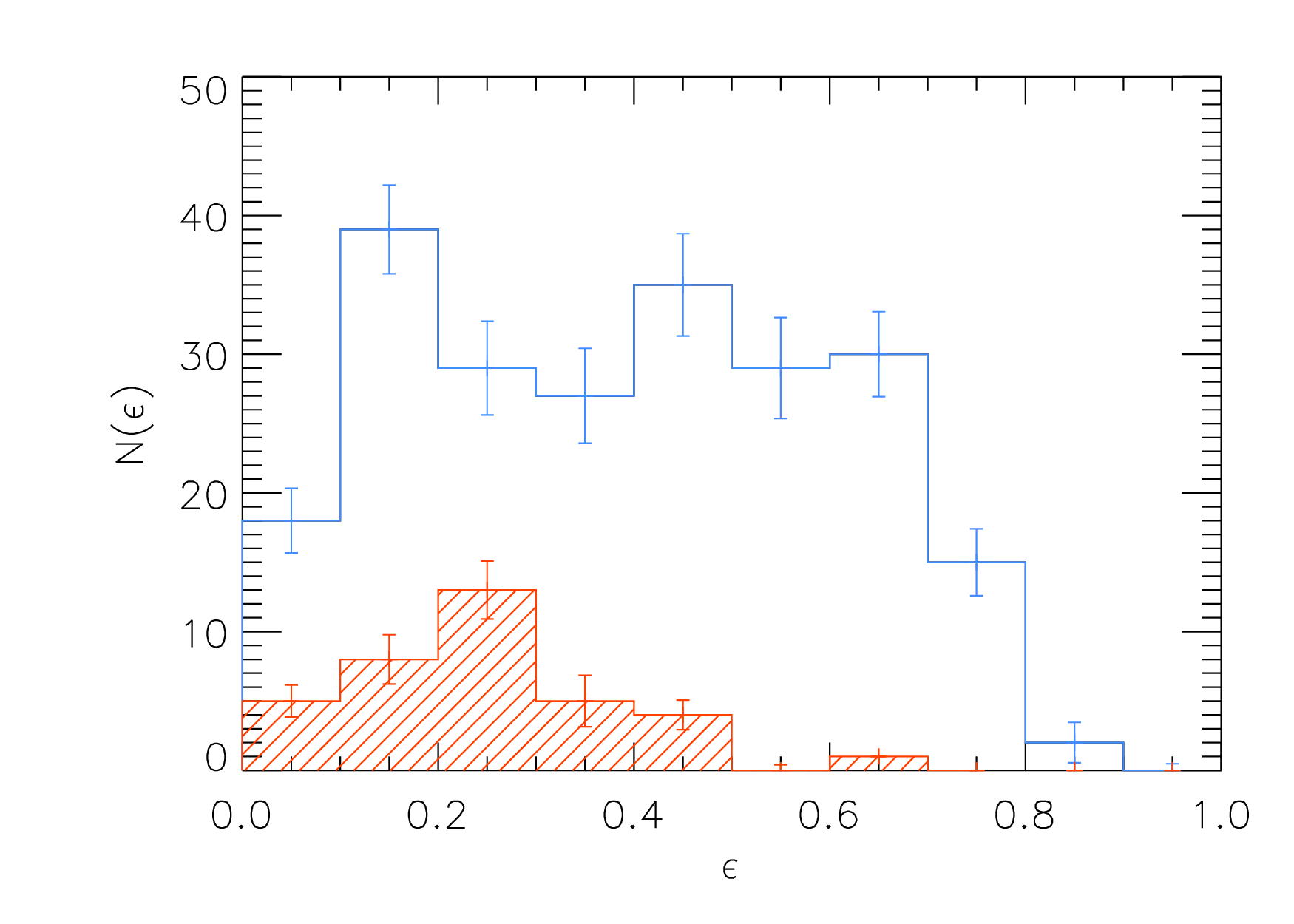,width=8cm} \\
\psfig{figure=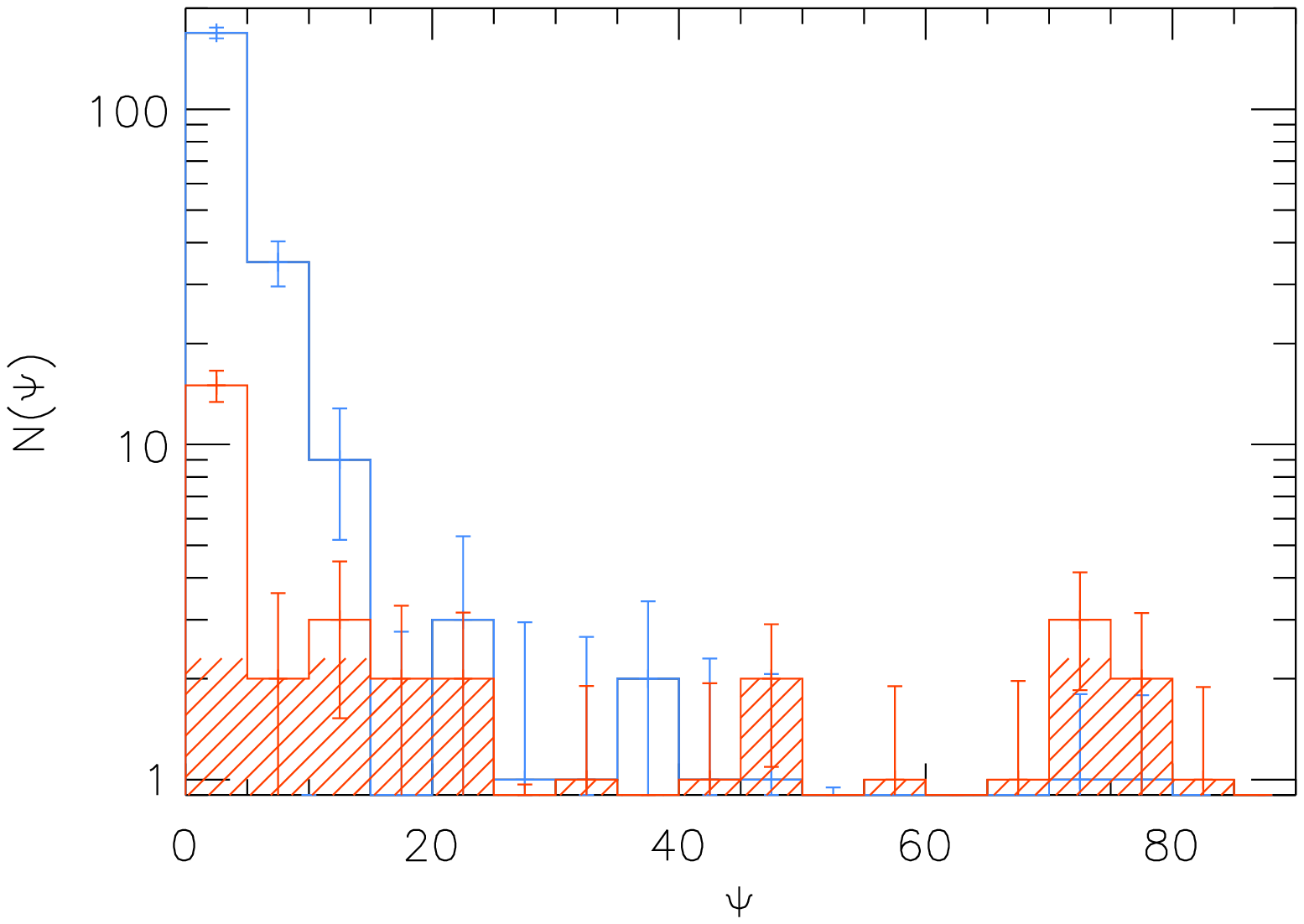,width=8.2cm} 
\end{tabular}
\caption{Top: histogram of observed ellipticities. The distribution of fast
  rotators is presented in blue (open histogram), while the slow rotators are denoted
  by the red, dashed histogram. The 1-$\sigma$ errorbars are based
on Monte Carlo simulations, taking the individual measurement errors
for each galaxy into account. Bottom: same as top panel, but now for
observed kinematic misalignments. The $y$-axis is now given in
log-scale.}
\label{fig:hist_eps}
\end{figure}

Uncertainties were determined by repeating the ellipticity
measurements for each galaxy at different thresholds (0.5, 1, 3 and 6
times the sky rms) and the standard deviations of these measurements
were adopted as errors. We show the resulting observed ellipticity
distribution for our sample in Figure~\ref{fig:hist_eps}, both for the
fast and slow rotators. The 1-$\sigma$ errors in the histograms have
been determined using Monte Carlo simulations, based on the errors in
ellipticity of the individual galaxies. These individual values can
be found in Table 1 of Paper II.

Paper II also provides values for the photometric and kinematic
position angles, used to measure the kinematic misalignment $\Psi$ between
the projected rotation axis and the minor axis of a galaxy. The
photometric position angle was measured on the SDSS, INT or 2MASS imaging, using the
same method as described above for the ellipticity. The kinematic
position angle was measured on the SAURON velocity maps using the
method outlined in Appendix C of Krajnovi\'c et
al. (2006)\nocite{2006MNRAS.366..787K}. Both these position angles, as well as the
kinematic misalignment are tabulated in Table 1 of Paper II, and we
show the histogram of observed kinematic misalignments for fast and
slow rotators in Figure~\ref{fig:hist_eps}. The majority of the fast
rotators have small kinematic misalignments, with 76 per cent having
misalignments smaller than 5$^\circ$. The slow rotators on the other
hand show more kinematic misalignment, with less than half of them (44
per cent) having $\Psi < 5^\circ$.


\section{Intrinsic shape distributions for fast and slow rotators}

Fast and slow rotators are two distinct classes of early-type
galaxies, as was shown by Emsellem et
al. (2007)\nocite{2007MNRAS.379..401E} and Cappellari et
al. (2007)\nocite{2007MNRAS.379..418C}. They defined slow rotators to
have a specific angular momentum $\lambda_R < 0.1$, while fast
rotators in their classification have $\lambda_R > 0.1$. Later, this
classification was refined in Paper III\nocite{2011MNRAS.414..888E},
considering the regularity of the velocity maps (Paper
II\nocite{2011MNRAS.414.2923K}). In the resulting classification, the
separation between slow and fast rotators takes the projected
ellipticity of the systems into account, with slow rotators having
$\lambda_R < 0.31\sqrt{\epsilon}$, and fast rotators $\lambda_R >
0.31\sqrt{\epsilon}$. Figure 6 in Paper
III\nocite{2011MNRAS.414..888E} illustrates that this new division of
the early-type galaxy population into fast and slow rotators nicely
follows the kinematic classification based on the velocity maps. This
figure also shows that $\lambda_R$ is a more reliable separator
between fast and slow rotators than the $V/\sigma_e$ quantity, with
$V$ the velocity amplitude, and $\sigma_e$ the velocity dispersion
measured within one $R_e$. We refer to paper III for more details on
this classification scheme. Important for our analysis is that
the separation of our sample in slow and fast rotators does not
introduce any biases in viewing directions: this is discussed in Paper
III  (see their sections 5.1 and 5.2), but also shown in
simulations performed independently by Jesseit et
al. (2009)\nocite{2009MNRAS.397.1202J} and Bois et al. (2011, Paper
VI)\nocite{2011MNRAS.416.1654B}. In particular, Jesseit et
al. (2009)\nocite{2009MNRAS.397.1202J} perform an extensive study of
variations in $\lambda_R$ with inclination, and find that $\lambda_R$
does not deviate significantly from its maximum value for a large
range of viewing angles. This makes $\lambda_R$ a reliable and robust
estimator of the intrinsic angular momentum. Jesseit et
al. (2009)\nocite{2009MNRAS.397.1202J} quote a confusion probabiliy of
4.6 per cent of mistakingly classifying a fast rotating galaxy as a
slow rotator. They add that this probability will be even lower in
practise, as their simulated merger sample has a significantly larger
number of prolate shaped galaxies than observed in galaxy surveys, and
most of the wrongly classified galaxies in their sample fall into this
category. In our \atlas sample we only have two clear examples
  of prolate galaxies: one of them is classified as a fast rotator,
  but both have non-regular rotation (Paper II). We therefore are
  confident that any contaminations in our sample due to
  misclassification of fast and slow rotators is negligible for our
  intended purposes.

Based on papers II\nocite{2011MNRAS.414.2923K} and
III\nocite{2011MNRAS.414..888E} fast rotators are galaxies with
regular, aligned velocity fields that often possess discs and bars,
while slow rotators are often kinematically misaligned, have
kinematically distinct cores (KDCs) and are located on the more
massive end of the luminosity function. In addition, Cappellari et
al. (2011b, hereafter Paper VII)\nocite{2011MNRAS.416.1680C} show that
slow rotators are predominantly found in the high-density environment,
which for our sample is the core of the Virgo cluster, and are almost
non-existent in the field.

These all are hints that fast and slow rotators have different
formation scenarios (see also Paper VI). It is therefore unlikely that
these two classes of objects have a similar shape distribution, and
indeed a simple Kolmogorov-Smirnov test confirms at the 5 per cent
significance level that the ellipticity distributions of the fast and
slow rotators in our sample are not drawn from the same underlying
distribution ($p_{\mathrm{KS}}=3\times10^{-5}$, see also
Figure~\ref{fig:hist_eps}). A Mann-Whitney U-test also rejects the
notion that the two ellipticity distributions have the same mean
($p_\mathrm{{MW}} = 1.7\times10^{-4}$). We therefore consider the fast
and slow rotators separately, when inverting their shape
distributions. As explained below, we assume an axisymmetric
underlying shape distribution in this section, and we will explore
deviations from this axisymmetric assumption later on in this paper in
\S~\ref{sec:triax}.

\subsection{Intrinsic and observed shape distributions}

\begin{figure}
\begin{center} 
\begin{tabular}{|l|l|}

\psfig{figure=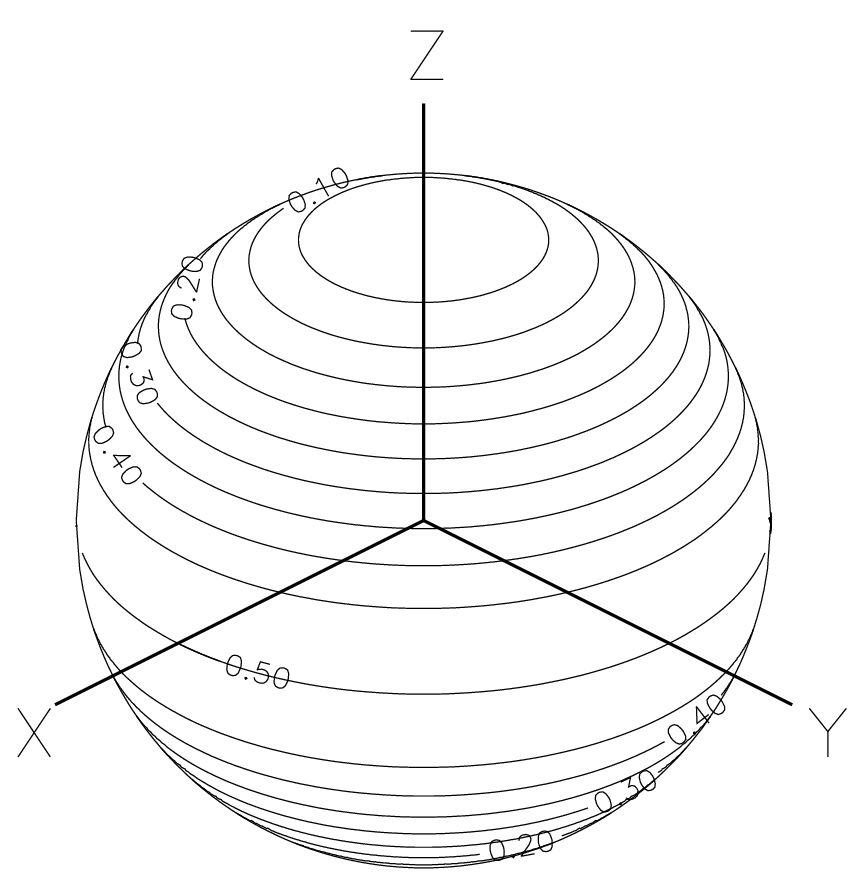,width=4cm} & 
\psfig{figure=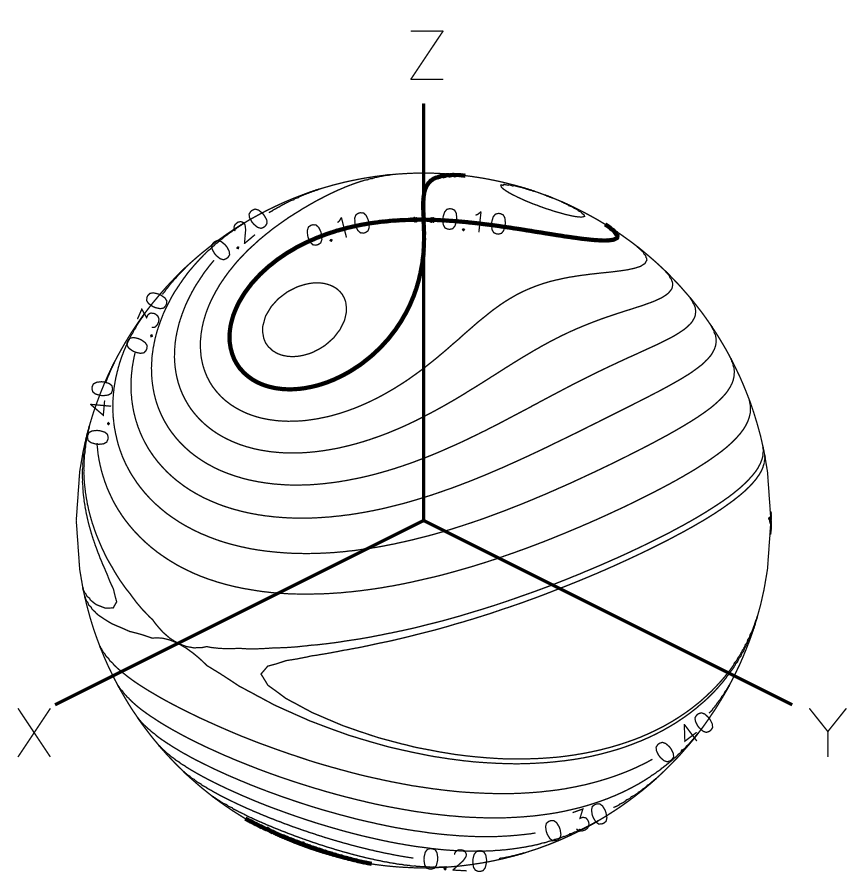,width=4cm}

\end{tabular}
\end{center}
\caption{Contours of constant ellipticity on the sphere of viewing
  angles, for an oblate galaxy ($p=1$ and $q=0.5$, left) and a triaxial galaxy
  ($p=0.9$ and $q=0.5$, right). The ellipticity varies between 0 and $1-q$.}
\label{fig:sphere_eps}
\end{figure}

The intrinsic shape of a galaxy can be modeled as an ellipsoid, with
intrinsic axis ratios $p$ and $q$, such that $1 \ge p \ge q \ge
0$. The observed shape or ellipticity $\epsilon$ of a galaxy then depends on its intrinsic shape
and on the viewing angles (inclination
$\vartheta$ and azimuthal angle $\varphi$), such that $\epsilon =
\epsilon(p, q, \vartheta, \varphi)$, see Figure~\ref{fig:sphere_eps} for
an example of ellipticity plotted on the sphere of viewing angles. It is therefore impossible to
deduce the intrinsic shape ($p,q$) for an individual galaxy, based on
its observed ellipticity only. Early work (e.g. Hubble
1926\nocite{1926ApJ....64..321H}; Sandage et al. 1970\nocite{1970ApJ...160..831S}) therefore used the observed
distribution $F(\epsilon)$ for a sample of galaxies, assuming that the
galaxies were axisymmetric ($p=1$ for oblate galaxies, $p=q$ for
prolate galaxies) to determine the intrinsic shape distribution
$f(q)$. This distribution is then uniquely determined, assuming that
the galaxies are oriented randomly in space (random viewing
angles). 

For triaxial galaxies ($p \ne 1$) this is no longer the case, as
$F(\epsilon)$ cannot uniquely determine $f(p,q)$ (e.g. Binggeli
1980\nocite{1980A&A....82..289B}; Binney \& de Vaucouleurs
1981\nocite{1981MNRAS.194..679B}). Binney
(1985)\nocite{1985MNRAS.212..767B} and subsequently Franx et
al. (1991)\nocite{1991ApJ...383..112F} showed that progress could be
made by use of the kinematic information of the galaxies, namely by
incorporating the kinematics misalignment angle ($\Psi$) between the observed minor axis
and the projected rotation axis in the probability
distribution\footnote{We present a shape analysis based on
    kinematic misalignment in Appendix B.}. $F(\Psi, \epsilon)$ is however also not able to uniquely
define $f(p,q)$, as $\Psi$ also depends on the intrinsic rotation
axis, which for a triaxial galaxy can lie anywhere in the plane
containing the short and long axis of the galaxy (see for example
Franx et al. 1991\nocite{1991ApJ...383..112F}). In an oblate galaxy,
however, the rotation axis coincides with the short axis of the
system, and no kinematic misalignment will be observed. 

Our integral-field observations show that the fast rotators in our
sample have zero or at most very small misalignments, and for this
reason we first assume that the fast rotators are exactly oblate. This
assumption allows us to invert the observed ellipticity distribution
$F(\epsilon)$ to obtain the distribution of intrinsic flattening
$f(q)$ of the fast rotators. For this inversion we use Lucy's method
(1974)\nocite{1974AJ.....79..745L}, which is an iterative technique to
solve for the underlying distribution function. We relax this
assumption of oblateness in \S~\ref{sec:triax}, where we put an
upper limit on the deviations from perfectly oblate shapes, using the
observed kinematic misalignments as an extra constraint. Note that in
the studies mentioned above, and in the analysis we present in this
paper, galaxies are approximated by triaxial spheroids, while in
reality many of them consist of separate bulge and disc
components. By measuring our ellipticities at large radius, we assume
that for disc-dominated galaxies we can ignore any bulge (and bar)
contributions, and that we are mostly probing the outer disc, while
for bulge-dominated galaxies we will be mostly sensitive to the shape
of the spheroid.

\subsection{The intrinsic shapes of fast rotators}
\label{sec:fast}

\begin{figure}
\begin{tabular}{c}

\psfig{figure=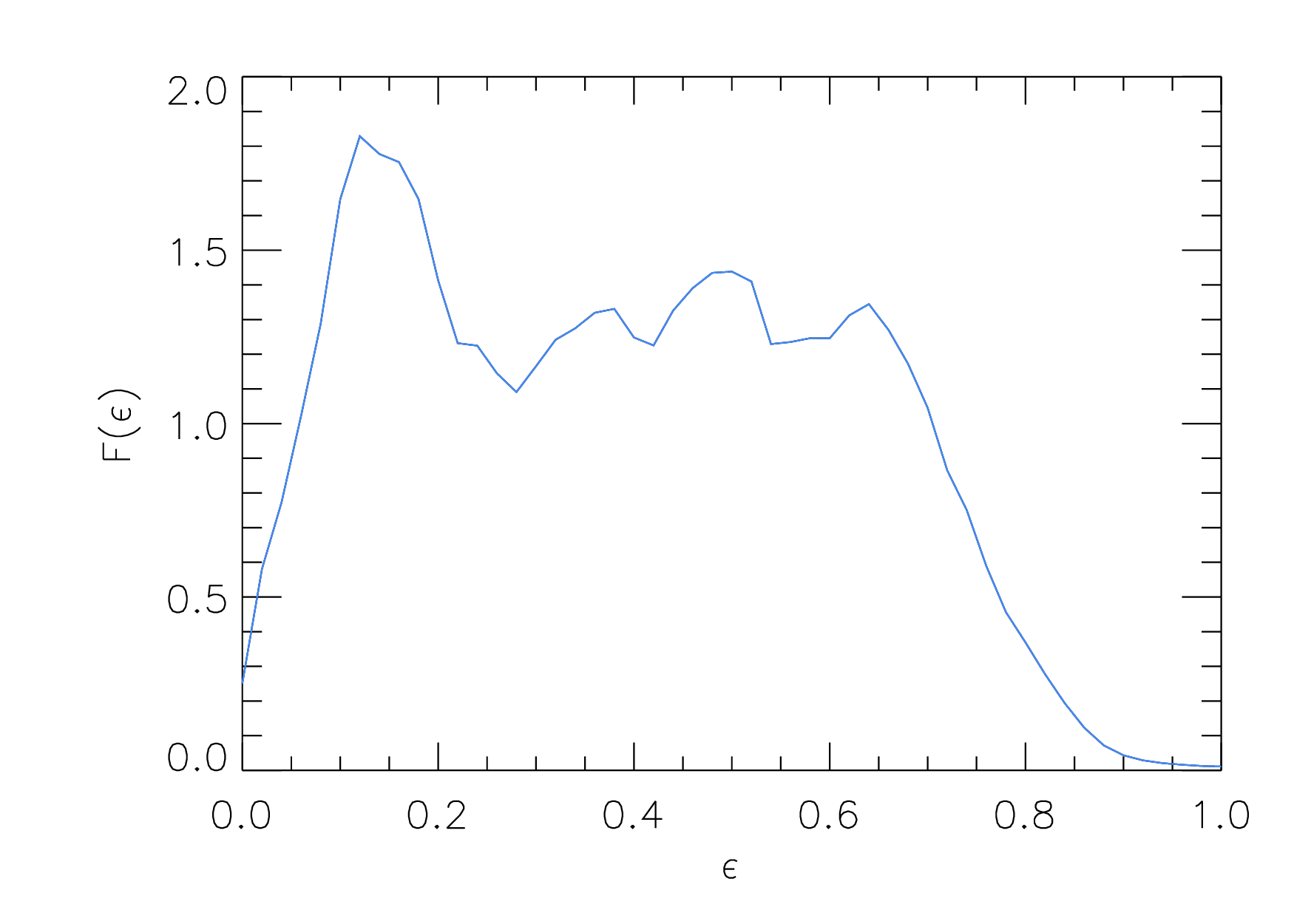,width=8cm} \\
\psfig{figure=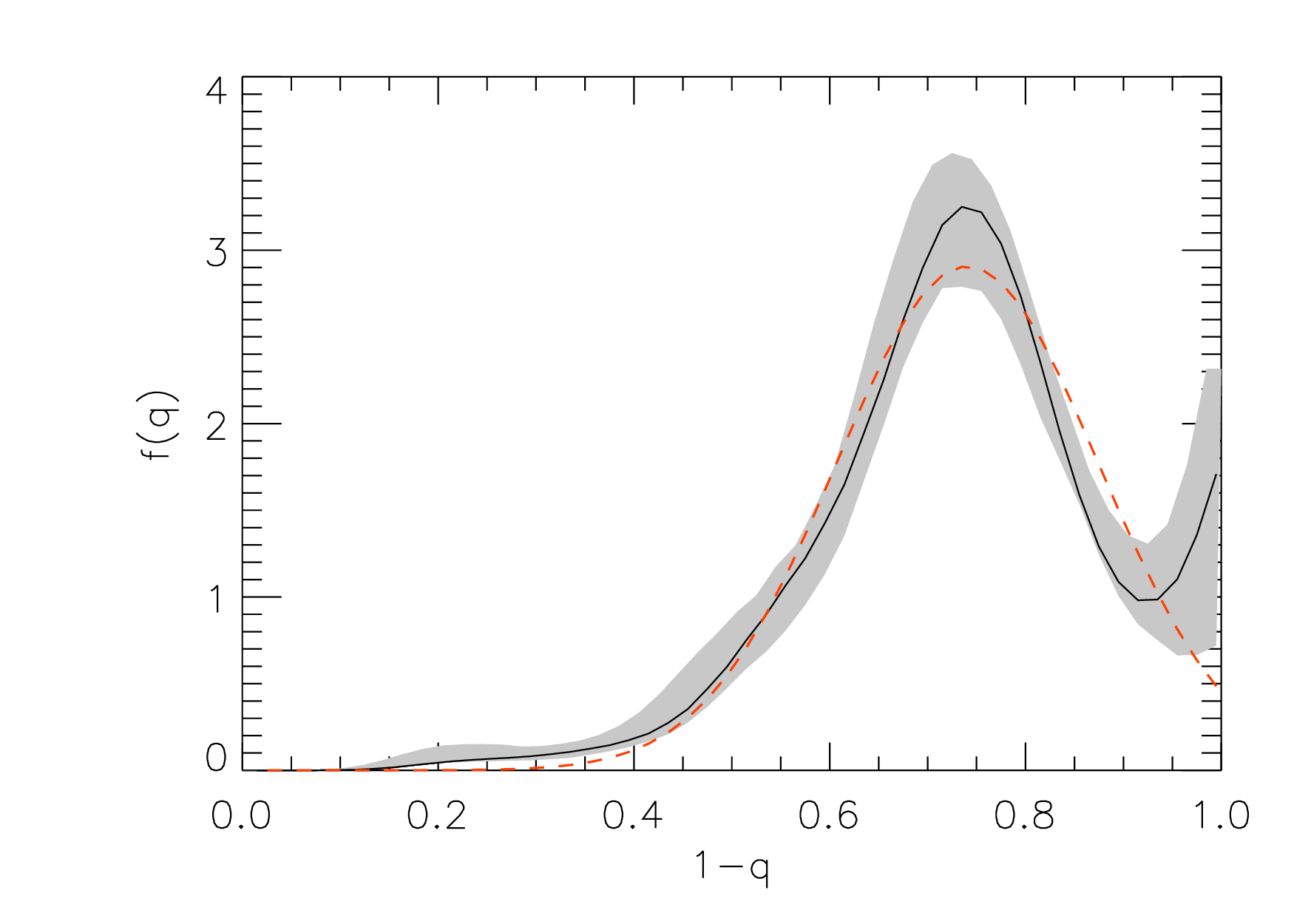,width=8cm} \\

\end{tabular}
\caption{Top panel: observed distribution $F(\epsilon)$ for the 224 fast
  rotators in the \atlas sample, obtained by approximating each galaxy
as a Gaussian function with mean given by its measured ellipticity,
and width (standard-deviation) by its 1-$\sigma$ measurement
error. Some mild smoothing is applied. Lower panel: the inverted intrinsic shape distribution $f(q)$
for the fast rotators, shown by the black solid line. We plot $1-q$ on
the horizontal axis such that round objects are on the left and
flattened ones on the right, to be consistent with our observed ellipticity
plots. The red solid line shows a Gaussian fit to the intrinsic shape
distribution. The intrinsic shape distribution peaks around $q=0.25$, but
has an extended tail towards rounder shapes. The grey area shows the
area enclosing 95 per cent of inversions for our Monte Carlo
simulations (see text for details).}
\label{fig:lucy_obl_fast}
\end{figure}

We first consider our fast rotators to be oblate systems, as
 validated by their small kinematic misalignment.
For oblate galaxies,
observed ellipticity is a function of intrinsic flattening $q$ and
inclination $\vartheta$ only:

\begin{equation}
e = (1-\epsilon)^2 = \cos^2\vartheta + q^2\sin^2\vartheta,
\label{eq:eps_obl}
\end{equation}

\noindent
with $e$ the eccentricity, introduced here to simplify some of our
notations. Assuming random orientations, integrating $\vartheta$ over the
sphere of viewing angles then yields a probability function
$P(\epsilon|q)$ such that:

\begin{equation}
P(\epsilon | q) = \frac{\sqrt{e}}{\sqrt{1-q^2}\sqrt{e-q^2}}.
\label{eq:prob_obl_eps}
\end{equation}

\noindent
With Lucy's method (1974)\nocite{1974AJ.....79..745L} we solve for the intrinsic shape distribution
$f(q)$:

\begin{equation}
F(\epsilon) = \int f(q) P(\epsilon|q) dq.
\label{eq:lucy_obl_eps}
\end{equation}

\noindent
For the observed distribution $F(\epsilon)$ we approximate each fast
rotator galaxy with a Gaussian distribution function, centred at its measured ellipticity, with
a standard deviation given by its measurement error. $F(\epsilon)$ is
then the superposition of these 224 Gaussian functions (one for each fast rotator), see
top panel of Figure~\ref{fig:lucy_obl_fast}. We applied some mild
smoothing with a boxcar before inverting this curve.  We checked that Lucy's method converges within 25 iterations,
and the resulting inverted distribution $f(q)$ is shown in the
lower panel of Figure~\ref{fig:lucy_obl_fast}, as the black
solid line. The intrinsic flattening distribution $f(q)$ can be approximated
by a Gaussian function (red solid line), with mean $\mu_q = 0.26$ and
standard-deviation $\sigma_q = 0.13$. Interestingly, this mean value is
very similar to the intrinsic flattenings found in similar studies for
spiral galaxies (e.g. Lambas et al. 1992\nocite{1992MNRAS.258..404L}; Ryden
2006\nocite{2006ApJ...641..773R}; Padilla \& Strauss 2008\nocite{2008MNRAS.388.1321P}),
and we will discuss this in more detail in \S~\ref{sec:spiral}. 

Although our inversion does technically take the measurement errors of
our observed ellipticities into account by approximating each
measurement as a Gaussian, we should ask ourselves how sensitive our
inversion is to small deviations in the so obtained observed
distribution $F(q)$. We therefore repeated our inversion another 100
times with a Monte Carlo simulation: we again approximated our
observed ellipticities with a Gaussian function and applied some mild
smoothing, but for its mean we drew from a Gaussian distribution,
centred on the observed ellipticity and with a standard deviation
given by the measurement error. We show the central 95 per cent of the
resulting inversion curves $f(q)$ in Figure~\ref{fig:lucy_obl_fast} with
the grey shaded area. This figure shows that our inversion is fairly
robust: we fitted Gaussians to all of $f(q)$ resulting from this Monte
Carlo exercise, and found that the best fitting Gaussian of the
overall intrinsic shape distribution parametrized with mean $\mu_q =
0.25 \pm 0.01$ and $\sigma_q = 0.14 \pm 0.02$. 

\begin{figure}
\begin{tabular}{c}

\psfig{figure=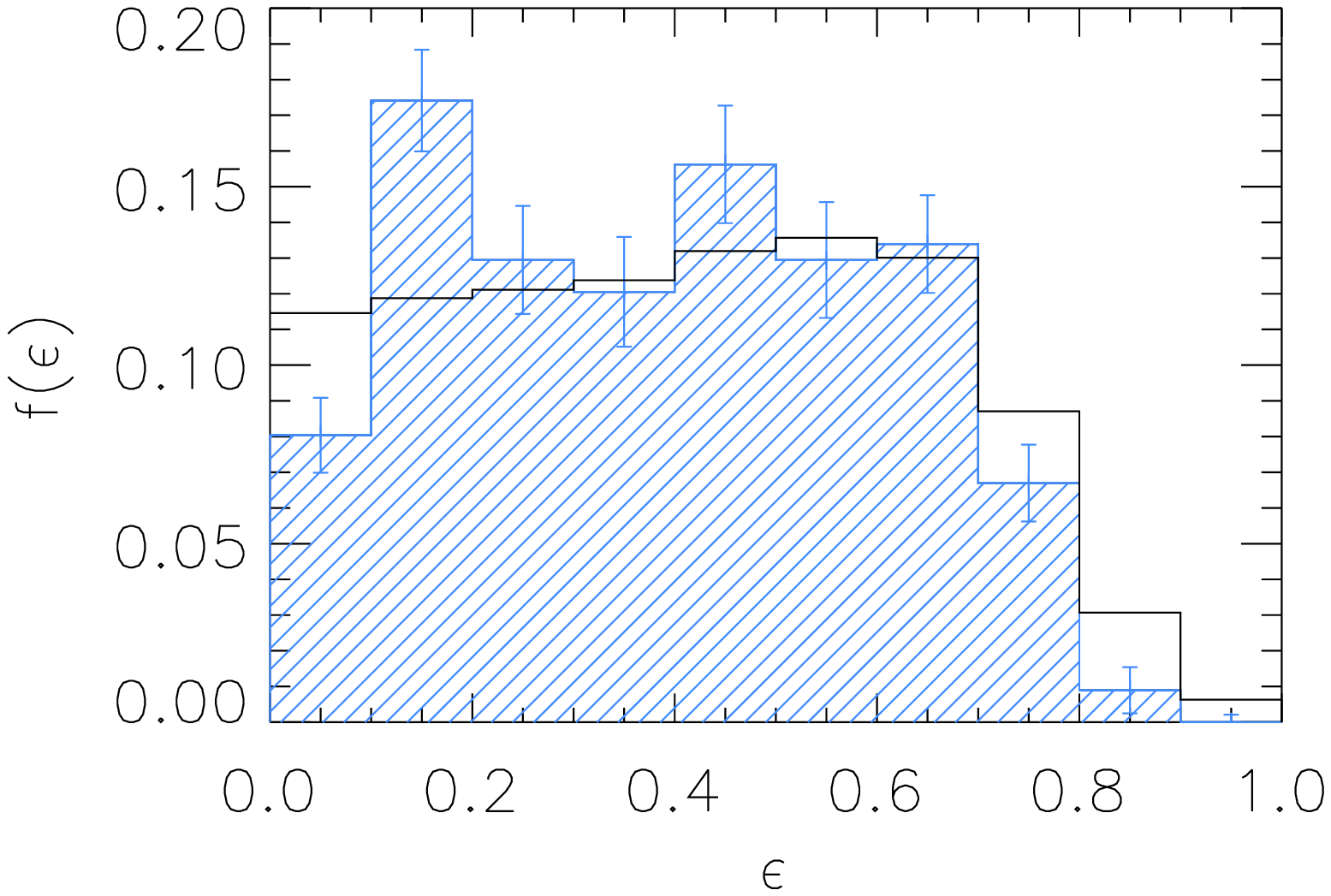,width=8cm} \\
\psfig{figure=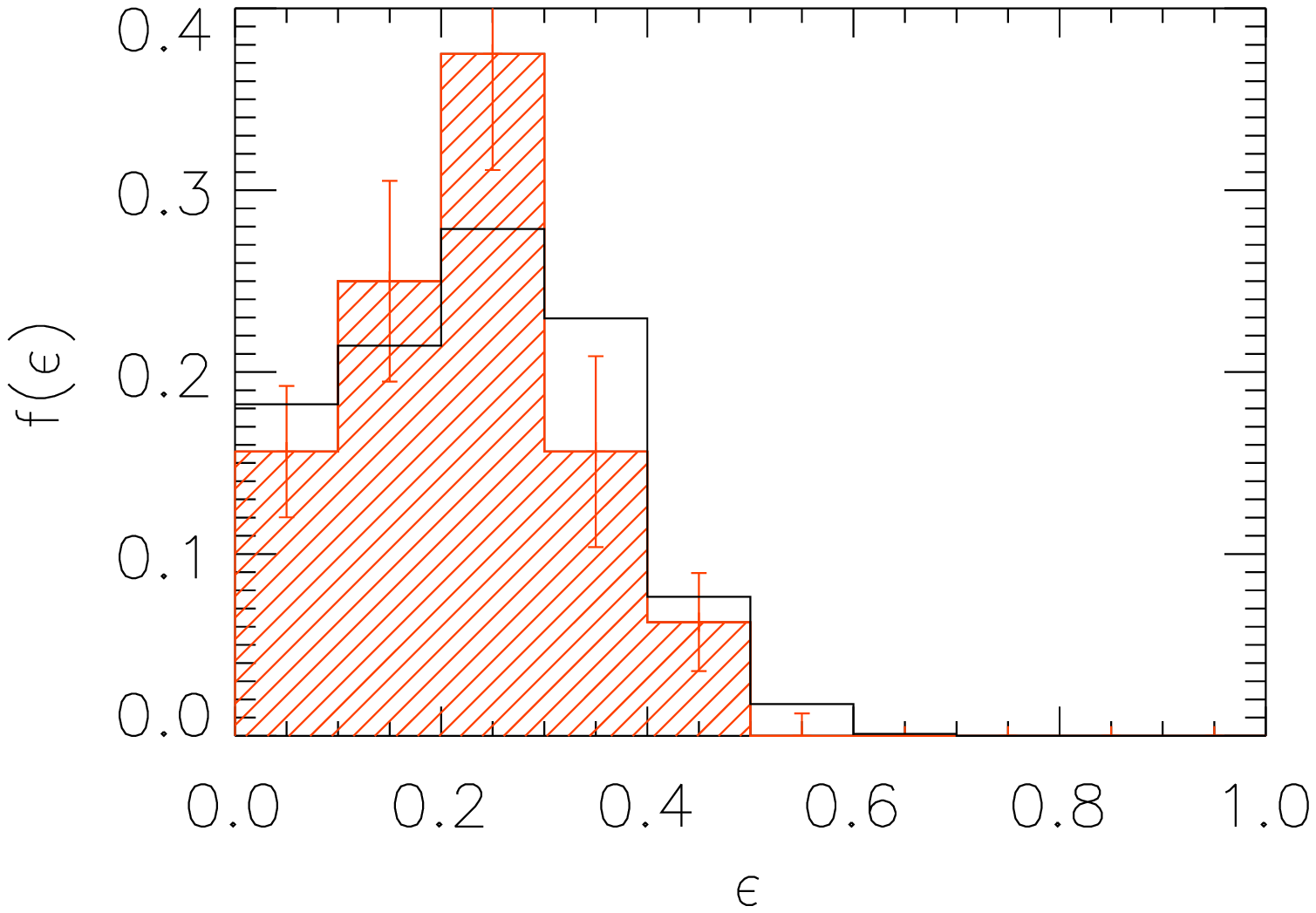,width=8cm} \\

\end{tabular}
\caption{Top panel: observed ellipticity distribution of the fast
  rotators in the \atlas sample (blue dashed histogram), compared to a mock ellipticity
  distribution of $10^6$ galaxies (black open histogram), drawn from the intrinsic shape
  distribution $f(q)$ with random viewing angles. Bottom panel: same
  as above, but now for the slow rotators (red dashed histogram).}
\label{fig:histcompare_fast}
\end{figure}

In the top panel of Figure~\ref{fig:histcompare_fast} we compare the
predicted ellipticity distribution from our model with our
observations, by generating a mock sample of $10^6$ galaxies, drawn
from the intrinsic shape distribution $f(q)$. The predicted
ellipticity distribution does deviate somewhat from our observed
distribution, but a one-sided KS-test shows that these deviations are
not significant ($p_\mathrm{KS}=0.19$) given the relatively small
sample size of our observed sample.

\begin{figure*}
\begin{tabular}{cc}
\psfig{figure=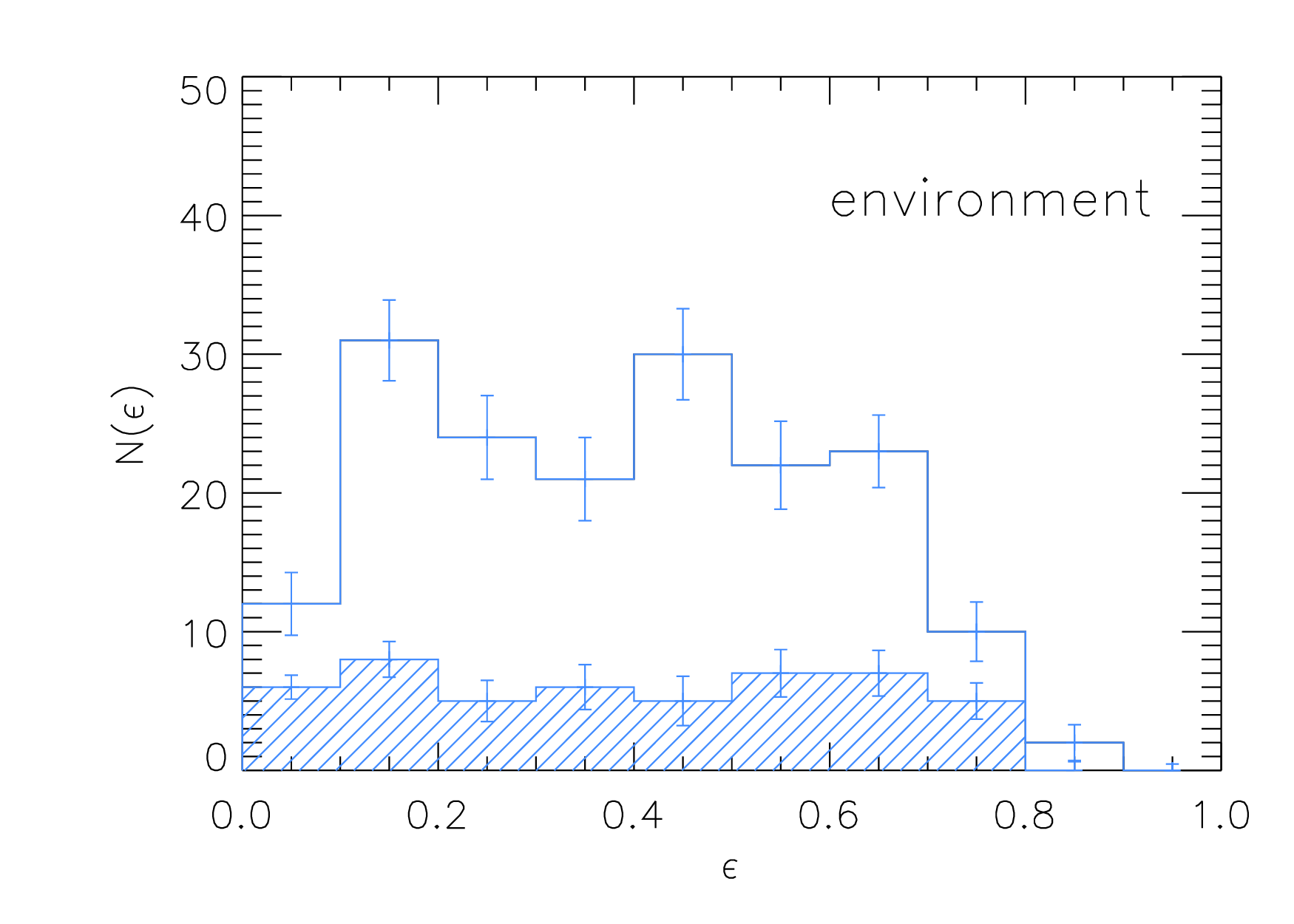,width=8cm} & 
\psfig{figure=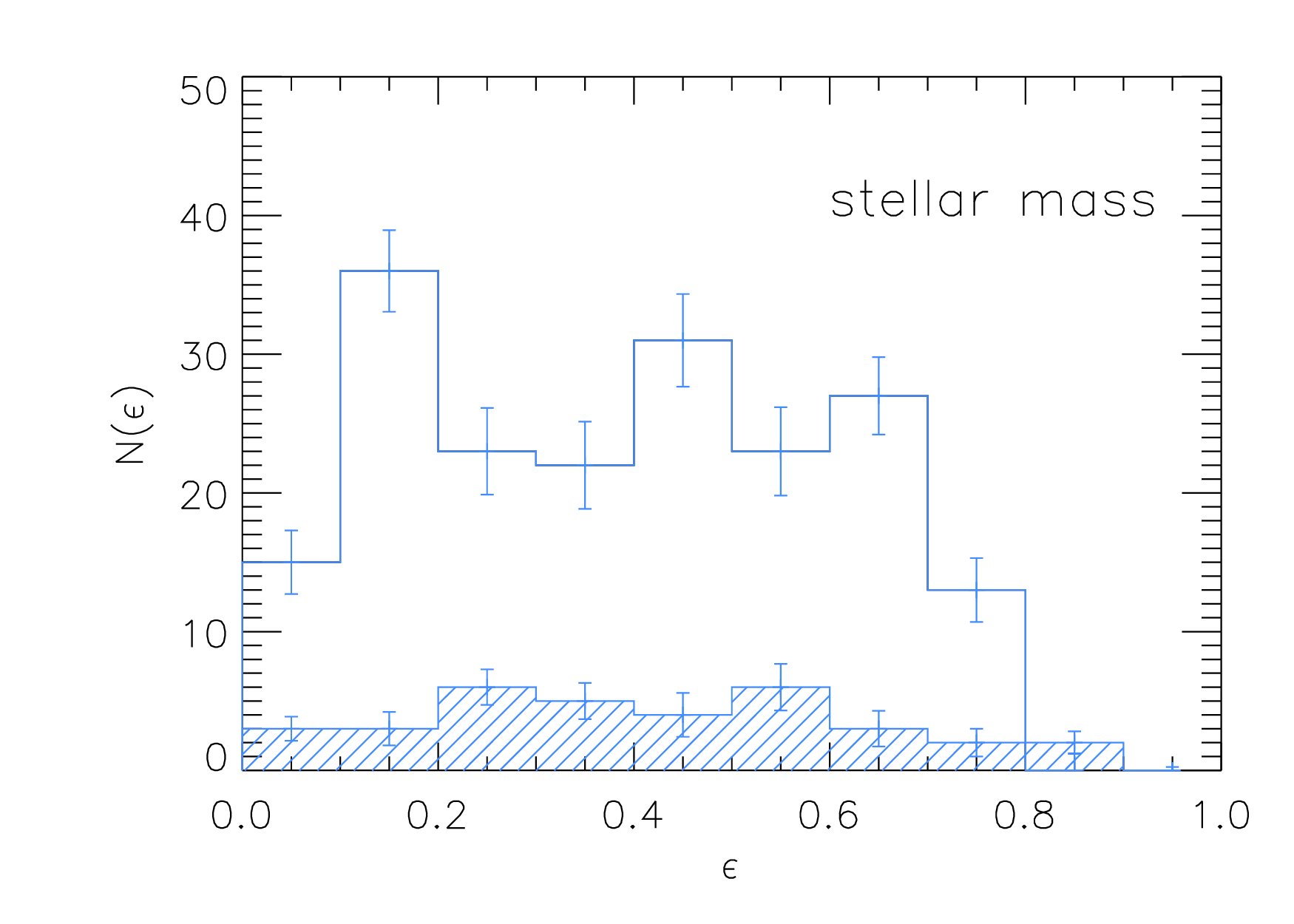,width=8cm} \\ 

\end{tabular}
\caption{Histogram of observed ellipticities for fast rotators,
  divided based on environment (left) and mass $M_\mathrm{JAM} $ (right). The left
  plot shows the 49 fast rotators in Virgo (dashed histogram) versus the
  175 field fast rotators (open histogram) for our sample. The right
  plot shows the fast rotators with $M_\mathrm{JAM}  > 10^{11} M_\odot$ (34
  galaxies, dashed
  histogram) versus the lower mass fast rotators with $M_\mathrm{JAM}  < 10^{11}
  M_\odot$ (190 galaxies, open histogram). The 1-$\sigma$ errorbars are based
on Monte Carlo simulations, taking the individual measurement errors
for each galaxy into account.  The division in environment and
in stellar mass do not result in statistically significant different
shape distributions. See text for
details.}
\label{fig:hist_eps_env}
\end{figure*}

We also investigated whether we could find differences in intrinsic
shape distributions based on environment. Since different formation
processes are at play in clusters than in the field (see e.g. Blanton
\& Moustakas 2009\nocite{2009ARA&A..47..159B} for a review), we
could expect that therefore the intrinsic shape distribution of fast
rotators in the Virgo cluster would be different from that for fast
rotators in less dense environments. We did however not detect any
significant deviations in shape distribution between these two sets of
galaxies, as could indeed already have been inferred from a
Kolmogorov-Smirnov test on the observed ellipticity distributions. The
hypothesis that the ellipticity distributions of both field and Virgo
fast rotators are drawn from different underlying distributions is
rejected at the 5 per cent significance level with $p_{\mathrm{KS}} = 0.96$, while also the Mann-Whitney
U-test rejects the hypothesis of different means for the distributions
with $p_{\mathrm{MW}} = 0.46$, at the same significance level (see left panel of
Figure~\ref{fig:hist_eps_env}). Similarly, we also did not find any
differences in shape distributions and means of distributions if we divide our sample based on
mass\footnote{Mass was taken from Paper XV as
  $M_\mathrm{JAM} = L \times (M/L)_e \approx 2 \times M_{1/2}$, with
  $(M/L)_e$ the total mass-to-light
  ratio measured within one half-light radius $R_e$, with self-consistent
  Jeans Anisotropic modelling, and $M_{1/2}$ the total mass within a
  sphere of radius $R_e$, enclosing half of the galaxy light. The contribution of dark matter to $(M/L)_e$ within one
  $R_e$ is small (see Paper XV for details), so $M_\mathrm{JAM}$ can be
  interpretated as a dynamical estimate of stellar mass. Throughout this paper,
we will therefore refer to $M_\mathrm{JAM}$ as a stellar mass estimate.} ($M_\mathrm{JAM} <
10^{11} M_\odot$ versus $M_\mathrm{JAM}  > 10^{11} M_\odot$), with
$p_\mathrm{{KS}}=0.79$ and $p_{\mathrm{MW}} = 0.29$ (right panel Figure~\ref{fig:hist_eps_env}). These
masses were determined based on dynamical  modeling, and the values for individual
galaxies are listed in Cappellari et al. (2013a, paper
XV)\nocite{2013MNRAS.432.1862C}.

\subsection{The intrinsic shapes of slow rotators}
\label{sec:slow}

\begin{figure*}
\begin{tabular}{cc}

\psfig{figure=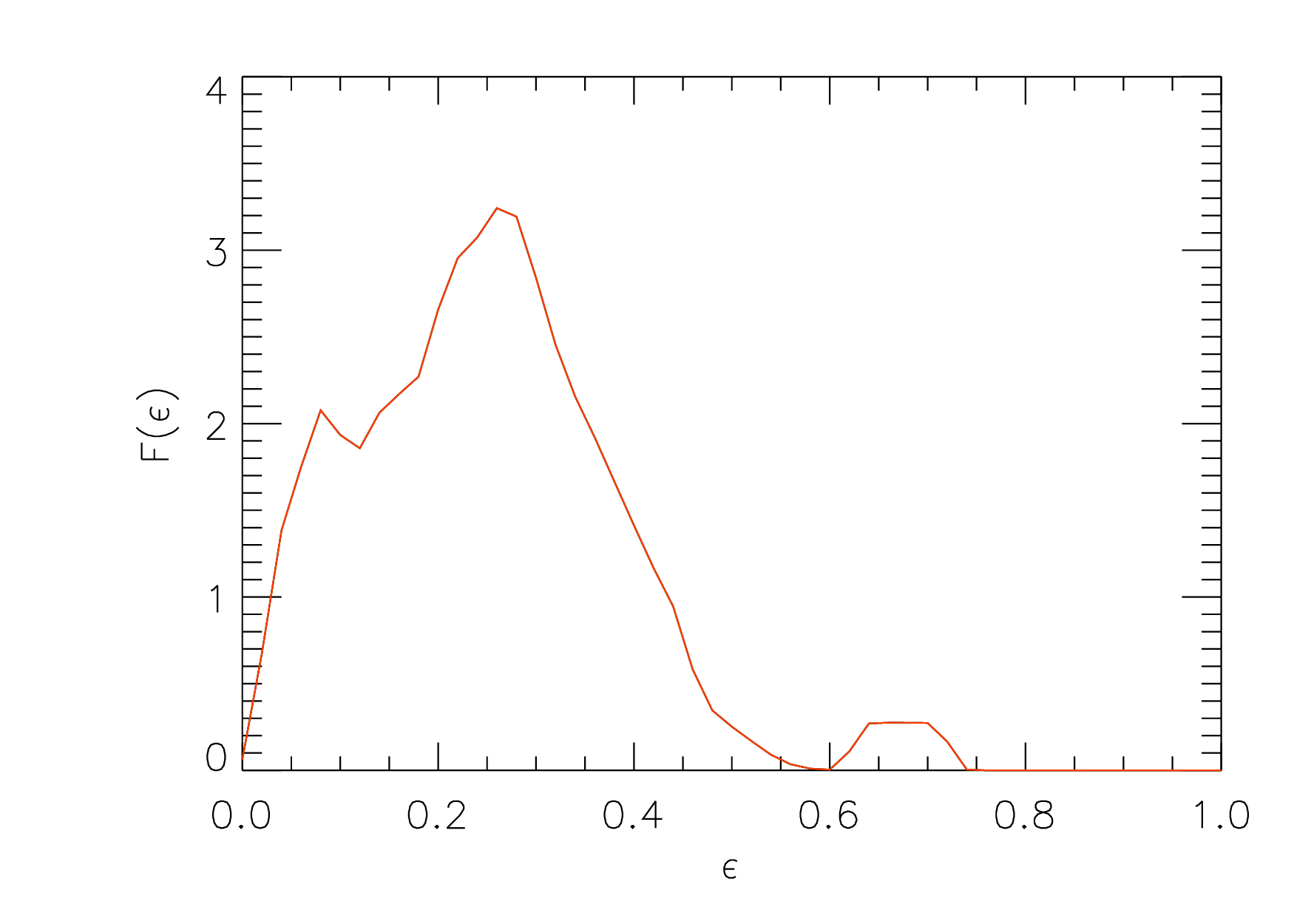,width=8cm} &
\psfig{figure=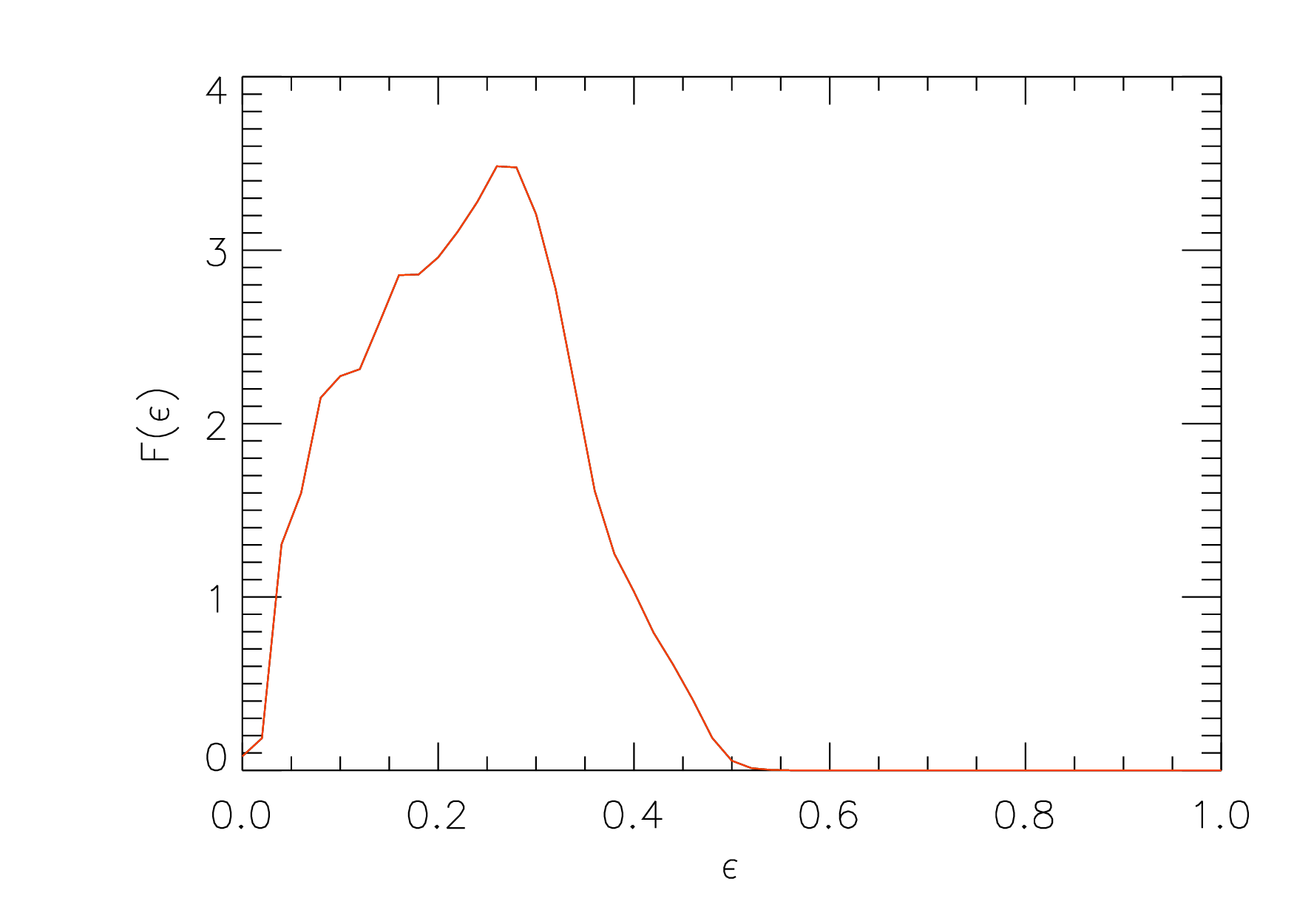,width=8cm}\\
\psfig{figure=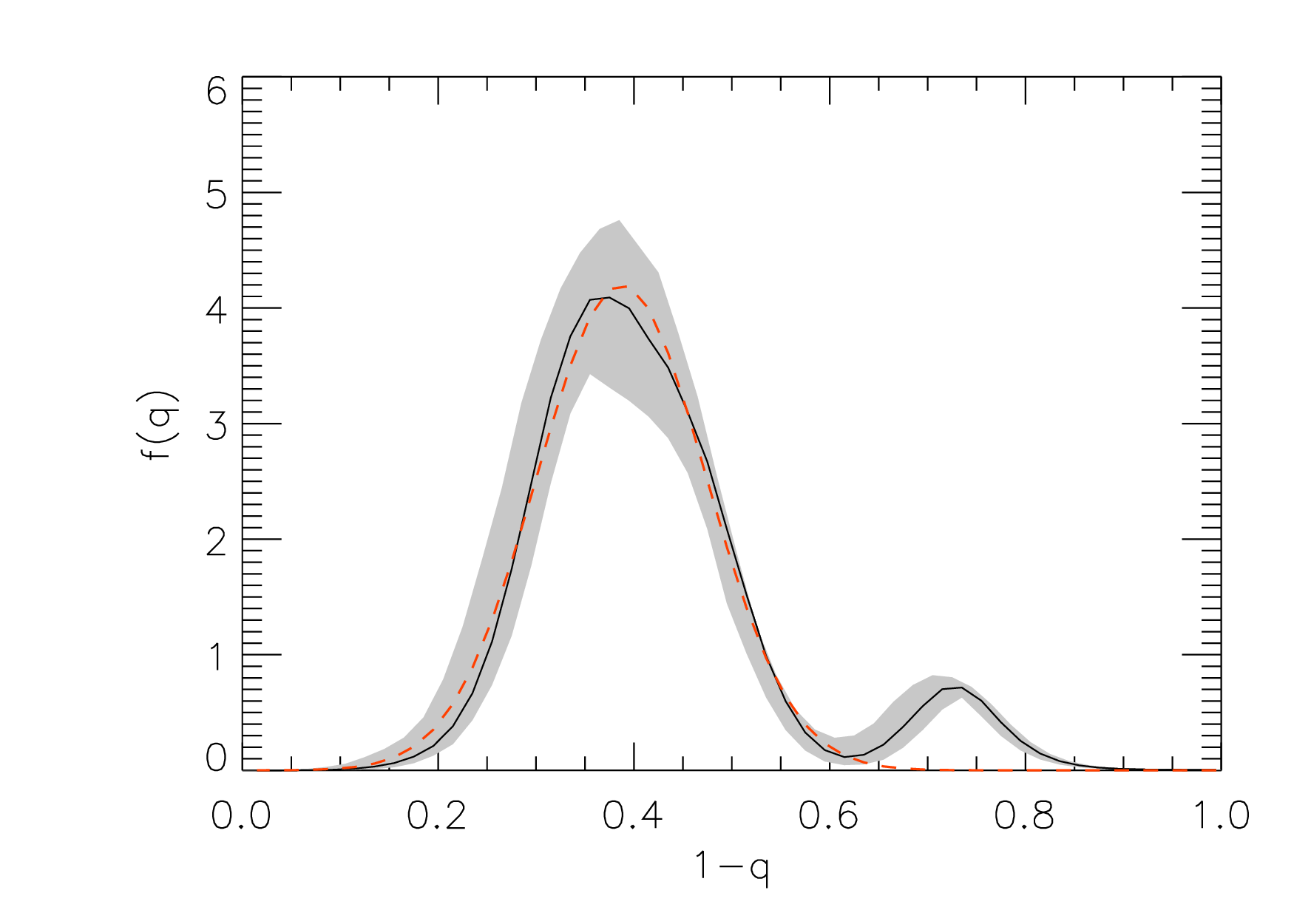,width=8cm}&
\psfig{figure=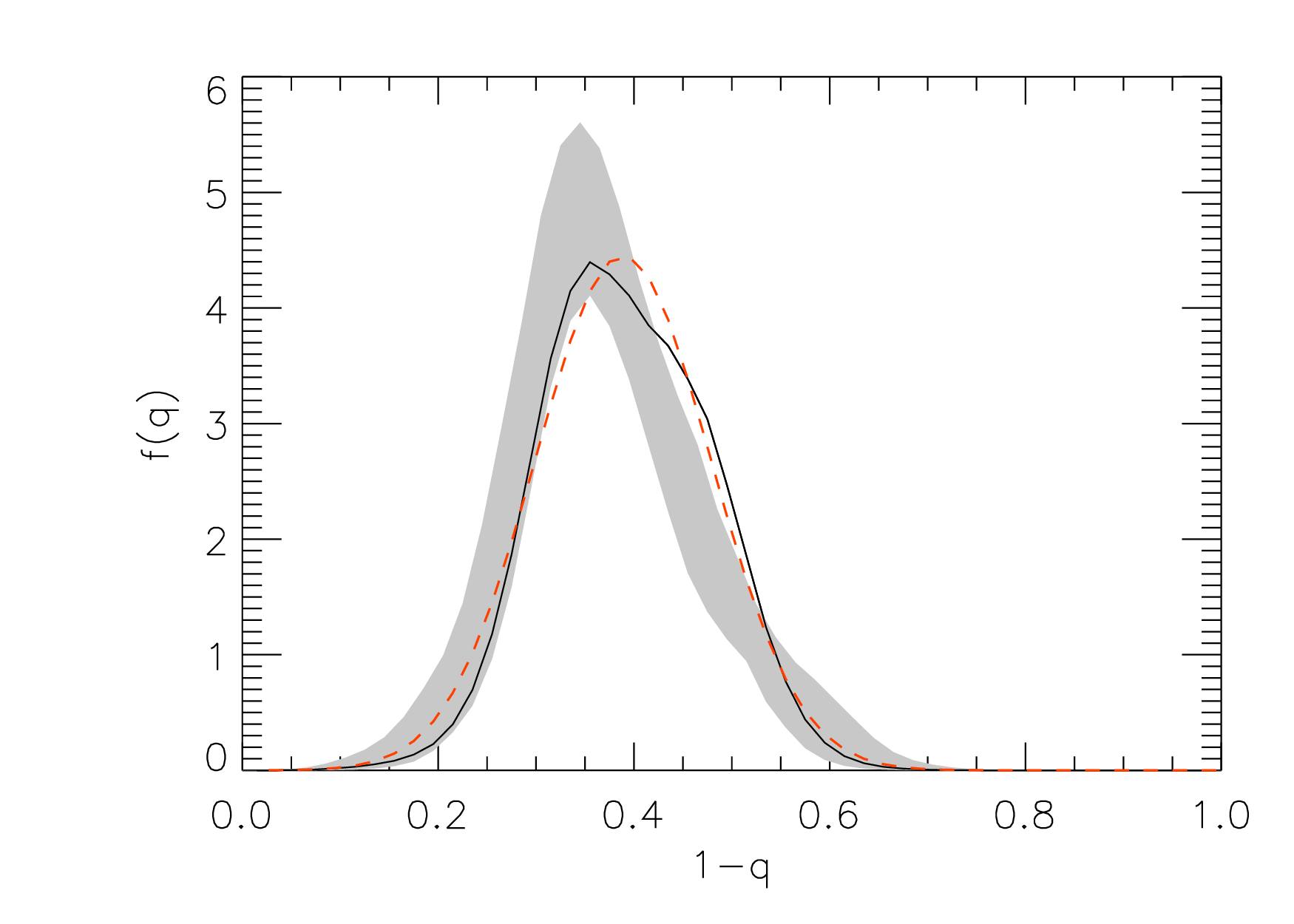,width=8cm}\\

\end{tabular}
\caption{Top left panel: observed distribution $F(\epsilon)$ for the 36
  slow rotators, similar to top panel of Figure~\ref{fig:lucy_obl_fast}. Bottom left panel: inverted shape distribution $f(q)$ for the 36 slow
rotators in our sample (solid black line). The red dashed line shows a Gaussian fit to the distribution, and the grey area indicates a
95 per cent spread around our Monte-Carlo simulations (see text for
more detail). Right panels: same as
left panels, but now the four $2\sigma$-galaxies with counter-rotating
discs have been removed from the slow rotator sample, resulting in a
cleaner and overall slightly rounder shape distribution. We use
  this instrinsic distribution for our subsequent discussions and analysis.}
\label{fig:lucy_obl_slow}
\end{figure*}

The slow rotators in our sample show clear signs of triaxiality, such
as kinematic misalignment. For the moment however we approximate these
systems as oblate, so that we can invert their observed ellipticity
distribution to obtain an estimate of their intrinsic flattening.

Following the same technique as described for the fast rotators, we
then arrive at the intrinsic shape distribution shown in the
bottom-left panel of Figure~\ref{fig:lucy_obl_slow}. The distribution
is clearly double peaked, with the larger peak being well approximated
with a Gaussian centred at $\mu_q = 0.61$ with $\sigma_q= 0.09$. The
smaller peak around $q=0.3$ coincides with the shape distribution of
the fast rotators. It therefore looks like our sample of slow rotators
consists of two populations, with the majority being roundish objects,
supplemented with a second smaller population of more flattened
galaxies. Indeed, 4 of our 36 slow rotators are flattened,
counter-rotating disk galaxies (so-called 2$\sigma$-galaxies
exhibiting a double peaked profile in velocity dispersion, see Paper
II\nocite{2011MNRAS.414.2923K} for details). These are NGC3796,
NGC4191, NGC4528 and NGC4550 with the latter the most extreme case
with $\epsilon = 0.68$. Removing these galaxies from our slow rotator
sample did not change the larger peak significantly (the best-fit
Gaussian remained the same), but did remove the secondary peak. In
fact, removing just NGC4550 from the slow rotator sample resulted in
the disappearance of the secondary peak altogether, showing the
sensitivity of our inversion method. Though the parameters of the
best-fit Gaussian remain the same when removing the
$2\sigma$-galaxies, most of the inverted distributions $f(q)$ from the
Monte Carlo simulations that define the grey 95 per cent area in the
lower right panel of Figure~\ref{fig:lucy_obl_slow} are shifted
towards rounder shapes: the Gaussians fit to these Monte-Carlo
inversions are $\mu_q = 0.63 \pm 0.01$ and $\sigma_q = 0.09 \pm 0.01$.

As for the fast rotators, we compare the ellipticity distribution of a
mock galaxy sample drawn from the intrinsic distribution $f(q)$
derived above, to the observed ellipticities in the \atlas sample. The
results are shown in the bottom panel of
Figure~\ref{fig:histcompare_fast}. A one-sided KS-test indicates that
we can indeed accept the hypothesis that the observed distribution
($p_\mathrm{KS} = 0.29$) was drawn from the proposed intrinsic
distribution.

\begin{figure*}
\begin{tabular}{cc}

\psfig{figure=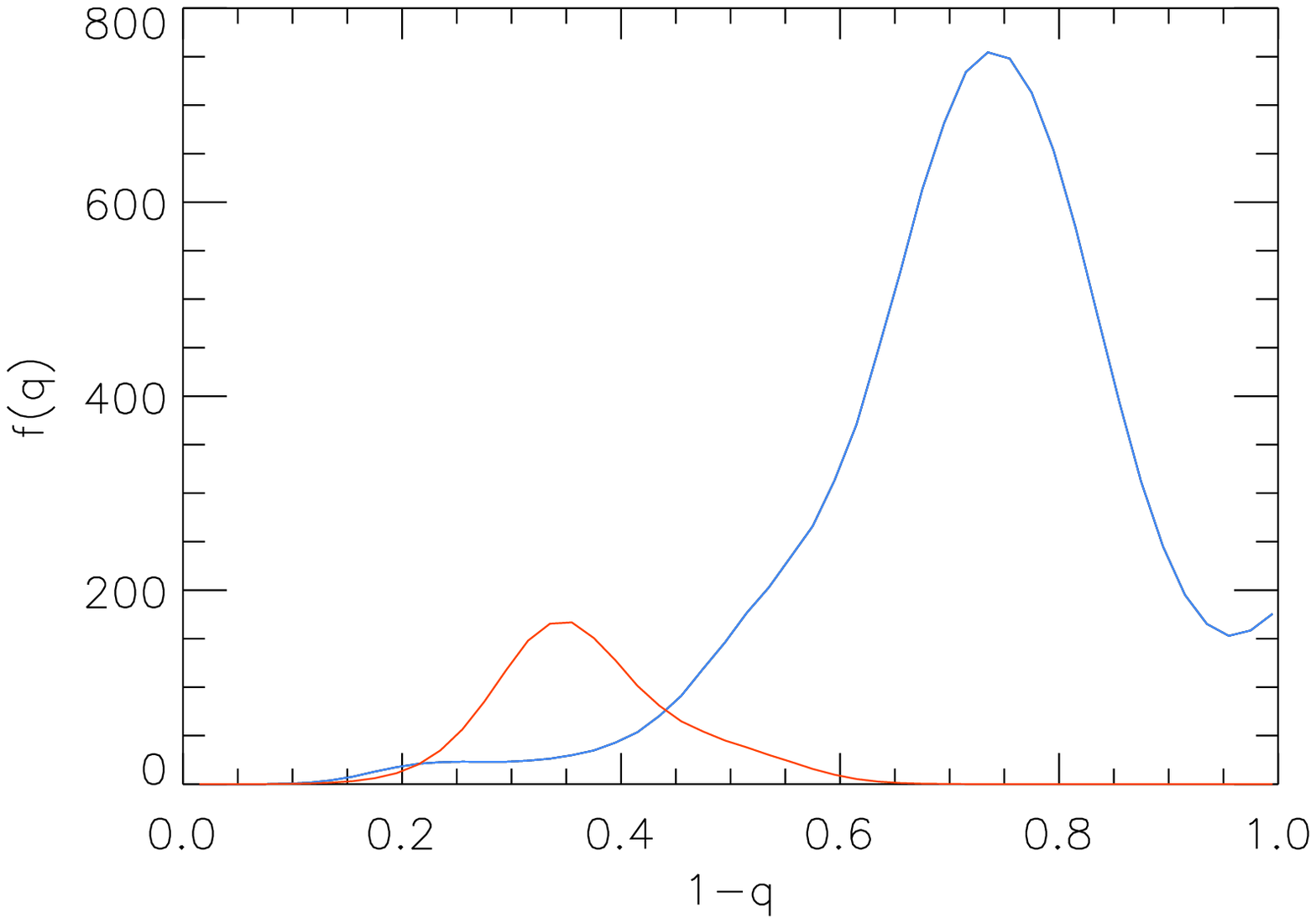,width=8cm} &
\psfig{figure=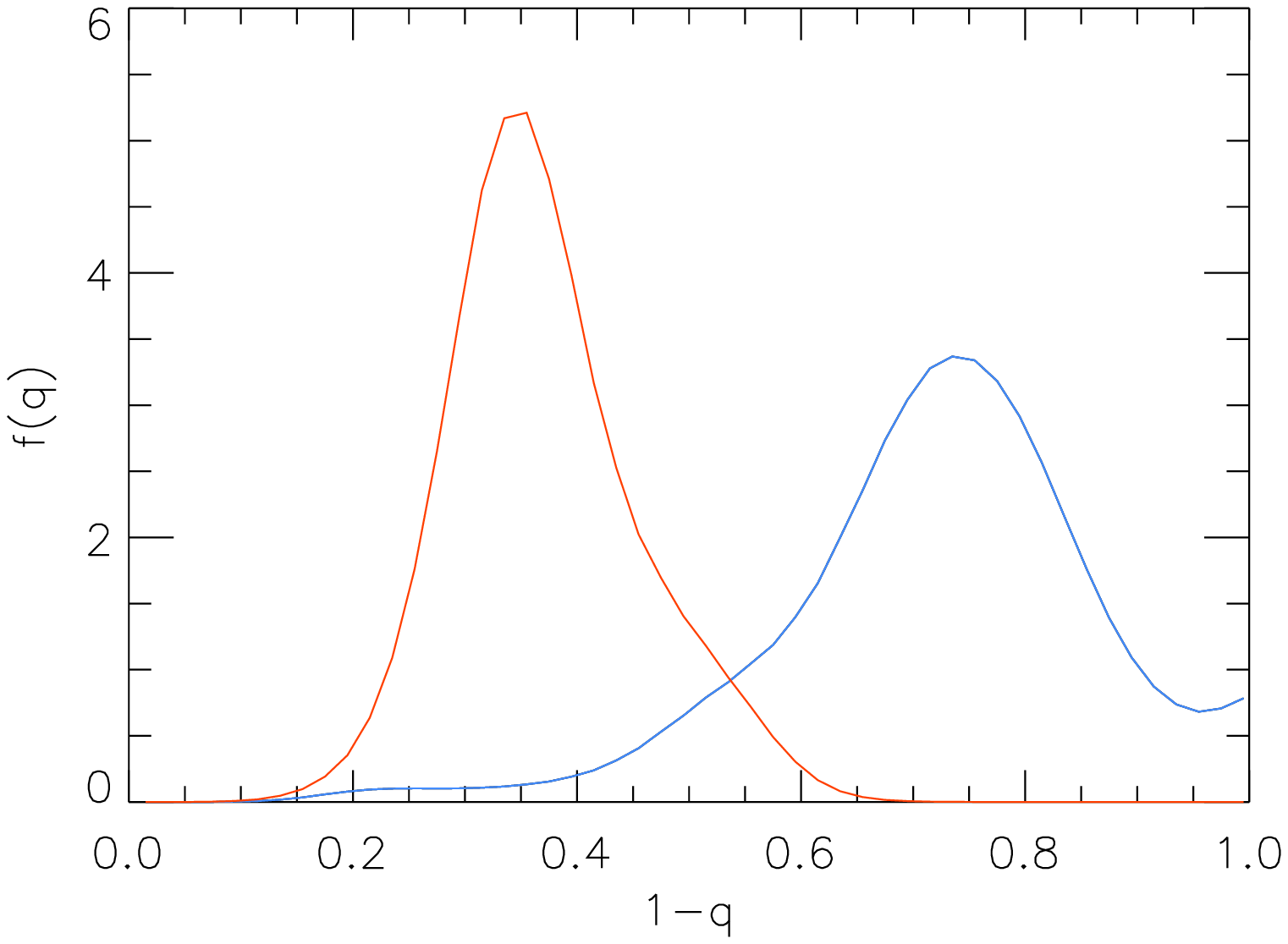,width=8cm}\\

\end{tabular}
\caption{Comparing the oblate intrinsic shape distributions $f(q)$ of fast
  (blue solid line) and slow rotators (red solid line). Left:
  distribution scaled with absolute number of galaxies in each sample
  (224 fast rotators versus 32 slow rotators). Right: normalized distributions.}
\label{fig:lucy_obl_fast_slow}
\end{figure*}

In Figure~\ref{fig:lucy_obl_fast_slow} we contrast the intrinsic
flattening of fast rotators and slow rotators in our \atlas\
sample. It is obvious that on average the fast rotators are
much more flattened than the slow rotators, as already emphasized in
our morphological classification 'comb' diagram in Figure~2 of Paper
VII,  though it is interesting
to see that there is also a large overlap between the two
distributions, with the tail towards rounder shapes of the fast
rotator distribution overlapping with the one of the slow rotators.

\section{Discussion}

\subsection{Fast rotators and spirals}
\label{sec:spiral}
Our fast rotators are significantly flatter than the slow rotators in
our sample, and are in fact close to the intrinsic flatness observed
in spiral galaxy populations, although we do observe a tail
  towards rounder shapes. Lambas et
al. (1992)\nocite{{1992MNRAS.258..404L}} for instance find $\mu_q =
0.25$ for their sample of 13,482 spiral galaxies, based on imaging of
the APM Bright Galaxy Survey, which is consistent with the intrinsic
flattening that we found in \S~\ref{sec:fast} for the fast
rotators. In contrast, the 2135 elliptical and 4782 lenticular galaxies in their
sample are best described with intrinsic flattening $\mu_q = 0.55$ and
$\mu_q = 0.59$, respectively. They note that all three galaxy
populations need to be slightly triaxial, which is something we will
explore in \S~\ref{sec:triax}. More recently, Padilla \& Strauss
(2008)\nocite{2008MNRAS.388.1321P}  reported similar results based on
SDSS Data Release 6 imaging, with their 282,203 spirals having $\mu_q = 0.21 \pm
0.02$, although their 303,390 ellipticals are flatter than the Lambas
et al. result, with $\mu_q = 0.43 \pm
0.06$. We note that the slow rotators in our sample with $\mu_q =
0.63$ are slightly rounder than the elliptical samples in both these
previous studies. These deviations could be caused by our smaller
sample sizes, but could also be indicative of the fact that we classified our early-type galaxies kinematically, while the
early-type galaxies samples based on imaging only contain a mixture of
fast and slow rotators. Indeed, in Paper III we show that 66 per cent
of the galaxies in the \atlas\ sample classified as elliptical (E) are
in fact fast rotators. Another recent study of axis ratio measurements
at both local and higher redshift ($1 < z < 2.5$) 
finds that the total population of early-type galaxies in both samples
is well-described with an intrinsic shape distribution consisting of a
triaxial, round component, and an oblate, flattened ($q \sim 0.3$)
component, with the fractions of these two populations varying as a
function of stellar mass and redshift (Chang et
al. 2013\nocite{2013ApJ...773..149C}). These results would agree with
our observations of the different shape distributions for our slow and
fast rotator sample.

That the fast rotators have a similar shape distribution to spiral
galaxies is in line with previous studies that have shown that spiral
galaxies display a large range of disc-to-total (D/T) ratios
(e.g. Graham 2001\nocite{2001AJ....121..820G}; Weinzirl et al. 2009\nocite{2009ApJ...696..411W}), which is also found to hold
true for the galaxies in our sample: Krajnovi\'c et
al. (2013, paper XVII)\nocite{2013MNRAS.432.1768K} performed bulge-disc
decompositions for the \atlas\ sample and found that 83 per cent of
the non-barred galaxies in the sample have disc-like components. The
resemblance between spiral and early-type galaxies was most notably
pointed out by Van den Bergh (1976)\nocite{1976ApJ...206.883V}, who redesigned the Hubble
tuning fork to include a parallel sequence of lenticular galaxies
(S0) to the spiral galaxies, with decreasing D/T ratios when moving
from S0c to S0a closer to the elliptical galaxies. 

In Paper VII\nocite{2011MNRAS.416.1680C} we revisited Van den Bergh's
classification scheme by showing that it are the fast rotators who
form a parallel sequence to the spiral galaxies, re-emphasizing the
importance of this parallelism to understand how galaxies
form, see also
Laurikainen et al. (2011)\nocite{2011AdAst2011E..18L} and Kormendy \& Bender
(2012)\nocite{2012ApJS..198.2K}.

\subsection{Shape as a function of stellar mass}

\begin{figure}
\psfig{figure=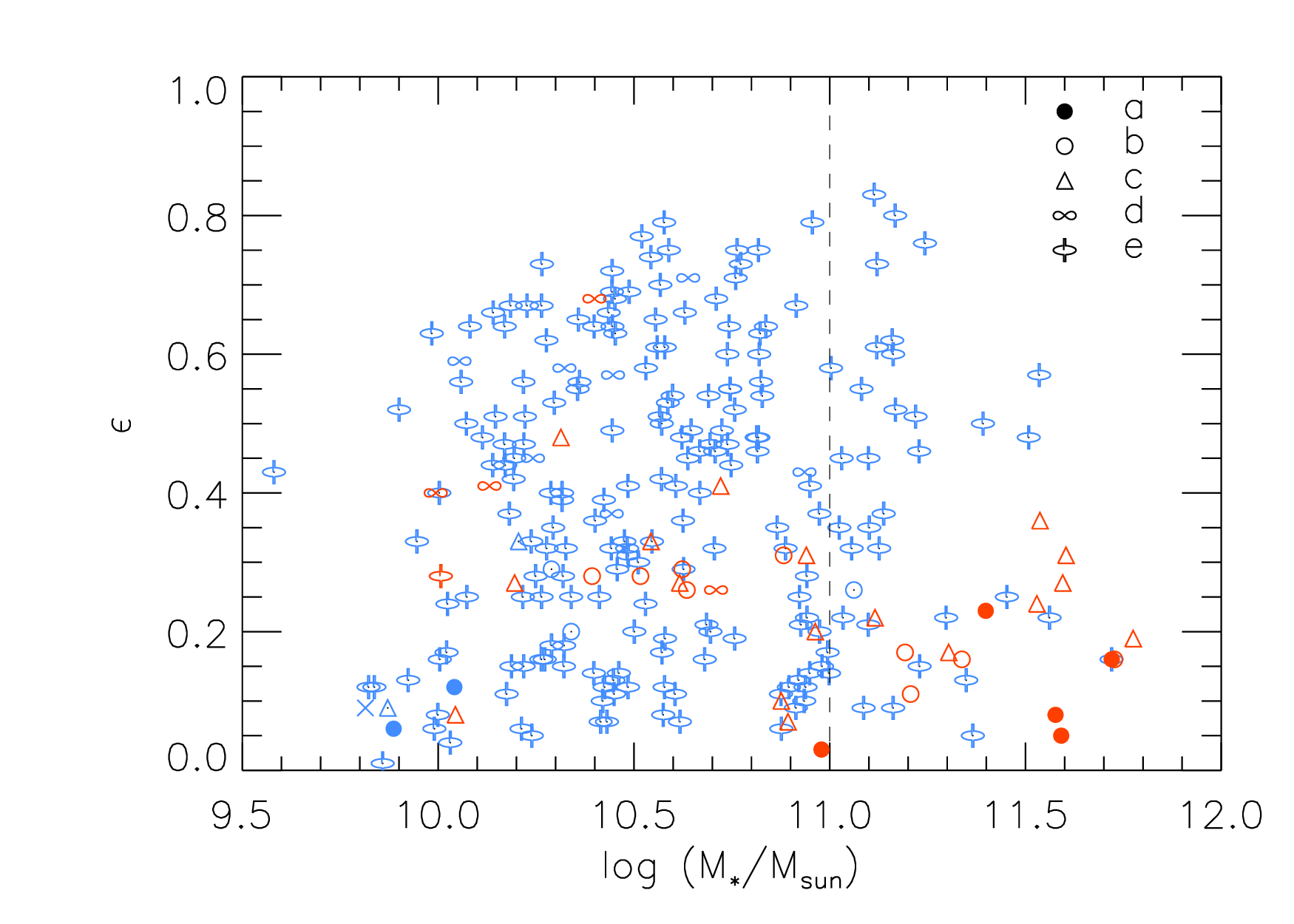,width=8cm}
\caption{Ellipticity as a function of stellar mass (as introduced in
  \S~\ref{sec:fast}), for
  fast (blue symbols) and slow rotators (red symbols). Fast rotators
  show a large spread in ellipticity over all mass ranges, while the
  most massive galaxies are predominantly round slow rotators. The symbols
  labeled $a$-$e$ define different kinematical classes, and are
  explained in the text. The galaxy marked with a cross (X) could not
  be kinematically classified.}
\label{fig:mass_fast_slow}
\end{figure}

In \S~\ref{sec:fast} we showed that there is no clear
difference between shape distributions of fast rotators above and
below a stellar mass of $10^{11} M_\odot$ (see
Figure~\ref{fig:hist_eps_env}, right-hand plot). At first sight, this
seems in contradiction with Tremblay \& Merritt
  (1996)\nocite{1996AJ....111.2243T}, and more recently, with van der Wel et
al. (2009)\nocite{2009ApJ...706L.120V} and Holden et al. (2012)\nocite{2012ApJ...749...96H}, who based on a sample of
quiescent galaxies selected from the Sloan Digital Sky Survey, find
that galaxies with stellar mass $M_*  > 10^{11}M_\odot$ are predominantly round,
while galaxies with lower masses have a large range in
ellipticity. This change in shape at a characteristic mass of $M_\mathrm{JAM} 
\sim 2\times 10^{11} M_\odot$ is also evident in our sample when
studying the mass-size relation (Figure~7 of Cappellari et
al. 2013b\nocite{2013MNRAS.432.1709C}, hereafter Paper XX). However,
as already illustrated in Figure~14 of Paper XX, the picture changes
when we include the kinematical information. In
Figure~\ref{fig:mass_fast_slow} we show the ellipticities of both fast
and slow rotators as a function of stellar mass, and we also indicate
different kinematical classes as defined in Paper II: class $a$
includes galaxies which do not show any significant rotation
(non-rotators), class $b$ comprises galaxies with complex velocity
maps, but without any distinct features, class $c$ consists of
galaxies with kinematically distinct cores (including counter-rotating
cores), class $d$ has galaxies with double peaks in their dispersion
maps (the $2\sigma$-galaxies, consisting of counter-rotating discs)
and finally, class $e$ is the group of galaxies with regularly
rotating velocity maps. Taking this subdivision into account, we note
that above $M_\mathrm{JAM}  \sim 10^{11}M_\odot$ the number of fast rotators
quickly declines, and the highest mass galaxies are predominantly
round non-rotators (class $a$). This indicates that the observed trend
with more massive galaxies being on average rounder than less massive
ones can be explained by the increasing fraction of slow rotators at
high masses, and that the orbital make-up drives the dependency of
shape on mass.

\begin{figure}
\psfig{figure=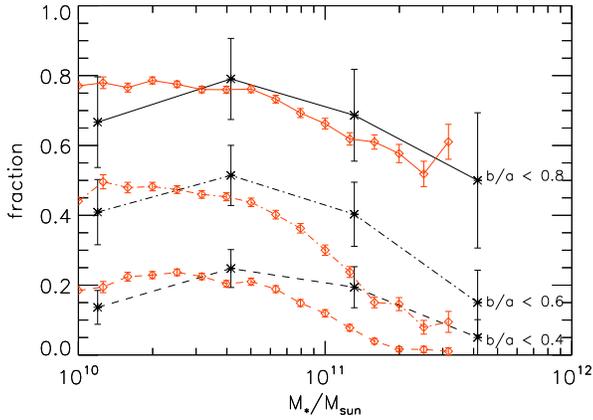,width=8cm}
\caption{Fraction of galaxies with axis ratio smaller than 0.8 (solid
  line), 0.6 (dotted-dashed line)
  and 0.4 (dashed line), for the \atlas sample (black stars) and van
  der Wel sample (red diamonds), as a function of stellar mass
  $M_*$. For the \atlas sample, the bins boundaries (in $\log{M_\odot}$) are given by
  $10.3$, $10.8$ and $11.3$. The van der Wel sample shows a clear
  trend with more massive galaxies being rounder: this trend is also
  seen in the \atlas sample in the largest mass bin, which contains a relatively large fraction of
  slow rotators.}
\label{fig:arjen}
\end{figure}

Figure~\ref{fig:arjen} shows the fraction of galaxies with axis
ratios below 0.8, 0.6 and 0.4 as a function of stellar mass  for the
total \atlas galaxy sample, and compares these fractions with the
results from van der Wel et
al. (2009)\nocite{2009ApJ...706L.120V}. The \atlas\ fractions remain
constant up to $M_* \sim 10^{11.3}$, as our
sample is dominated by fast rotators
(216/240 galaxies) in that mass range. The fractions from the van der
Wel sample show a clear trend between axis ratio and stellar mass,
with more massive galaxies being rounder. We reproduce that trend in our sample in the
highest massbin ($M_* > 10^{11.3}$), which contains a relatively large
number of slow rotators (12/20 galaxies).

\subsection{A lack of round galaxies?}
For a family of perfect oblate objects, we expect the shape
distribution to peak at $\epsilon = 0$ (see Equation~\ref{eq:prob_obl_eps},
which behaves asymptotically at $q=1$). However, our observed
ellipticity distribution for fast rotators decreases towards round shapes (see Figure~\ref{fig:hist_eps}). This lack of
round galaxies has been observed before (e.g. Fasano \&
Vio 1991\nocite{1991MNRAS.249..629F}; Ryden 1996\nocite{1996ApJ...461..146R}), and before investigating
deviations from axisymmetry (\S~\ref{sec:triax}), we first explore
whether our selection or ellipticity measurement methods could be
responsible for this observation.

\begin{figure}
\psfig{figure=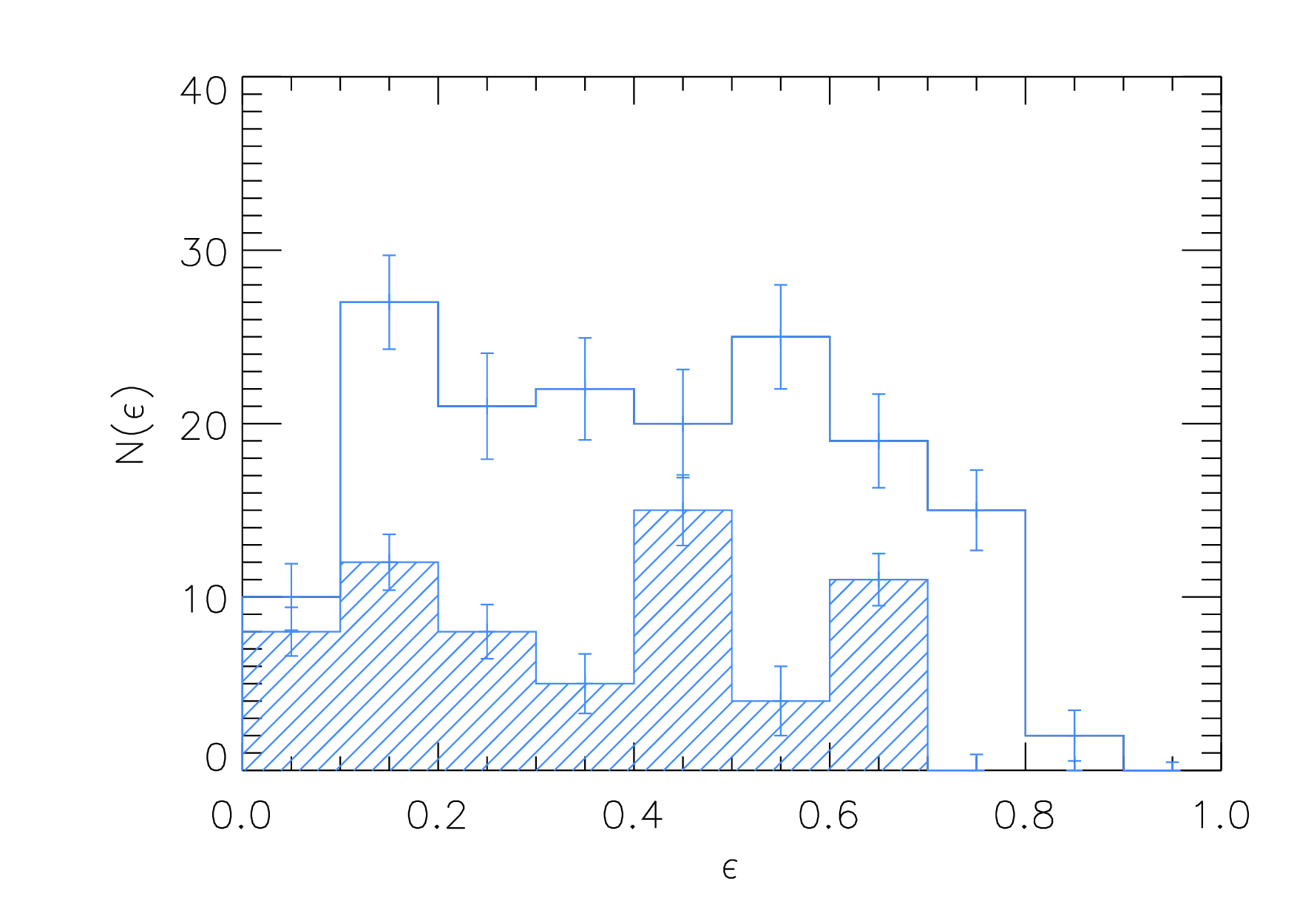,width=8cm}
\caption{Observed ellipticity histogram for the barred (dashed
  histogram) and non-barred (open histogram) fast rotators. The
  non-barred galaxies in our sample are intrinsically flatter than the
  barred galaxies, but this is a selection
  effect, as bars are easier detected in face-on than edge-on
  galaxies, see text for details.}
\label{fig:bar}
\end{figure}

Our measurements of ellipticity are based on moment of inertia, and a
positive bias is introduced for nearly round objects, as negative
ellipticities are not allowed. Tests conducted in Paper II show however that
this positive bias is of order 0.02, and therefore too small to expel
a significant number of galaxies out of the roundest ellipticity
bin. The influence of bars on our ellipticity measurements would be of
larger concern: although we obtain a global measurement of the
ellipticity by using moment of inertia as opposed to a
radius-dependent measurement, large bars could still significantly
increase the ellipticity of their round host galaxies. To
investigate this effect, we simulated perfectly oblate galaxies both
with and without bars, following the methods outlined in Lablanche et
al. (2012, paper XII)\nocite{2012MNRAS.424.1495L}, and observed these galaxies
face-on (so at $\epsilon = 0$). We found that bars indeed increased
the observed ellipticity to about 0.15, which is sufficient to move
round galaxies from the roundest ellipticity bin into the next
one. However, when splitting our sample of fast rotators into barred
and non-barred galaxies (following the classification of Paper II), we find that the barred galaxies are on
average rounder than the non-barred galaxies, contrary to what we expected
based on our simulations (see Figure~\ref{fig:bar}). This is however a selection effect:
bars are more easily identified in face-on (round) galaxies than in edge-on
(flattened) ones. It is therefore likely that there are still some
undetected bars present in our galaxy sample at higher ellipticities,
but this would not explain the possible deficiency of low ellipticity
galaxies. We therefore conclude that it is very unlikely that barred
galaxies are biasing our observed shape distribution towards flatter
systems. The perceived lack of round galaxies is therefore either
real, or has some other, more subtle cause. Despite this discrepancy
however, we show in the next section by including the observed kinematic
misalignment in our intrinsic shape analysis, that an oblate distribution is
indeed a very good description of our fast rotator sample.

\subsection{Deviations from axisymmetry}
\label{sec:triax}

So far we have assumed that the fast rotators in our sample are oblate
($p = 1$) systems, motivated by the observation that almost all fast
rotators have small or negligible misalignment. We now investigate
whether a triaxial ($p \neq 1$) distribution would be preferred above an
oblate one, using the observed kinematic misalignment $\Psi$ as an additional
constraint (Binney 1985\nocite{1985MNRAS.212..767B}; Franx et
al. 1991\nocite{1991ApJ...383..112F}). 

We cannot use a Lucy inversion as above to invert the observed
distribution, as we now have two observables ($\Psi$$, \epsilon$) and
three unknowns ($p, q$ and the intrinsic misalignment
$\theta_\mathrm{int}$, which is defined such that
$\theta_\mathrm{int}$=0 corresponds to alignment of the intrinsic
rotation axis with the short axis
of the galaxy). We therefore fit the observed two-dimensional
distribution $F(\Psi$$, \epsilon$) to simulated distributions, generated by assuming a Gaussian
distribution in $q$ with mean and standard deviation $\mu_q$ and
$\sigma_q$, and a log-normal distribution in $Y = \ln(1 - p)$ with mean and
standard deviation $\mu_Y$ and $\sigma_Y$, following
e.g. Padilla \& Strauss (2008)\nocite{2008MNRAS.388.1321P}. For
$\theta_\mathrm{int}$ we assume that this angle only depends on the
intrinsic shape, such that $\theta_\mathrm{int}$ coincides with the
viewing direction that generates a round observed ellipticity ($\epsilon =
0$, see right-hand plot of
Figure~\ref{fig:sphere_eps}). Mathematically, this 
corresponds to:

\begin{equation}
\tan \theta_\mathrm{int} = \sqrt{\frac{T}{1-T}},
\label{eq:thetaf}
\end{equation}

\noindent
with $T$ the triaxiality parameter defined by Franx et
al. (1991)\nocite{1991ApJ...383..112F} as:

\begin{equation}
T = \frac{1-p^2}{1-q^2}.
\label{eq:triax}
\end{equation}

\noindent
This assumption ensures that in systems close to oblateness,
$\theta_{\mathrm{int}}$ is small and close to the short axis, and only
increases for larger triaxiality. This assumption is valid for many
self-consistent models (e.g., Hunter \& de Zeeuw 1992\nocite{1992ApJ...389...79H};
Arnold, de Zeeuw \& Hunter 1994\nocite{1994MNRAS.271..924A}), and we will give a more detailed overview of the geometry and
probability distributions for such systems in appendix A.

To determine the best-fitting simulated distribution, we calculate
$\chi^2$ as:

\begin{equation}
\chi^2 (\mu_Y, \sigma_Y, \mu_q, \sigma_q) = \sum_{i,j} \frac{(O_{i,j} - M_{i,j})^2}{\delta O_{i,j}^2},
\label{eq:chi2}
\end{equation}

\noindent
where $O_{i,j}$ is the number of observed galaxies in each bin
$(\Psi_i, \epsilon_j)$, with $\Psi$ ranging from $0^\circ$ to
$90^\circ$ and $\epsilon$ from 0 to 1, in binsteps of $5^\circ$ and
0.1, respectively. The corresponding errors $\delta O_{i,j}$ are
determined with Monte Carlo simulations, similar to the errors for the
one-dimensional histograms in $\epsilon$ used before. For many of our
bins with large misalignment this error is zero, which raises problems
in our $\chi^2$ determination. We therefore replaced these zero errors
with artificially small values, corresponding to 0.1 times the minimal
error in the total histogram. As a result, our $\chi^2$ values are not
statistically valid, but as we are interested in
locating the best-fitting intrinsic distribution, we simply restrict
our analysis to finding the minimal $\chi^2$. 

$M_{i,j}$ is the number of galaxies predicted for each bin given by the model,
generated with the parameters $\mu_Y, \sigma_Y, \mu_q, \sigma_q$, and
under the assumption that $\theta_\mathrm{int}$ is given by Equation~\ref{eq:thetaf}. For
each combination of these four parameters, we generate 100,000 random viewing
angles and construct a distribution of an equal number of observed
galaxies, drawing their intrinsic axis ratios $p$ and $q$ from their
log-normal and Gaussian distributions, respectively. We then calculate
for each galaxy its observed ellipticity and misalignment, using the
formularium outlined in appendix A. 

\begin{figure*}
\begin{tabular}{cc}
\psfig{figure=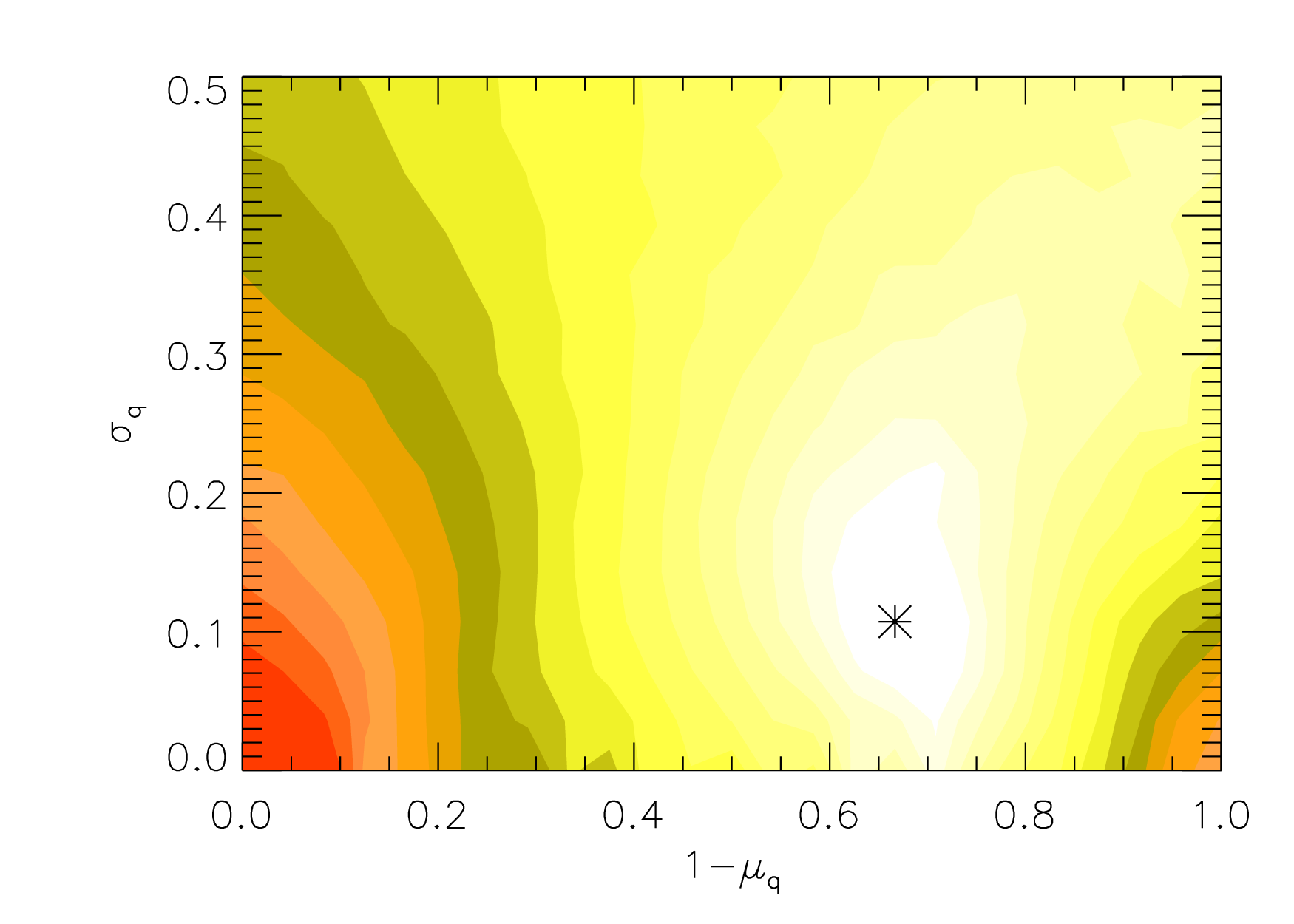,width=8cm} &
\psfig{figure=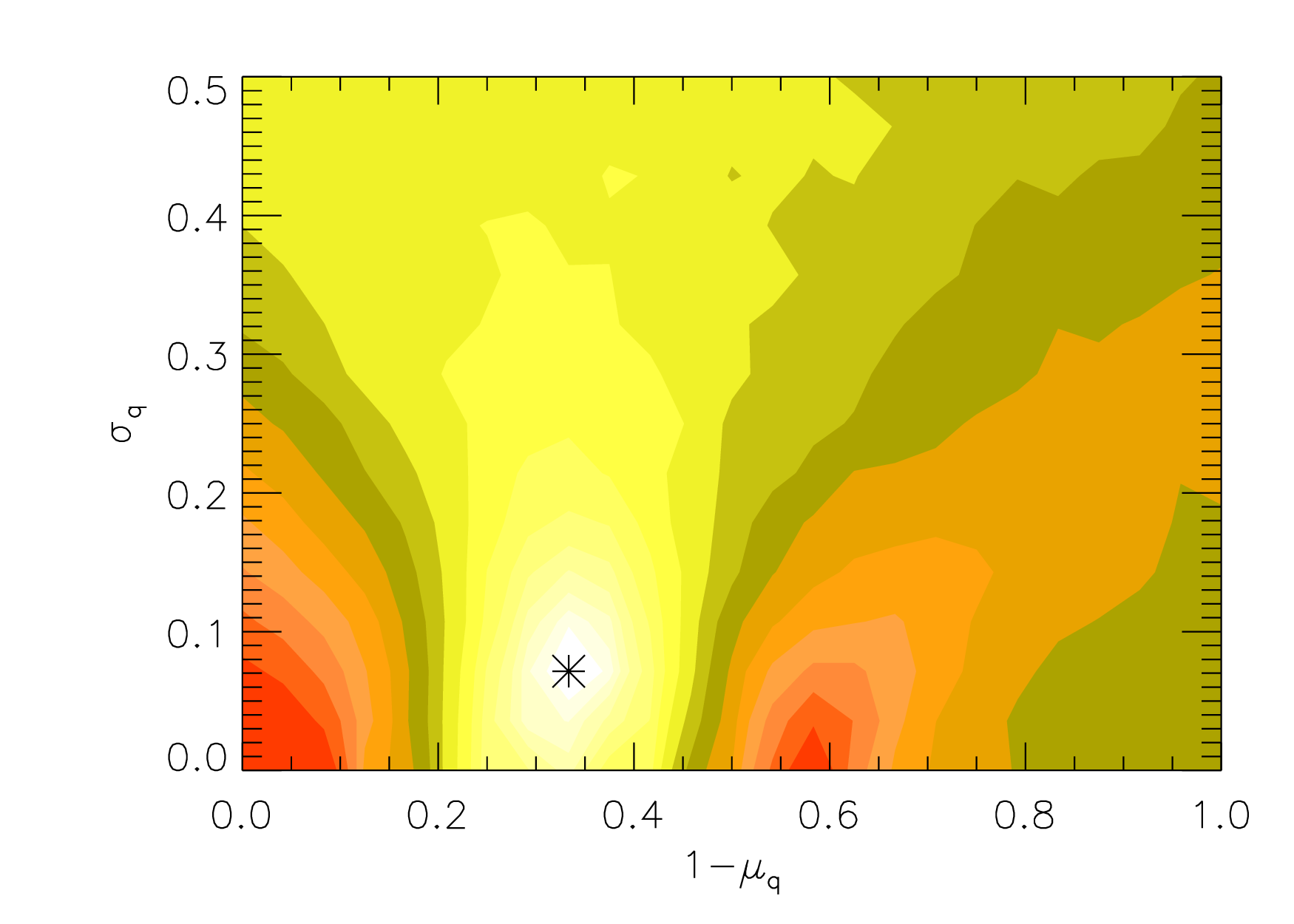,width=8cm} \\
\end{tabular}
\caption{Contours of constant $\chi^2$ assuming oblate intrinsic shapes
  with a Gaussian distribution in $q$ for the fast rotator (left) and slow rotator (right)
  samples. Contours increase logarithmically from light to darker
  colours, and the minimal $\chi^2$ value is indicated with a black asterisk.}
\label{fig:chi2_oblate}
\end{figure*}

Before exploring the full grid of $\mu_Y, \sigma_Y, \mu_q, \sigma_q$,
we first apply the above analysis to an oblate model, and only fit the
one-dimensional histogram in ellipticity presented in
Figure~\ref{fig:hist_eps}, ignoring the kinematic misalignment for the
moment. As such, we are repeating the analysis presented in the
previous section, though with a very different method. We plot the resulting $\chi^2$ contours in
Figure~\ref{fig:chi2_oblate}, both for our fast and slow rotator
samples. For the fast rotators, we find a minimal $\chi^2$ for $\mu_q
= 0.33$ and $\sigma_q = 0.11$, which is somewhat rounder than the
distribution we found with the direct inversion described in
\S~\ref{sec:fast}, although the Gaussian fit to the intrinsic
distribution does not take the tail towards higher $q$ into
account. For the slow rotators, we find $\mu_q = 0.66$ and $\sigma_q =
0.08$, which is very similar to the direct inversion described in
\S~\ref{sec:slow}. This shows that the results we presented for the intrinsic shape
distributions are not method dependent.

\begin{figure}
\begin{center}
\begin{tabular}{c}
\psfig{figure=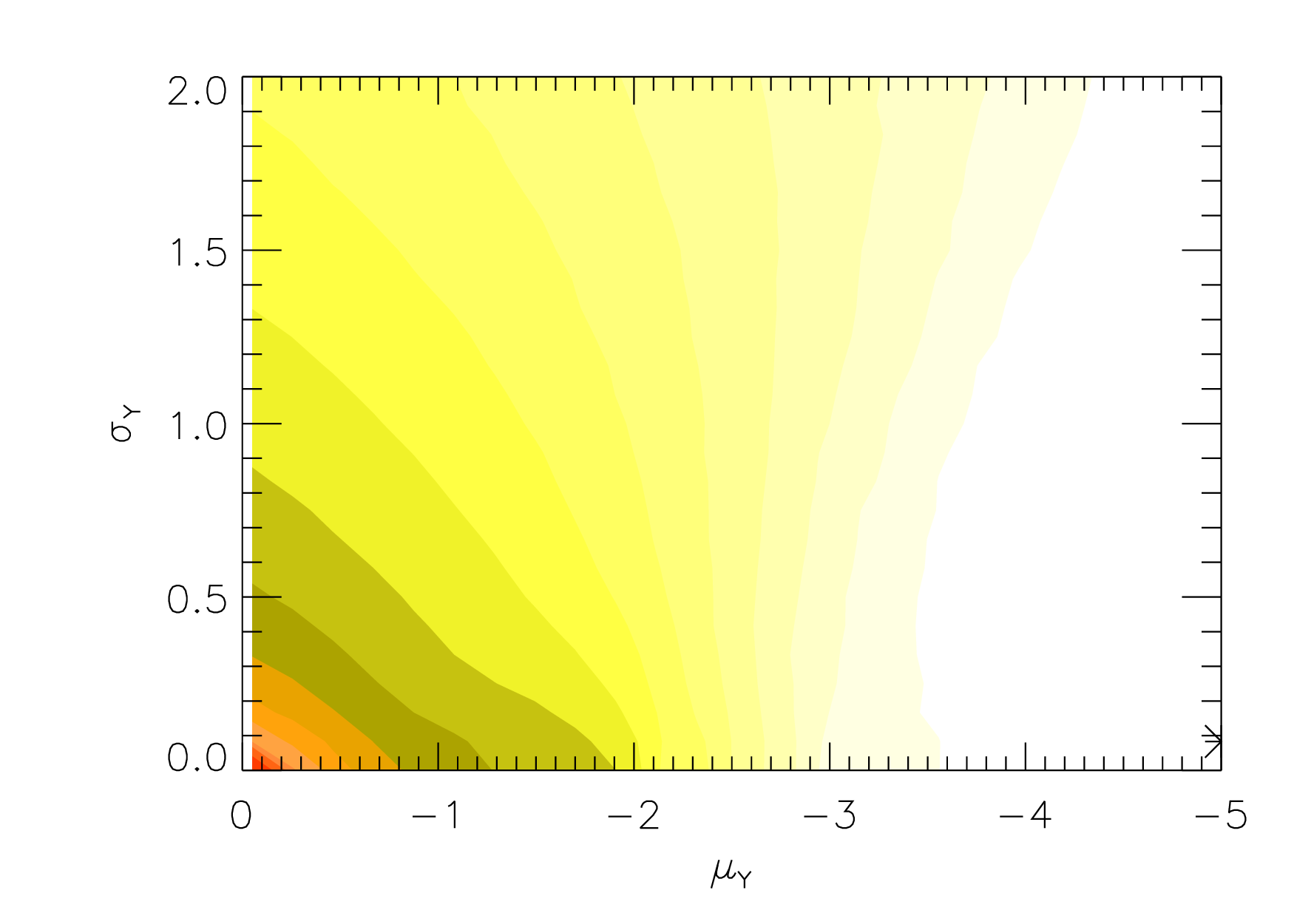,width=8cm} \\
\psfig{figure=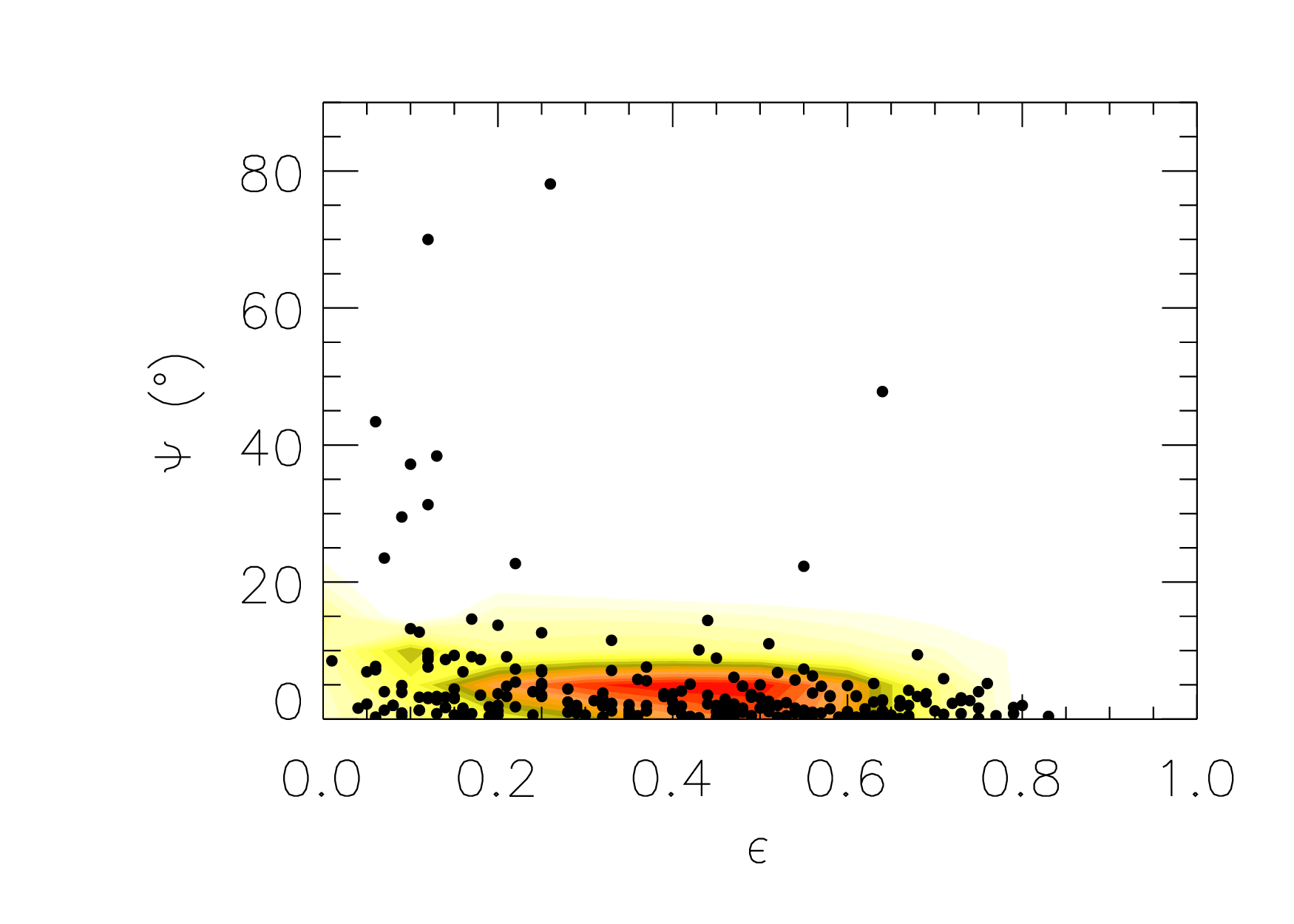,width=8.5cm}\\
\psfig{figure=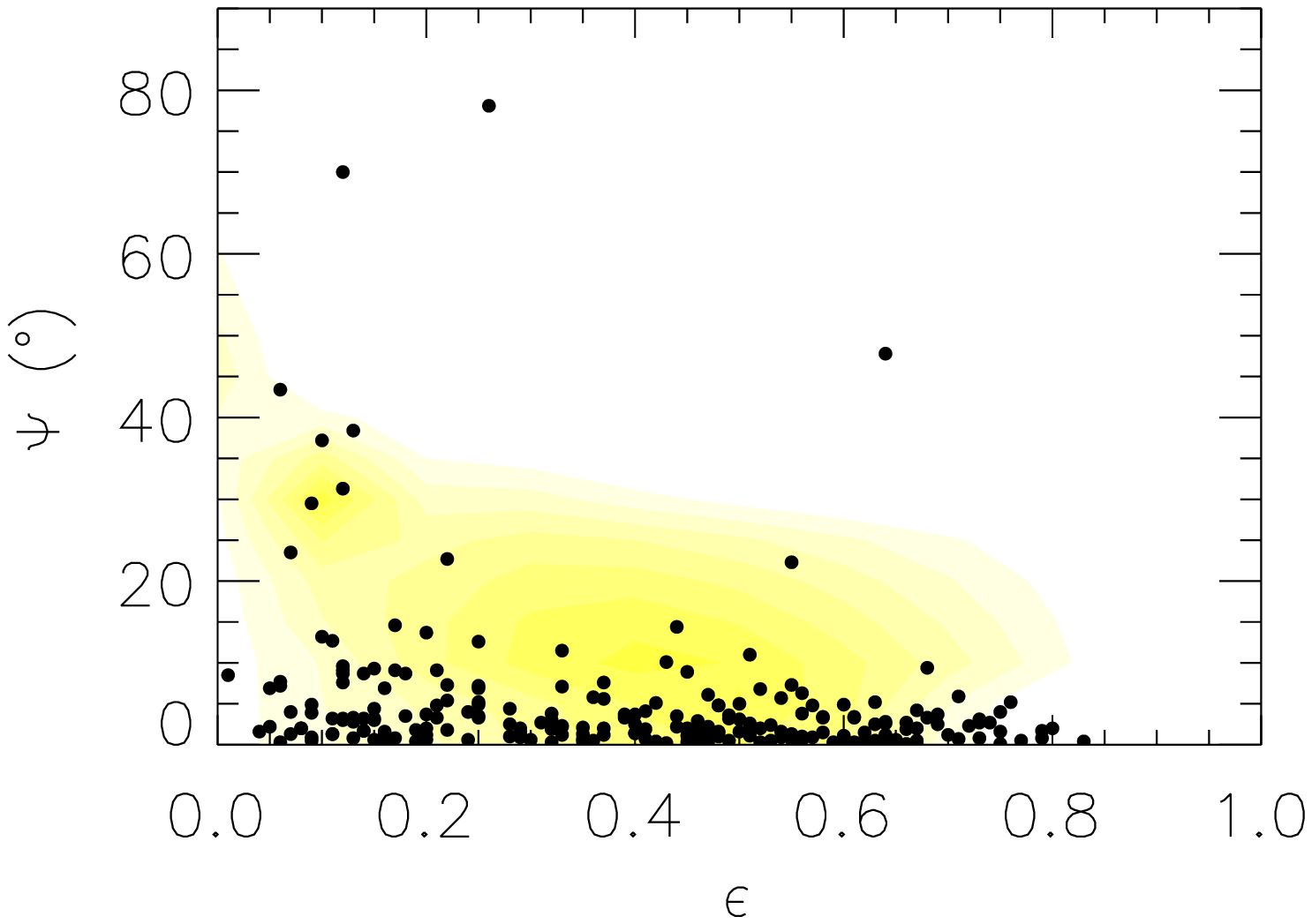,width=8.5cm}
\end{tabular}
\end{center}
\caption{Top: contours of constant $\chi^2$ for fast rotators as a
  function of $\mu_Y$ and $\sigma_Y$, assuming
  $\tan\theta_\mathrm{int} = \sqrt{T/(1-T)}$ and a Gaussian distribution in $q$,
  with parameters as indicated in the text. Contours are
increasing logarithmically from light to dark colours, and the minimum
in $\chi^2$ is indicated with a black asterisk. Middle: contours
indicate the distribution $F(\Psi, \epsilon)$, as predicted by the
best-fit model indicated by the asterisk in the top panel. Contours
increase linearly from light to dark colours. Overplotted in black dots
is our observed fast rotator sample, indicating the nice agreement
between model and observations. There are some galaxies with
significant larger misalignment than predicted by our best-fitting
model: these systems however all are dominated by strong bars or are
interacting systems. Bottom: same as middle, but here we
show a model with $\mu_Y = -3.0$. The corresponding distribution is
overprediciting the number of galaxies with larger ($\Psi >
10^{\circ}$) misalignment compared to the observations.}
\label{fig:chi2_theta}
\end{figure}

We now relax our assumption of oblateness on the fast rotator sample and explore the full grid
$\mu_Y, \sigma_Y, \mu_q, \sigma_q$, and fit the two-dimensional
histogram in $\Psi$ and $\epsilon$, with
$\theta_\mathrm{int}$ given by Equation~\ref{eq:thetaf}, as described above. The best-fit in this triaxial
model space has values for $\mu_q$ and  $\sigma_q$ that are very
close to the best-fit values for the oblate model discussed above, and
therefore to limit the parameter search, we run a finer grid in
$\mu_Y$ and $\sigma_Y$, keeping $\mu_q$ and  $\sigma_q$ fixed to 0.33
and 0.11, respectively. The best-fit model that we so obtain is very
close to oblate, with $\mu_Y = -5.0$ (which corresponds to $p \sim
0.99$), and $\sigma_Y = 0.08$. In fact, $\mu_Y = -5.0$ is one of the
boundaries in our grid, meaning that the best-fit model is as oblate
as allowed by our grid choice. The resulting $\chi^2$ contours are
shown in Figure~\ref{fig:chi2_theta}. Unfortunately, we cannot put any
statistical significance to these contours, but we do note that models
close to oblate (large negative $\mu_Y$) are strongly preferred, while
$\sigma_Y$ is largely unconstrained.

Interestingly, the deviation from axisymmetry of our fast
rotators is smaller than that of the spiral galaxies studied by Ryden
(2006)\nocite{2006ApJ...641..773R}, who used the same methods to
obtain a triaxial intrinsic shape distributions of her sample. She finds for her
early-type spirals (Hubble type Sbc and earlier) a median value for
$p$ of 0.82 (in B-band). For 
her late-type spirals (Sc and later), she reports a median value of $p \sim 0.93$, which
is more in agreement with the results we find for our fast
rotators, although our sample is again closer to axisymmetry. This may
not be so surprising though, given that the shape measurements of our sample of early-type
galaxies do not suffer from additional structures introduced by spiral waves
and dust, which are commonly present in spiral galaxies. Another
possible explantion for the non-circularity of disc galaxies could
come from lopsidedness (e.g. Rudnick \& Rix 1998\nocite{1998AJ....116.1163R}). We also compare our results with Padilla \& Strauss
(2008)\nocite{2008MNRAS.388.1321P}, who for their elliptical galaxies
report $\mu_Y = -2.2\pm0.1, \mu_q = 0.43\pm0.06$ and for their spirals
$\mu_Y = -2.33\pm0.13, \mu_q = 0.21\pm0.02$. Again, in comparison to
both galaxy populations, our fast rotators
are intrinsically closer to axisymmetry.

To check that our best-fit model is a reasonable fit to the data (and
not simply the best of a set of only bad models), we plot in the
bottom panel of Figure~\ref{fig:chi2_theta} the expected observed
distribution $F(\Psi, \epsilon)$ given our best-fit $f(\mu_Y,
\sigma_Y, \mu_q, \sigma_q)$, generated with Monte Carlo simulations,
and we overplot the observed $(\Psi, \epsilon)$ values for our fast
rotator galaxy sample. Errorbars have been omitted, but can be found
in Paper II: the median error in ellipticity is 0.03, while the
median error in kinematic misalignment is 6$^\circ$. Apart from a few
(mostly barred or interacting) outliers with
high $\Psi$, the predicted distribution by our best-fit model closely
follows the observed distribution.

Finally, we fit an intrinsic aligned model ($\theta_{\mathrm{int}} =
0$) to our data. In triaxial systems, alignment occurs when the
long-axis tube orbits cancel each other out, or when the system is
dominated by short-axis tube and box orbits instead. The observed
kinematic misalignment therefore cannot originate from intrinsic
misalignment, and has to be caused by projection of the triaxial
intrinsic shape only (see Appendix \ref{sec:kinmis}, and in particular
Equations \ref{eq:def-Psi}-\ref{eq:def-Thmin} for details). This model
would therefore set a firm upper limit on the allowed amount of
triaxiality in our galaxy population. We find again a best fit for
$\mu_Y = -5$ (or equivalently, $p \sim 0.99$), although with a larger
best-fit standard-deviation $\sigma_Y = 0.42$.  We therefore conclude
that the fast rotators are indeed oblate systems, and that if there
are any deviations from axisymmetry, these would have to be small.

\begin{figure}
\begin{center}
\begin{tabular}{c}
\psfig{figure=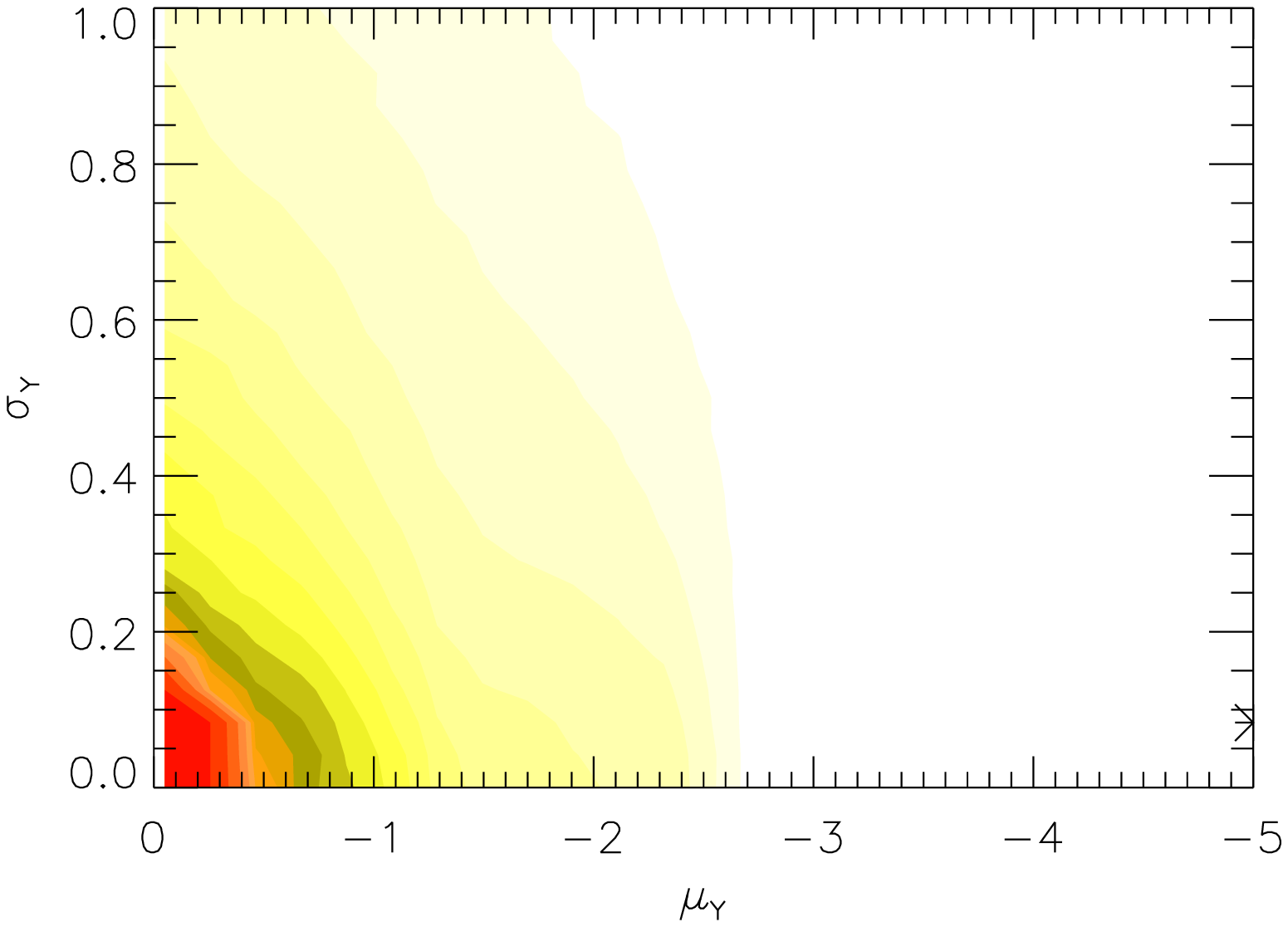,width=8cm} \\
\psfig{figure=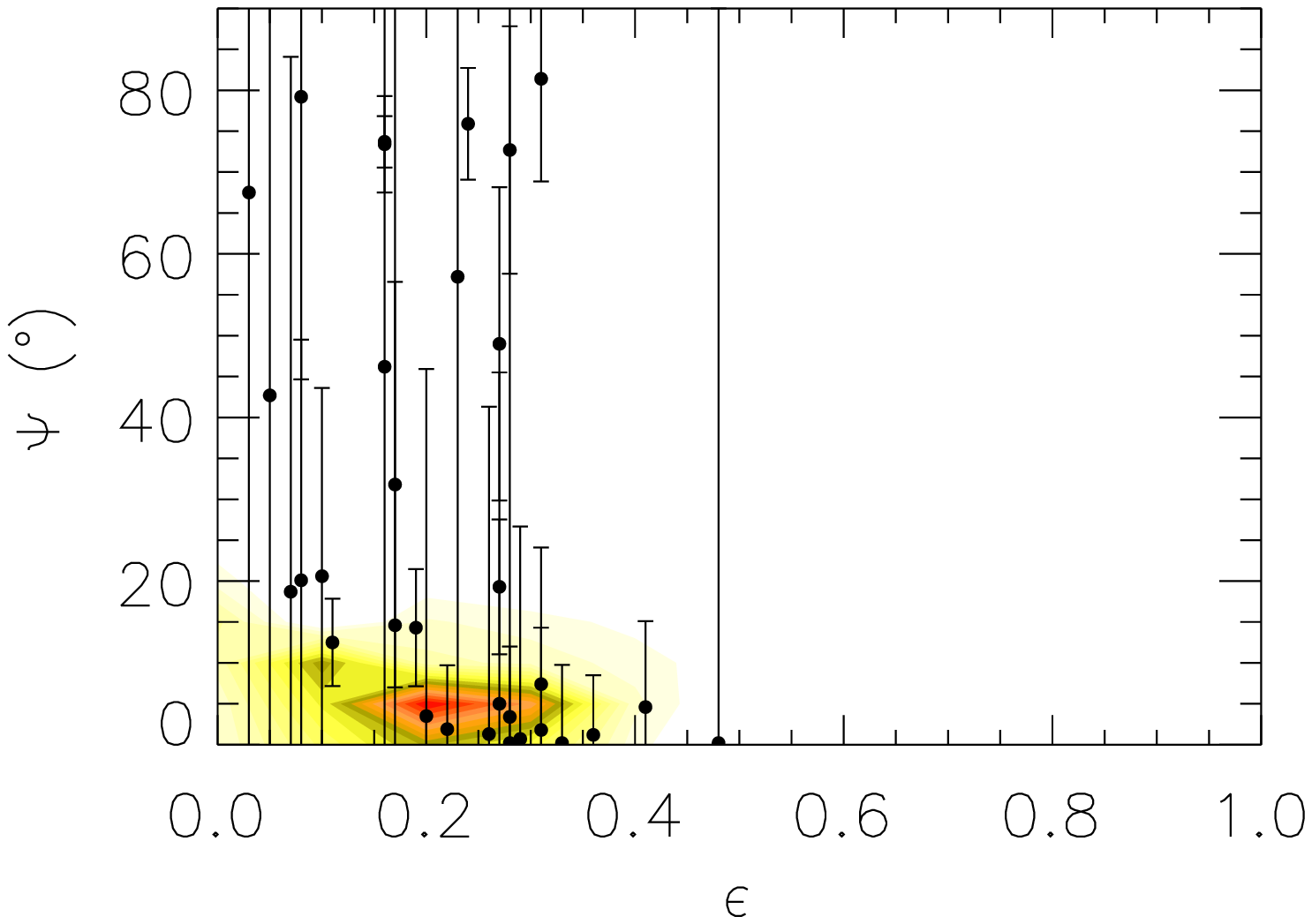,width=8.5cm} \\
\psfig{figure=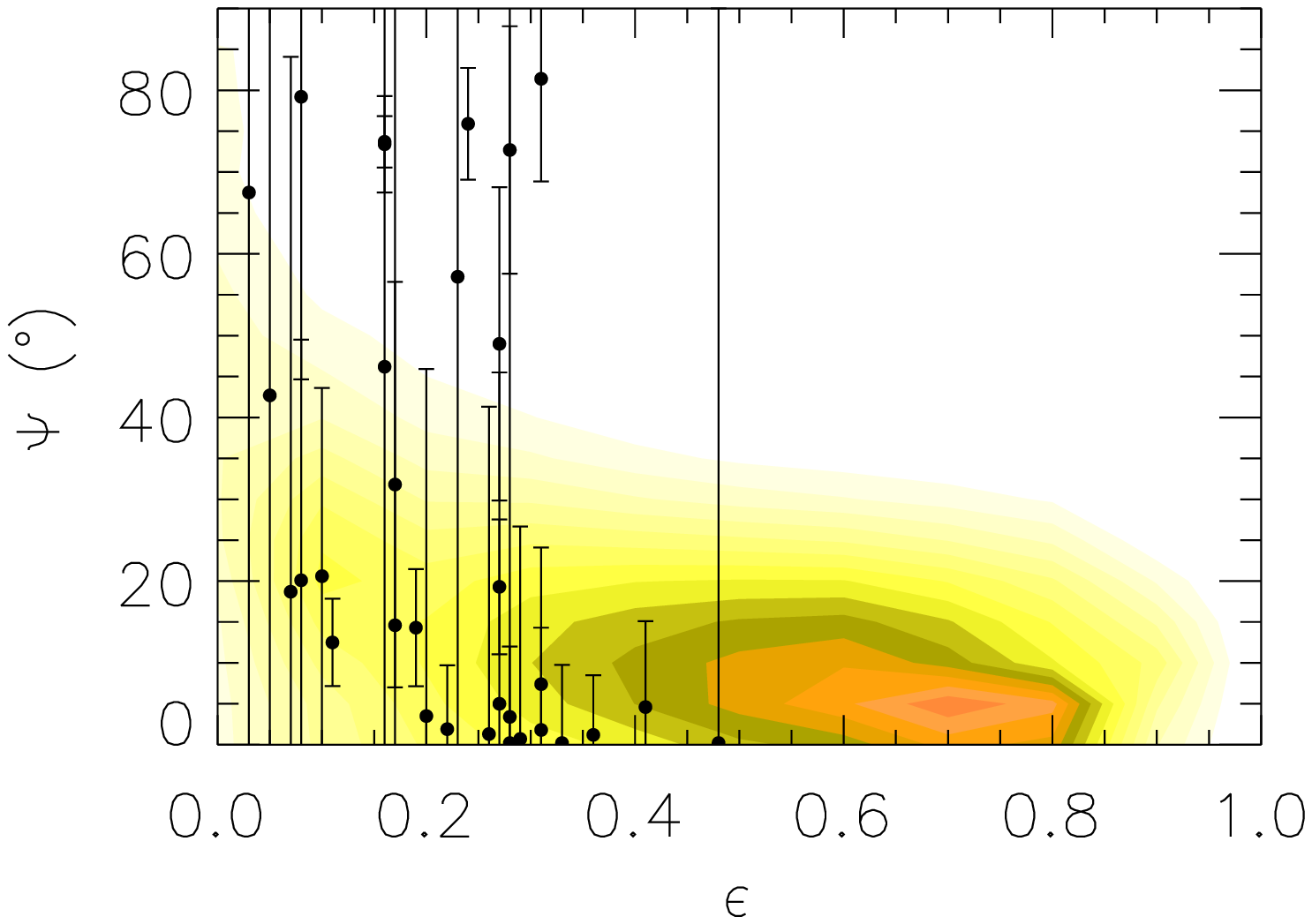,width=8.5cm}
\end{tabular}
\end{center}
\caption{Same as Figure~\ref{fig:chi2_theta}, but now for the slow
  rotators in our sample. Error bars have been added for $\Psi$, taken
from Paper II. In the bottom plot we show again the predicted
distribution for $\mu_Y  = -3.0$. This model is too triaxial, as it
overpredicts the number of flattened objects compared with the observations.}
\label{fig:chi2_theta_triax}
\end{figure}

Unfortunately, a similar analysis for the slow rotators in our sample
failed due to the small sample size compared to the parameter space,
as well as the lack of a clearly defined projected rotation axis in
many of the systems (most notably for the non-rotators, or class $a$
galaxies in our sample). Fixing the intrinsic flattening to $\mu_q =
0.66$ and $\sigma_q = 0.08$, as derived from the axisymmetric
distributions, we find for the model with $\theta_{\mathrm{int}}$ a
best-fit of $\mu_Y = -5.0$ and $\sigma_Y = 0.08$, which is an oblate
shape. However, as we show in Figure~\ref{fig:chi2_theta_triax}, the
minimum is not clearly defined, and the best-fit model is not able to
reproduce the observed kinematic misalignment.  We also note that a
model with a larger triaxiality $\mu_Y = -3.0$ does allow for the
larger observed misalignments, but does not reproduce the rounder observed
shapes. The derived numbers are therefore not trustworthy. A model
with no intrinsic misalignment ($\theta_\mathrm{int} = 0$) did prefer
a triaxial model, but also did not show a clear minimum in $\chi^2$,
and also was not able to reproduce the observed distributions.

\section{Summary and conclusion}

We inverted the observed ellipticity distributions of the early-type
galaxies in the \atlas\
sample to obtain their intrinsic shapes. Based on kinematical
classification, we divided our sample into fast and slow rotators, and
inverted these populations separately. We find that the fast rotator
population is significantly flatter than the slow rotator population
($\mu_q = 0.25$ versus $\mu_q = 0.63$, assuming axisymmetry), and that we cannot treat
early-type galaxies as one single population, but that we need to
consider fast and slow rotators separately. This is 
consistent with the conclusions in previous papers of this series: in
Paper II we noted that based on their kinematic alignment, fast
rotators are consistent with being axisymmetric, while slow rotators
are not. In Papers III and VII we pointed out the difference in
observed axial ratios between fast and slow rotators, while in Paper
XVII we uncovered a distinction between fast and slow rotators in
terms of the presence of discs from photometric
decomposition. Finally, in Paper XX we showed dynamical models to
deproject the galaxies, while in this paper we use a statistical
inversion to show the difference in intrinsic flattening between fast
and slow rotators. Given that
both lenticular and elliptical galaxies are present in the fast
rotator class, a purely morphological classification would not have
been sufficient for the shape study presented in this paper. 

We did not observe any trends of intrinsic shape with environment or
stellar mass for the fast rotators, but we did note a decrease in
observed ellipticity above stellar masses of $\sim 10^{11} M_\odot$
for the total early-type galaxy population, which is mainly driven by
round massive, non-rotating slow rotators. We showed with simulations that
our results are not affected by (weak) bars, which could potentially
increase the observed ellipticity of their host galaxies. 

Fast rotators have similar intrinsic flattening as spiral galaxies,
which is in line with the results of Paper XVII, where we showed that
fast rotators show a large span in disc-to-total ratios, and with the
classification scheme introduced by Van den Bergh (1976), and
revisited in Paper VII to emphasize the parallelism between fast
rotators and spirals (see also Laurikainen et
al. 2011\nocite{2011AdAst2011E..18L}; Kormendy \& Bender
2012\nocite{2012ApJS..198.2K}).  This observation could hint to a
similar evolutionary path of spirals and fast rotators, and it would
be interesting to study this further in the context of the
morphology-density relation, as mentioned in e.g. Paper VII, and
Cappellari (2013c)\nocite{2013ApJ...778L...2C}.

Next, we relaxed our assumption of axisymmetry and fitted triaxial
models to our observations. We again took advantage of having
integral-field data available for our dataset, by including the
kinematic misalignment as an extra constraint in this fit. Assuming
that the intrinsic misalignment is a function of intrinsic shape, we
show convincingly that fast rotators are very close to oblateness,
with only small deviations from axisymmetry allowed by our
observations. Due to their small numbers in our sample,
we could not repeat this analysis for the slow rotators, but based on
their observed kinematic misalignment, we do expect this population to
be more triaxial. That slow rotators are systematically rounder
  than fast rotators could also contribute to explain why, at a given
  mass, they appear to hold on better to their hot-gas medium and show
brighter X-ray haloes (Sarzi et al. 2013, Paper XIX\nocite{2013MNRAS.432.1845S}).

Despite the small size of the \atlas\ sample
compared to the larger SDSS samples used in various previous shape
studies, our sample has the big advantage of having kinematic
information available. This not only allowed us to separate the
early-type galaxy populations in two distinct kinematical classes,
which showed to have significantly different intrinsic shape
distributions, but also made it possible to include the kinematic
misalignment in our exploration of triaxial shape distributions. We
therefore conclude that integral-field data is crucial to refine
intrinsic shape studies, and to separate galaxy populations into
distinct kinematical classes.


\section*{Acknowledgements}

The authors thank Arjen van der Wel for kindly sharing his data,
  as well as fruitful discussions. The authors also thank the referee, for his/her constructive comments.
This work was supported by the rolling grants ‘Astrophysics at Oxford’
PP/E001114/1 and ST/H002456/1 and visitors grants PPA/V/S/2002/00553,
PP/E001564/1 and ST/H504862/1 from the UK Research Councils. RLD
acknowledges travel and computer grants from Christ Church, Oxford and
support from the Royal Society in the form of a Wolfson Merit Award
502011.K502/jd. RLD is also grateful for support from the Australian
Astronomical Observatory Distinguished Visitors programme, the ARC
Centre of Excellence for All Sky Astrophysics, and the University of
Sydney during a sabbatical visit.  MC acknowledges support from a
Royal Society University Research Fellowship.  SK acknowledges support
from the Royal Society Joint Projects Grant JP0869822.  RMcD is
supported by the Gemini Observatory, which is operated by the
Association of Universities for Research in Astronomy, Inc., on behalf
of the international Gemini partnership of Argentina, Australia,
Brazil, Canada, Chile, the United Kingdom, and the United States of
America.  TN and MBois acknowledge support from the DFG Cluster of
Excellence `Origin and Structure of the Universe'.  MS acknowledges
support from a STFC Advanced Fellowship ST/F009186/1.  PS acknowledges
support of a NWO/Veni grant.  TAD: The research leading to these
results has received funding from the European Community's Seventh
Framework Programme (/FP7/2007-2013/) under grant agreement No 229517.
MBois has received, during this research, funding from the European
Research Council under the Advanced Grant Program Num
267399-Momentum. LY acknowledges support from NSF AST-1109803. The
authors acknowledge financial support from ESO.  This paper is based
on observations obtained at the William Herschel Telescope and the
Isaac Newton Telescope, operated by the Isaac Newton Group in the
Spanish Observatorio del Roque de los Muchachos of the Instituto de
Astrof\'{\i}sica de Canarias. Funding for the SDSS and SDSS-II was
provided by the Alfred P. Sloan Foundation, the Participating
Institutions, the National Science Foundation, the U.S. Department of
Energy, the National Aeronautics and Space Administration, the
Japanese Monbukagakusho, the Max Planck Society, and the Higher
Education Funding Council for England. The SDSS was managed by the
Astrophysical Research Consortium for the Participating
Institutions. This publication makes use of data products from the
Wide-field Infrared Survey Explorer, which is a joint project of the
University of California, Los Angeles, and the Jet Propulsion
Laboratory/California Institute of Technology, funded by the National
Aeronautics and Space Administration.



\appendix
\section{Triaxial intrinsic shape distributions}
 
In this section we explore the triaxial shape distributions used in
\S\ref{sec:triax} in more detail. We first give the expression for
ellipticity and kinematic misalignment as function of intrinsic axis
ratio ($p, q$) and viewing angle ($\vartheta, \varphi$) that were used
to populate the simulated distributions, when we explored deviations
from axisymmetry in our galaxy sample. We then give analytical expressions for
the probability distributions $P(\Psi, \epsilon)$ and $P(\Psi)$, in the case of
intrinsic misalignment coinciding with the viewing direction that
yields an observed round galaxy (Equation~\ref{eq:thetaf}), which is
one of the assumptions we made in our triaxial analysis. We
  include these expressions here, as this case smoothly connects oblate
models with the intrinsic rotation axis along the short axis (in
agreement with their observed dynamics) with prolate models where the
instrinsic rotation axis coincides with the long axis (again, in
agreement with their observed dynamics). Many triaxial dynamical models
therefore follow this relation. In addition, somewhat surprisingly
given the need to calculate roots of polynomials, in this special
case both the expressions $P(\Psi, \epsilon)$ and $P(\Psi)$ are
elementary functions, and they were not previously recorded in Franx
et al. (1991)\nocite{1991ApJ...383..112F}.

\subsection{Ellipticity and kinematic misalignment in triaxial systems}
\label{sec:kinmis}

For oblate systems $(p=1)$, the observed ellipticity only depends on one
viewing angle: the inclination $\vartheta$ (see Equation~\ref{eq:eps_obl}). For triaxial systems ($p
\neq 1$) the observed ellipticity depends on both spherical viewing angles
$\vartheta$ and $\varphi$ (see also Figure~\ref{fig:sphere_eps}, which
shows observed ellipticity as function of viewing angle). The expression
for ellipticity is then given by (e.g. Contopoulos 1956\nocite{1956ZA.....39..126C}):

\begin{equation}
\label{eq:def-ellipticity}
e=(1-\epsilon)^2 = {a -\sqrt{b} \over a + \sqrt{b}},
\end{equation}
with
\begin{eqnarray}
\label{eq:def-auxaandb}
a\!\! &=& \!\!(1\!-\!q^2)\cos^2\vartheta  
          \!+\!(1\!-\!p^2)\sin^2\vartheta \sin^2\varphi\!+\!p^2\!\!+\!q^2, 
                                                \nonumber \\
b\!\! &=& \!\!\bigl[(1\!-\!q^2)\cos^2\vartheta 
          \!-\! (1\!-\!p^2)\sin^2\vartheta\sin^2\varphi 
             \!-\! p^2\!\!+\!q^2\bigr]^2 \nonumber \\
  &\phantom{=}& \quad\qquad +4(1\!-\!p^2)(1\!-\!q^2) \sin^2\vartheta
                            \cos^2\vartheta \sin^2\varphi. 
\end{eqnarray}

\noindent
For each triaxial shape there are four viewing directions that yield
an observed ellipticity equal to zero (see right-hand panel in Figure~\ref{fig:sphere_eps}); these viewing angles are given
by $\vartheta = \theta_f, \pi-\theta_f$ and $\varphi = 0, \pi$, with
$\theta_f$ given by:

\begin{equation}
\tan\theta_f = \sqrt{\frac{T}{1-T}}, 
\label{eq:round}
\end{equation}

\noindent
and $T$ the triaxiality parameter from Franx et al. (1991)\nocite{1991ApJ...383..112F},
as defined in Equation~\ref{eq:triax}.

Kinematic misalignment is the difference between the projected
rotation axis $\Theta_\mathrm{kin}$ and the projected short axis
$\Theta_\mathrm{min}$, and therefore defined as (e.g. Franx et
al. 1991\nocite{1991ApJ...383..112F}): 

\begin{equation}
\label{eq:def-Psi}
\sin\Psi = |\sin(\Theta_\mathrm{kin} - \Theta_\mathrm{min})|, \quad 0^\circ \le \Psi \le 90^\circ.
\end{equation}

\noindent
$\Theta_\mathrm{kin}$ is a function of the viewing angles as well as
the intrinsic misalignment $\theta_\mathrm{int}$. Measured with
  respect to the projected short axis, $\Theta_\mathrm{kin}$ can be
  calculated with a projection matrix (e.g. de Zeeuw \& Franx
  1989\nocite{1989ApJ...343..617D}):

\begin{equation}
\label{eq:def-Thkin}
\tan\Theta_\mathrm{kin} = {\sin\varphi \tan\theta_{\mathrm{int}} \over 
         \sin\vartheta -\cos\varphi \cos\vartheta \tan\theta_{\mathrm{int}}}.
\end{equation}

\begin{figure}
\begin{center} 
\begin{tabular}{|c|c|}
\psfig{figure=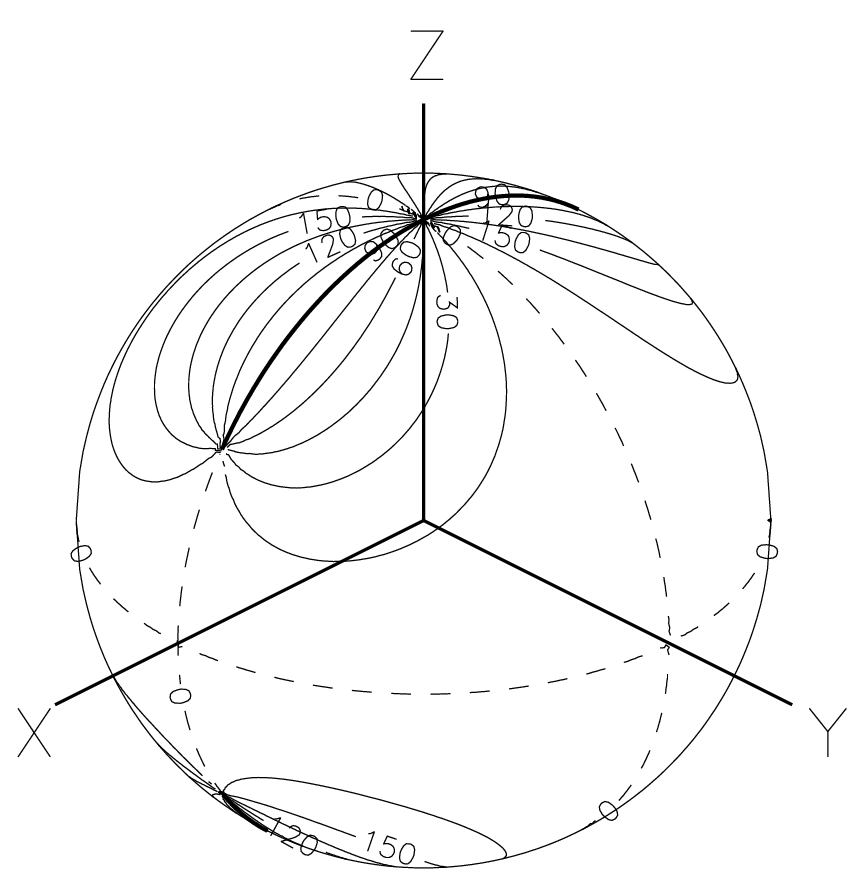,width=4cm} &
\psfig{figure=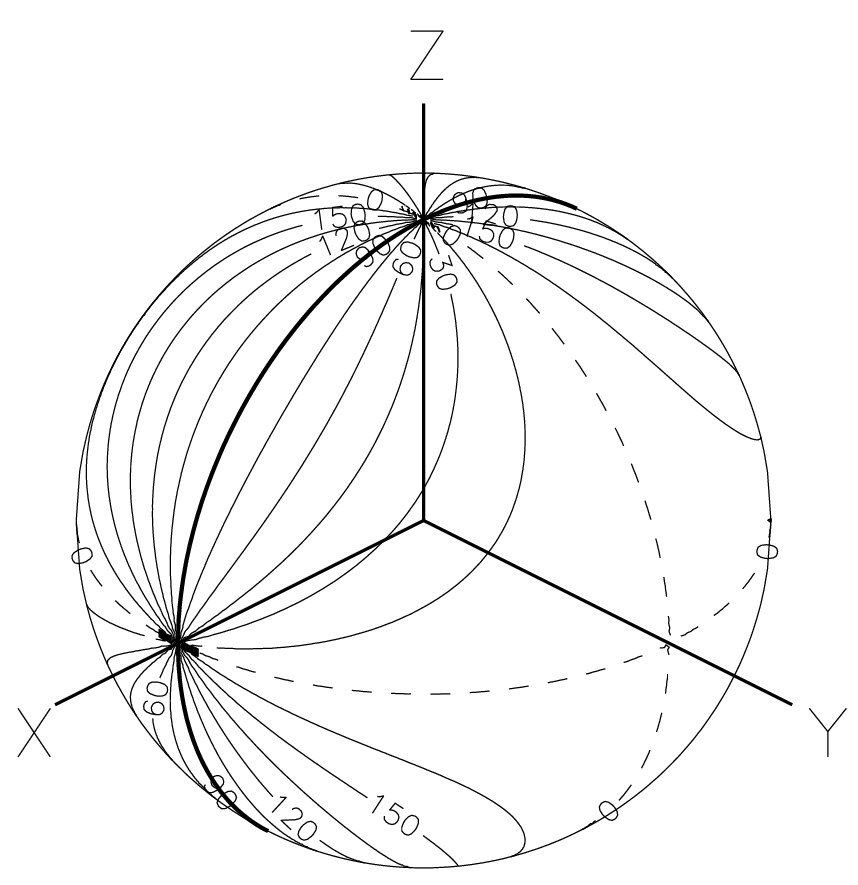,width=4cm}  \\
\end{tabular} 
\end{center}
\caption{Contours of constant $\Theta_\mathrm{min}$ as given in Equation~\ref{eq:def-Thmin} on the sphere of viewing
directions, defined by the angles $(\vartheta, \varphi)$. Left:
$\tan\theta_f = \sqrt{2}$. Right: $\theta_f = \pi/2$. The dashed contour is for $\Theta_\mathrm{min}=0$. }
\label{fig:sphere_min}
\end{figure}

\begin{figure}
\begin{center} 
\begin{tabular}{|c|c|}
\psfig{figure=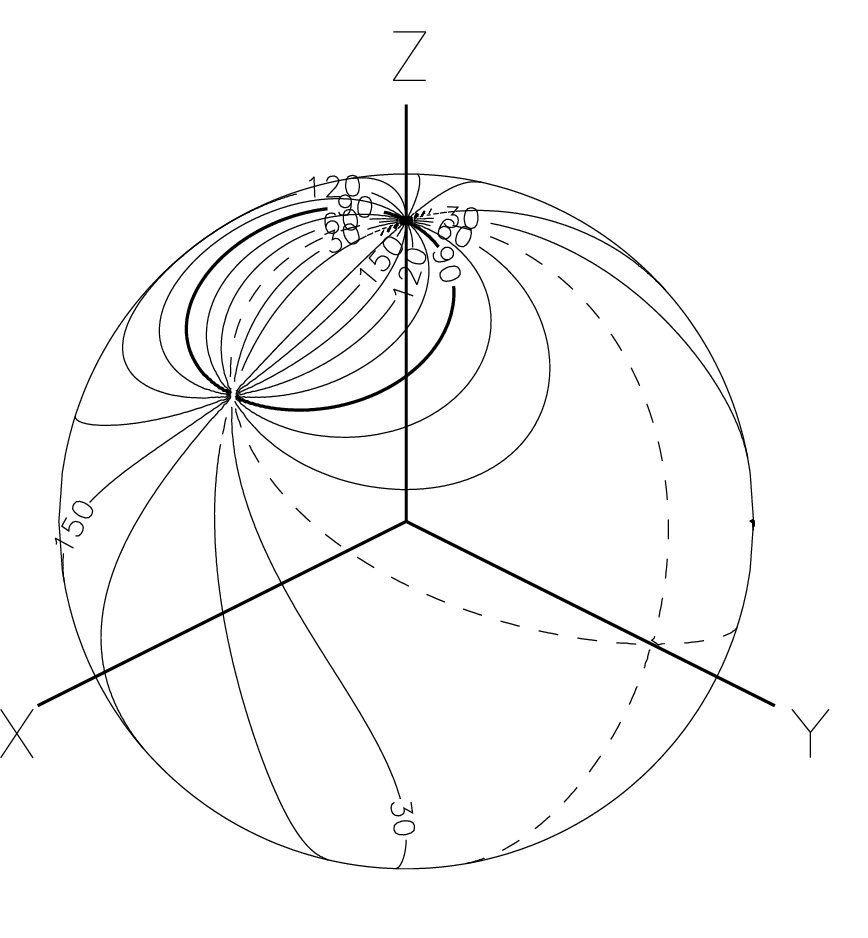,width=4cm} &
\psfig{figure=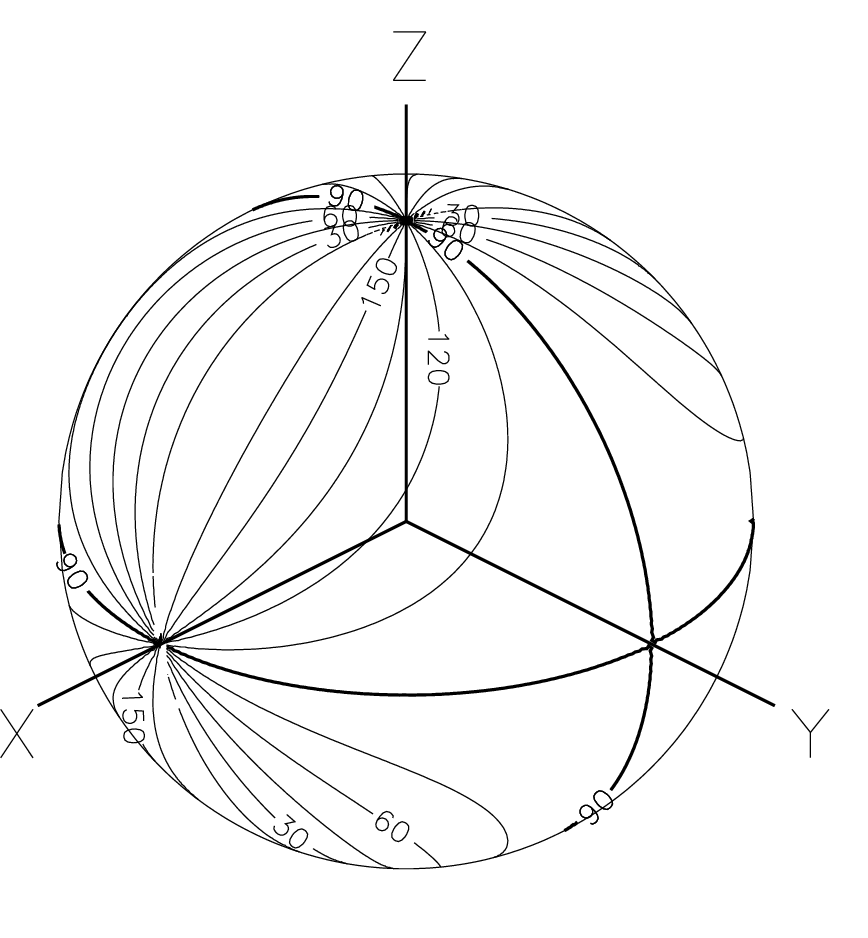,width=4cm} \\
\end{tabular} 
\end{center}
\caption{Contours of constant $\Theta_\mathrm{kin}$, defined in
Equation~(\ref{eq:def-Thkin}), on the sphere of viewing directions defined
by the angles $(\vartheta, \varphi)$. Left: $\theta_\mathrm{int}=\pi/4$. Right:
$\theta_\mathrm{int}=\pi/2$. While in the latter case all octants are similar, in
the former case two distinct sets of octants occur. The dashed contour indicates
$\Theta_\mathrm{kin}=\pi/2$. }
\label{fig:sphere_kin}
\end{figure}

\begin{figure}
\begin{center} 
\begin{tabular}{|c|c|}
\psfig{figure=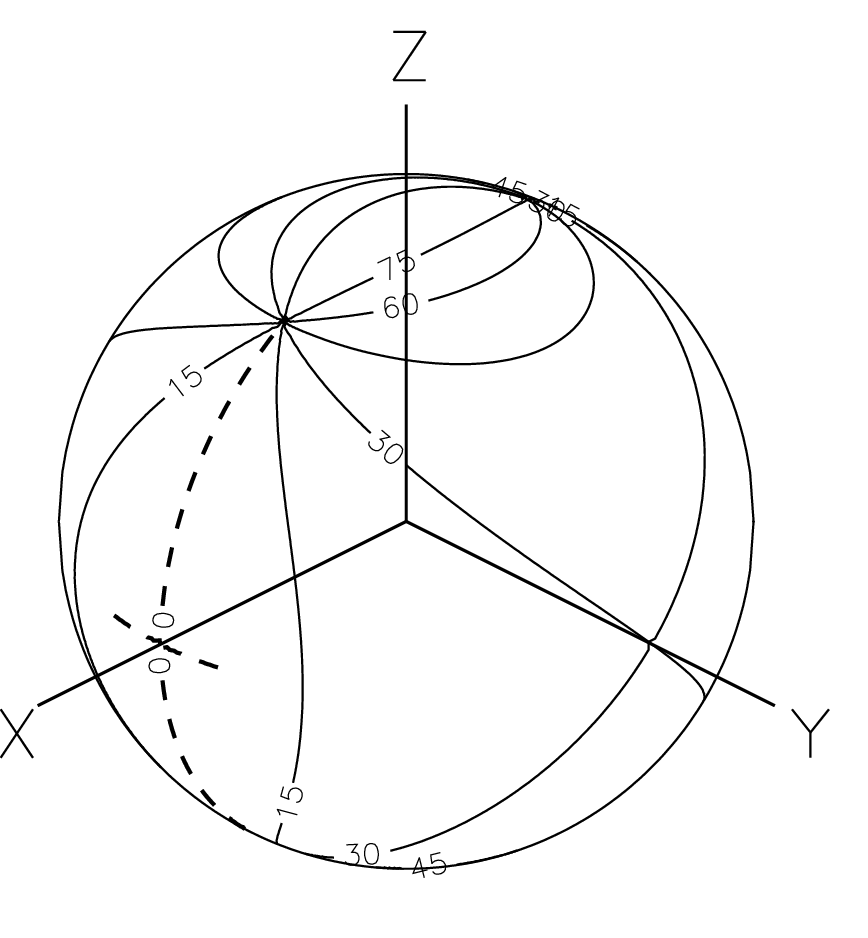,width=4cm} &
\psfig{figure=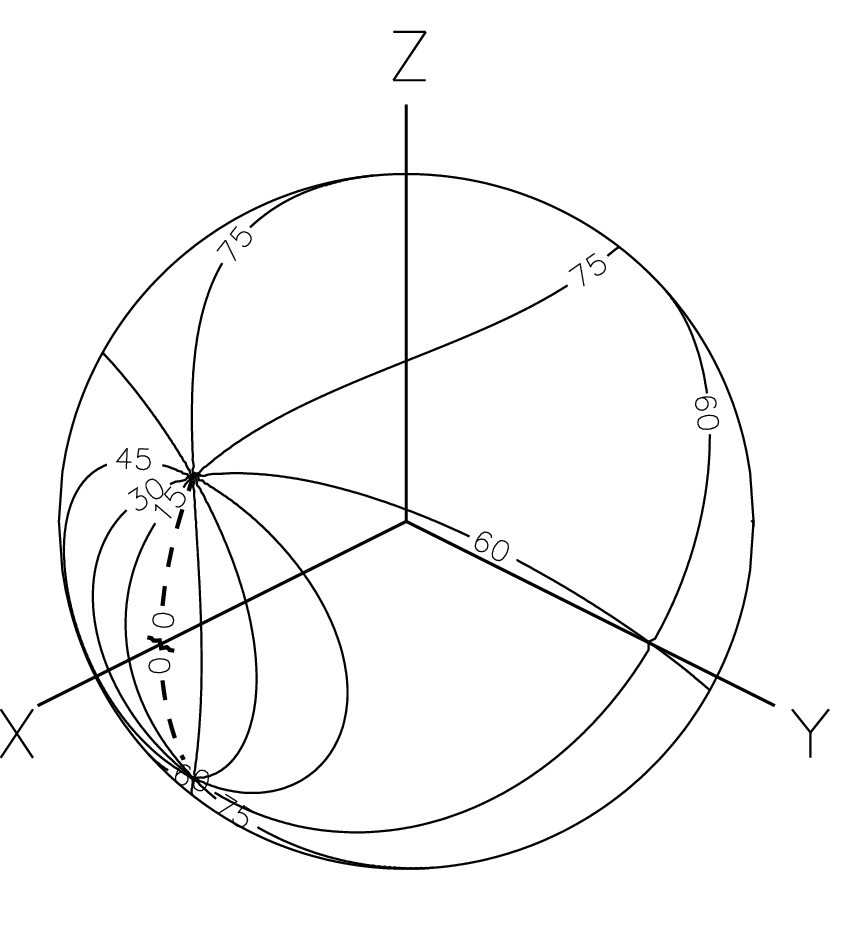,width=4cm}  \\
\end{tabular} 
\end{center}
\caption{Contours of constant misalignment angle $\Psi$, defined in
Equation~ \ref{eq:def-Psi}, on the sphere of viewing directions defined
by the angles $(\vartheta, \varphi)$. In both cases the model has
intrinsic misalignment $\theta_\mathrm{int} = \theta_f$. Left: $T = 1/4$, or $\theta_\mathrm{int}=\pi/6$.
Right:  $T = 3/4$, or $\theta_\mathrm{int}=\pi/3$. The dashed contour
corresponds to $\Psi=0$. }
\label{fig:sphere_psi}
\end{figure}

\noindent
$\Theta_\mathrm{min}$ depends on the intrinsic shape of the galaxy
through the triaxiality parameter $T$ as defined in
Equation~\ref{eq:triax}, and the viewing angles:

\begin{equation}
\label{eq:def-Thmin}
\tan 2\Theta_\mathrm{min} = {2T\sin\varphi \cos\varphi \cos \vartheta \over
        \sin^2\vartheta - T(\cos^2\varphi -\sin^2\varphi \cos^2\vartheta)}.
\end{equation}
\noindent
Examples of $\Theta_\mathrm{min}$, $\Theta_\mathrm{kin}$ and $\Psi$ on
the sphere of viewing angles are given in
Figures~\ref{fig:sphere_min}, \ref{fig:sphere_kin} and
\ref{fig:sphere_psi}, respectively.

\subsection{Probability distributions $P(\Psi, \epsilon)$ for
  $\theta_\mathrm{int} = \theta_f$}

Franx et al. (1991)\nocite{1991ApJ...383..112F} presented in their appendix probability
distributions $P(\Psi, \epsilon)$ for perfectly aligned triaxial
systems, or $\theta_\mathrm{int} = 0$. By following their analysis and
integrating over the sphere of viewing angles, we here give
expressions for $P(\Psi, \epsilon)$ with $\theta_\mathrm{int} =
\theta_f$, with $\theta_f$ corresponding to the viewing direction from
which the galaxy will appear round ($\epsilon = 0$), as defined in
Equation~\ref{eq:round}. 

We define $P(\vartheta, \varphi)\mathrm{d}\vartheta\mathrm{d}\varphi$ as the
probability of finding $\vartheta$ and $\varphi$ in the ranges
$(\vartheta, \vartheta+\mathrm{d}\vartheta)$ and $(\varphi,
\varphi+\mathrm{d}\varphi)$, respectively.  Then $P(\vartheta, \varphi)$ is
equal to the area element on the sphere of viewing angles, divided by
the total area of the sphere, and hence is given by:
\begin{equation}
\label{eq:probthetaphi}
P(\vartheta, \varphi) = {\sin\vartheta \over 4\pi}. 
\end{equation}

\noindent
Therefore, it follows that: 

\begin{equation}
\label{eq:ppsipeps-varangles}
P(\Psi, \epsilon) \, \mathrm{d}\Psi \mathrm{d}\epsilon = 
   \sum\limits_{(\vartheta_i, \varphi_i)}  {\sin\vartheta \over 4\pi} 
   \Big\vert {\partial(\Psi, \epsilon) \over 
              \partial (\vartheta, \varphi)} \Big\vert^{-1}
                                                      \, \mathrm{d}\Psi \mathrm{d}\epsilon,  
\end{equation}
\noindent
where the sum is over all pairs of angles $(\vartheta_i, \varphi_i)$
with $0 \leq \vartheta_i \leq \pi$ and $0 \leq \varphi_i \leq 2\pi$
for which $\Psi(\vartheta, \varphi)=\Psi$ and $\epsilon(\vartheta,
\varphi) =\epsilon)$. \footnote{Equation\ (A24) of Franx et al. (1991) erroneously replaces
$\sin\vartheta$ by $\cos\vartheta$. This is a typographical error with
no impact on their equations (A25)--(A29).}  

Franx (1988)\nocite{1988MNRAS.231..285F} showed that the properties
of projected triaxial ellipsoids are more effectively described in
terms of conical coordinates ($\mu, \nu$) instead of spherical
coordinates ($\vartheta, \varphi$), so we continue our analysis in
this coordinate system instead. The relation between conical and
spherical coordinates is given by (e.g. de Zeeuw \& Pfenniger
1988\nocite{1988MNRAS.235..949D}, their Equations 5.4-5.6):

\begin{eqnarray}
\label{eq:angles-conicals}%
\cos^2 \vartheta \!\!\! &=& \!\!\! 
                 {\displaystyle (\mu - q^2) (\nu - q^2) \over 
                  \displaystyle (1 -q^2) (p^2-q^2)}, 
                                                            \nonumber \\
\tan^2 \varphi \!\!\! &=& \!\!\! 
            {\displaystyle (\mu - p^2)  (p^2 - \nu ) (1-q^2) 
            \over \displaystyle (1- \mu ) (1 - \nu ) (p^2 - q^2)}, 
\end{eqnarray}

\noindent
such that each combination $(\mu, \nu)$ corresponds to eight directions, given
by $(\vartheta, \pm\varphi)$, $(\vartheta, \pm [\pi- \varphi])$,
$(\pi-\vartheta, \pm\varphi)$ and $(\pi-\vartheta, \pm [\pi-
\varphi])$. The area element $\mathrm{d} \Omega=\sin\theta \mathrm{d}\vartheta \mathrm{d}\varphi$ on the
unit sphere is given by:
\begin{equation}
\label{eq:area}
\mathrm{d} \Omega = {(\mu - \nu) \mathrm{d} \mu \mathrm{d} \nu 
         \over 4 \sqrt{-h(\mu)} \sqrt{h(\nu)}}, 
\end{equation}

\noindent
with
\begin{equation}
\label{eq:hfunction}
h(\tau) = (\tau -1 )(\tau -p^2) (\tau -q^2). 
\end{equation}
\noindent
Combining Equations~\ref{eq:probthetaphi} and \ref{eq:area}, it then
follows that the probability of finding $\mu$ and $\nu$ on the sphere
of viewing angles in the ranges $(\mu, \mu\!+\!\mathrm{d}\mu)$ and $(\nu, \nu\!+\!\mathrm{d}\nu)$,
respectively, is equal to:

\begin{equation}
\label{eq:def-fundprob}
P(\mu, \nu) = {(\mu-\nu) \over 16\pi \sqrt{-h(\mu)} \sqrt{h(\nu)}}, 
\end{equation}

\noindent
such that 

\begin{equation}
\label{eq:ppsieps-general}
P(\Psi, \epsilon)\mathrm{d}\Psi \mathrm{d}\epsilon = \sum\limits_{\mu_i, \nu_i} 
       P(\mu, \nu) \left| {\partial(\mu, \nu) \over
                  \partial(\Psi, \epsilon)} \right| \mathrm{d}\Psi \mathrm{d}\epsilon,
\end{equation}

\noindent
where $\mu_i$ and $\nu_i$ are all the pairs of solutions of $\Psi(\mu,
\nu) = \Psi$ and $\epsilon(\mu, \nu)=\epsilon$. 

To continue, we have to know expressions for our observables $\epsilon$
and $\Psi$ similar to Equations~\ref{eq:def-ellipticity} and \ref{eq:def-Psi}, but now in
conical coordinates $\mu, \nu$. For $\epsilon$, we combine
Equations~\ref{eq:def-ellipticity}, \ref{eq:def-auxaandb} and
\ref{eq:angles-conicals} to arrive at (see also de Zeeuw \& Pfenniger
1988\nocite{1988MNRAS.235..949D}, their Equation 5.4):

\begin{equation}
\label{eq:epsilon-conical}
\epsilon=1-\sqrt{\nu\over\mu}, \quad \hbox{or} 
\quad e={\nu\over\mu}.
\end{equation}
\noindent
For $\Psi$, it can be shown by combining Equations~\ref{eq:def-Psi}, \ref{eq:def-Thkin},
\ref{eq:def-Thmin} and \ref{eq:angles-conicals} that:

\begin{equation}
\label{eq:Psi-conical}
\tan\Psi   = 
               {(R_1\mp AR_2) \sqrt{\mu-p^2}
               \over 
               (R_2\mp AR_1) \sqrt{p^2-\nu}}, 
\end{equation}
\noindent
where we have defined the auxiliary functions
\begin{equation}
\label{eq:def-rfunctions}
R_1 = \sqrt{(1-\mu)(\nu-q^2)}, 
              \quad R_2 = \sqrt{(\mu-q^2)(1-\nu)},
\end{equation}
\noindent
and
\begin{equation}
\label{eq:def-candd}
A=\sqrt{1-T \over T} \tan\theta_\mathrm{int}.
\end{equation}
\noindent
Note that for the case that we are studying $\theta_\mathrm{int} =
\theta_f$, and therefore $A = 1$. We now introduce $t = \tan \Psi$, such that:

\begin{equation}
\label{eq:def-t}
\mathrm{d}\Psi = \frac{\mathrm{d}t^2}{2t(1+t^2)},
\end{equation}

\noindent
and therefore:

\begin{equation}
\label{eq:jacobian}
\left| {\partial(\mu, \nu) \over
                  \partial(\Psi, \epsilon)} \right|^{-1} 
= {1 \over 4t(1\!+\!t^2) \, \mu^{3/2}\nu^{1/2}} 
  \left| \mu {\partial t^2\over \partial \mu} +
         \nu {\partial t^2\over \partial \nu} \right| . 
\end{equation}

\noindent
We simply the above expression by substituting $\nu = e\mu$
(Equation~\ref{eq:epsilon-conical}), and combining the result with
Equation~\ref{eq:ppsieps-general}, we arrive at:

\begin{equation}
\label{eq:ppsieps-specific}
P(\Psi, \epsilon)=
  \sum\limits_i {\displaystyle{(1\!-\!e)\sqrt{e} \, t(1\!+\!t^2) \, \mu_i^2   
           \over 4\pi \sqrt{-h(\mu_i)h(e\mu_i)}}} \, 
   \left| {\displaystyle {\mathrm{d} t^2\over \mathrm{d}\mu}} \right|_{\mu=\mu_i}^{-1}\!,  
\end{equation}
\noindent
where the sum is over all octants, and over all physical roots
$p^2\leq \mu_i \leq 1$ of the equation $t^2(\mu_i, e\mu_i)=t^2$, and
$h(\mu)$ is defined in Equation~\ref{eq:hfunction}.

\begin{figure*}
\begin{center} 
\begin{tabular}{|l|l|l}

\psfig{figure=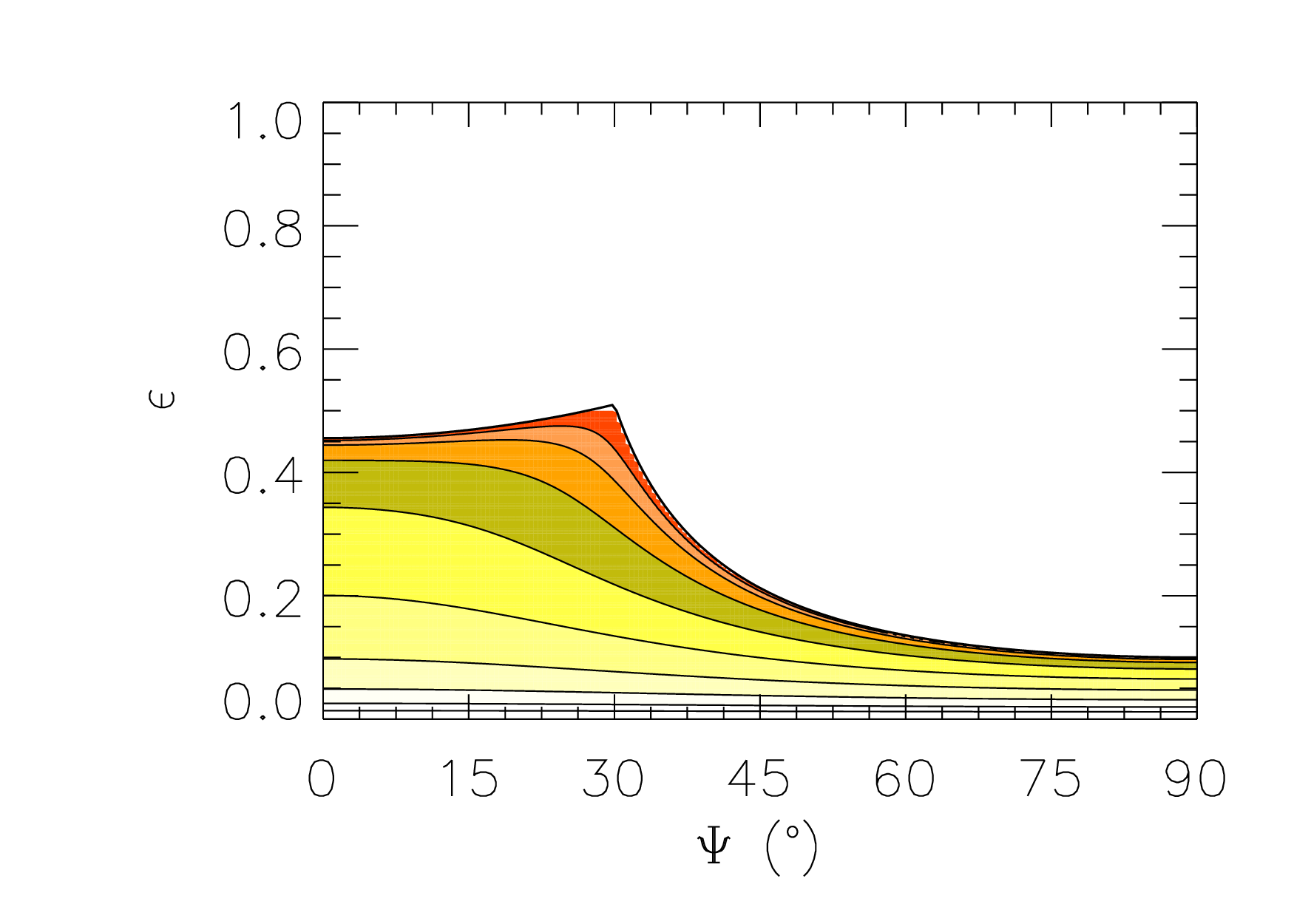,width=5.2cm} &
\psfig{figure=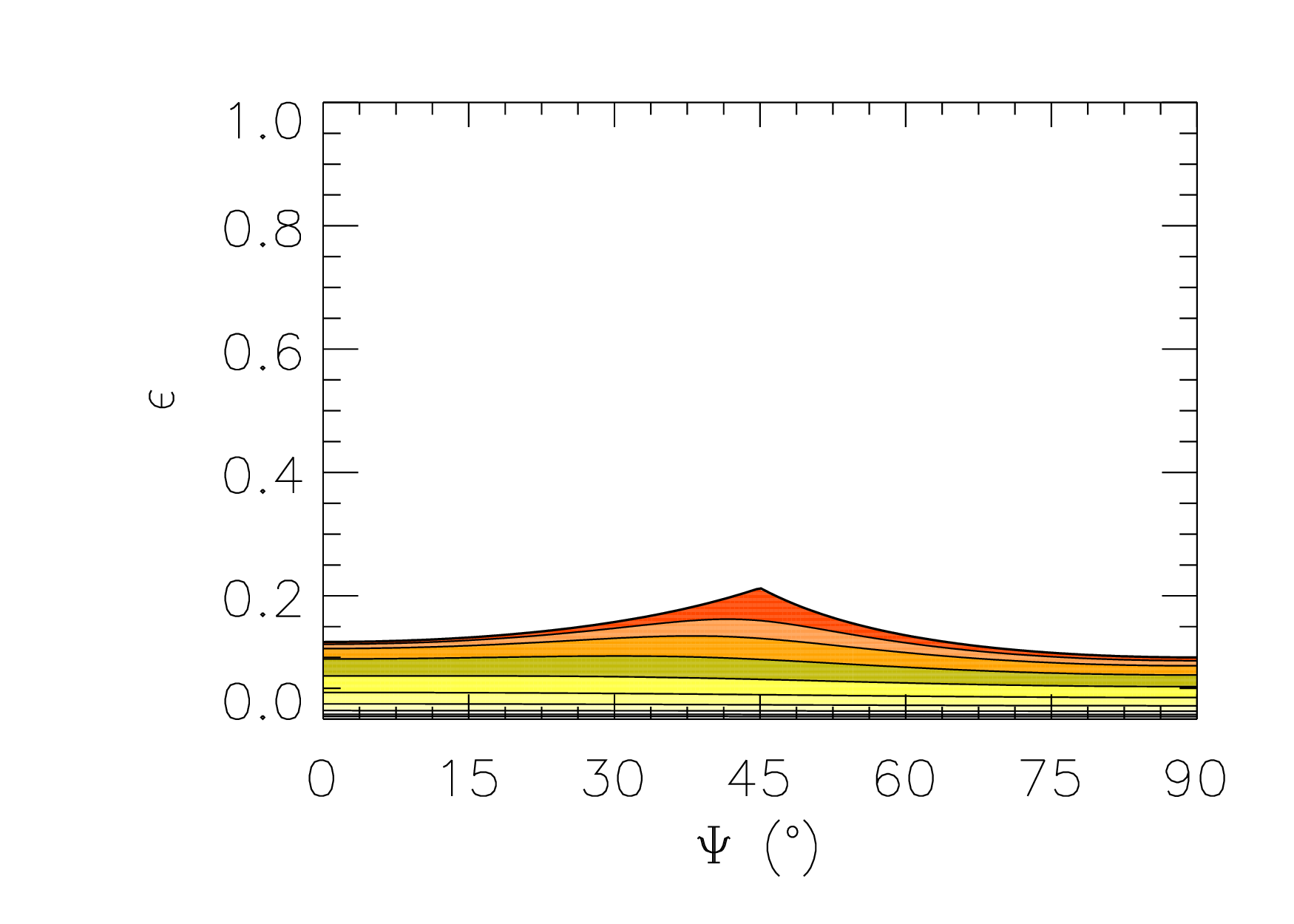,width=5.2cm} &
\psfig{figure=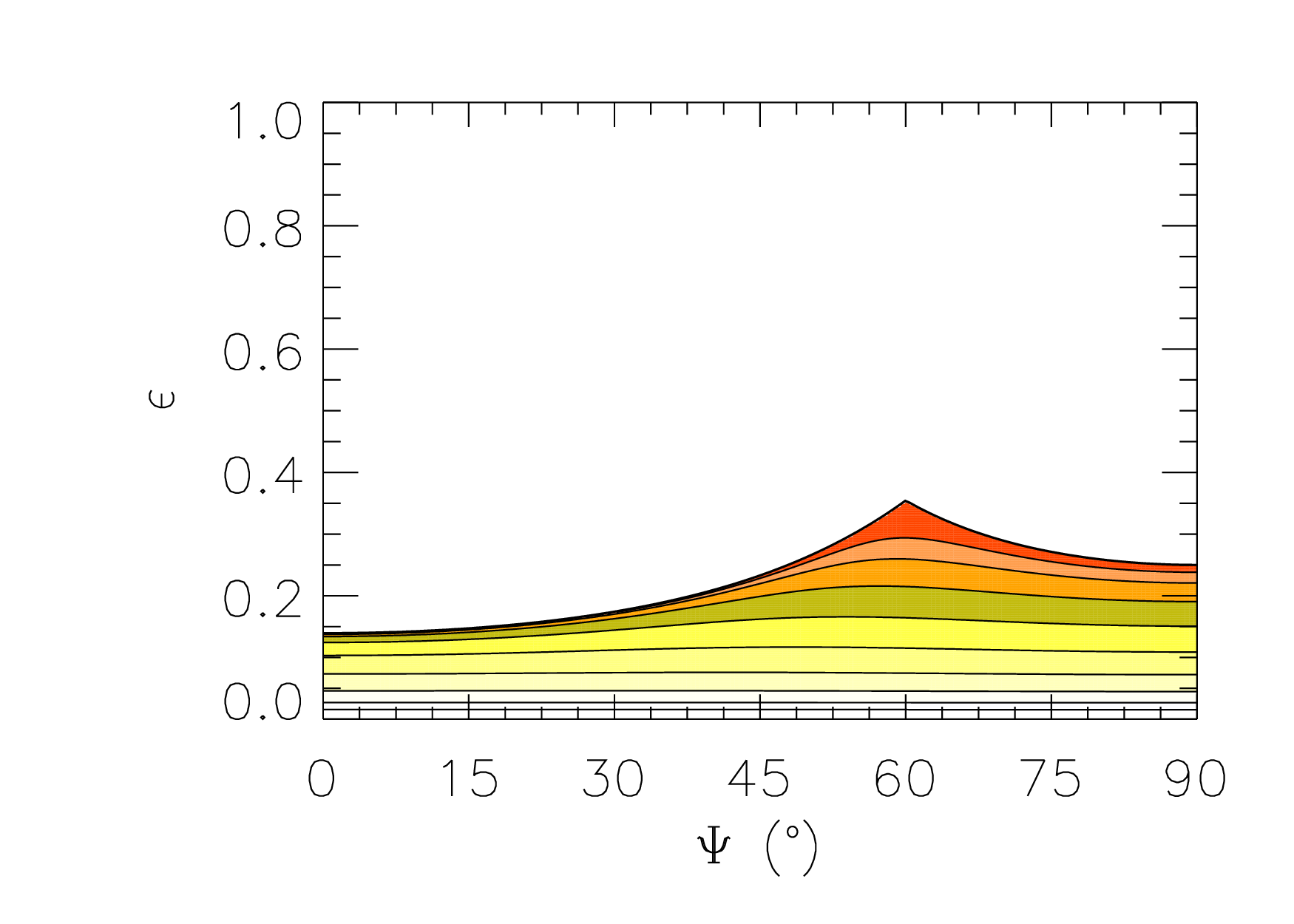,width=5.2cm} 
\end{tabular}
\end{center}
\caption{Probability distributions $P(\Psi, \epsilon)$ for
  $\theta_\mathrm{int} = \theta_f$. Left: $\theta_{\mathrm{int}}
  =30^\circ$,
  $p=0.9$. Middle: $\theta_{\mathrm{int}} =45^\circ$,
  $p=0.9$. Right: $\theta_{\mathrm{int}} =60^\circ$,
  $p=0.75$. The triaxiality increases in these plots from $T=0.25$
  (left),  $T=0.50$ (middle) to  $T=0.75$ (right). Contours are spaced logarithmically and increase with
  darker colours. $P(\Psi, \epsilon)$ is singular on the boundary
  curves.}
\label{fig:psieps}
\end{figure*}

We now concentrate on the case $\theta_\mathrm{int} = \theta_f$, and
the expressions for $t^2$ and $|\mathrm{d} t^2/\mathrm{d} \mu |$
simplify to (see Equations~\ref{eq:Psi-conical} - \ref{eq:def-candd}:

\begin{equation}
\label{eq:tdtmu-sinthint-t}
t^2 = {\mu-p^2 \over p^2 -e\mu}, \qquad 
{\mathrm{d} t^2 \over \mathrm{d} \mu} = {(1-e)p^2 \over (p^2-e\mu)^2},
\end{equation}
\noindent
which is valid in all octants. Solving for $\mu$ using the left-hand
expression in Equation~\ref{eq:tdtmu-sinthint-t} results in a single
root $\mu_1$ contributing to $P(\Psi, \epsilon)$:

\begin{equation}
\label{eq:mu-sinthint-t}
\mu_1= {p^2 (1\!+\!t^2)\over (1\!+\!et^2)} 
     = {p^2 \over \cos^2\Psi +e\sin^2\Psi}.
\end{equation}

\noindent
Substituting this root $\mu_1$ into Equation~\ref{eq:ppsieps-specific} then
leads us finally to an expression for the probability distribution
$P(\Psi, \epsilon)$ with $\theta_{\mathrm{int}} = \theta_f$:

\begin{equation}
\label{eq:ppsieps-sinthint-t}
P(\Psi,\epsilon)= {2(1-e)\sqrt{e} \, \mu_1^3 \over
      \pi p^2 \sqrt{1\!-\!\mu_1}\sqrt{e\mu_1\!-\!q^2} 
                    \sqrt{1\!-\!e\mu_1}\sqrt{\mu_1\!-\!q^2}}.
\end{equation}

\noindent
The area in the $(\Psi,\epsilon)$-plane where $P(\Psi, \epsilon)$ is
non-zero is bounded by $\epsilon=0$, $\Psi=0$, $\Psi=\pi/2$, and two
boundary curves, $e=e_{\rm I}(\Psi)$ and $e=e_{\rm II}(\Psi)$ with
\begin{eqnarray}
\label{eq:ppsieps-sinthint-t-boundaries}
e_{\rm I} 
      \!\! &=& \!\! {\displaystyle {q^2\over p^2+(p^2-q^2)t^2}}, \qquad 
        (0 \leq \Psi \leq \theta_f), \nonumber \\
e_{\rm II} 
      \!\! &=& \!\! p^2 -{\displaystyle{(1-p^2) \over t^2}}, \quad
         \qquad(\theta_f \leq \Psi \leq {\textstyle {\pi \over 2}}).  
\end{eqnarray}

\noindent
$P(\Psi, \epsilon)$ diverges on both these curves, which join at
$e=q^2$ and $\Psi=\theta_\mathrm{int}=\theta_f$. $P(\Psi, \epsilon)$
vanishes in the limit $\epsilon\downarrow 0$, but is finite for
$\Psi=0$ and $\pi/2$. In Figure~\ref{fig:psieps} we show several
examples of $P(\Psi, \epsilon)$.

\subsubsection{$P(\Psi)$ for $\theta_\mathrm{int} = \theta_f$}

For completeness, we also derive the probability distribution
$P(\Psi)$ for the case that $\theta_\mathrm{int} = \theta_f$. This
expression can be obtained by integrating $P(\Psi, \epsilon)$ as given
in Equation~\ref{eq:ppsieps-specific} over $\mathrm{d}\epsilon =
\mathrm{d}e/2\sqrt{e}$, which results in:

\begin{equation}
\label{eq:ppsi-general}
P(\Psi) = {t(1\!+\!t^2) \over 8\pi} \sum\limits_i 
           \int\limits_{e^-}^{e^+} \! \mathrm{d} e \, 
           {(1\!-\!e) \mu_i^2 \over \sqrt{-h(\mu_i)h(e\mu_i)}} 
        \left| {\displaystyle \mathrm{d} t^2\over 
                    \displaystyle \mathrm{d}\mu} \right|_{\mu=\mu_i}^{-1}\!\!\!,
\end{equation}

\noindent
where the sum is taken over all octants, and all physical roots $p^2
\leq \mu_i \leq 1$ of $t^2(\mu_i, e\mu_i) = t^2$. The integration
limits $e_-$ and $e_+$ depend on $p$, $q$ and
$\theta_\mathrm{int}$. For our purposes, it is convenient to
substitute $\mu$ back for $e$ into this equation, leading to:

\begin{equation}
\label{eq:ppsi-alternate}
P(\Psi) = {t(1\!+\!t^2) \over 8\pi} \sum\limits_i 
           \int\limits_{\mu^-}^{\mu^+} \! \mathrm{d} \mu
           {(\mu\!-\!\nu_i) \over \sqrt{-h(\mu)h(\nu_i)}} 
        \left| {\displaystyle \mathrm{d} t^2\over 
                    \displaystyle \mathrm{d}\nu} \right|_{\nu=\nu_i}^{-1}\!,
\end{equation}
where the sum is over all octants, and $q^2 \leq \nu_i(\mu, t^2) \leq
p^2$ is a root of $t^2(\mu, \nu)=t^2$. This expression can also be derived
directly from the fundamental probability distribution
(\ref{eq:def-fundprob}) by the transformation $(\mu, \nu) \to (\mu,
\Psi)$, and has the advantage that all quantities are functions of $T$
only, which is not the case for expression (\ref{eq:ppsi-general})
which contains $e$. 

It further is useful to transform from $(\mu, \nu)$ to the rescaled
conical coordinates $({\bar\mu}, {\bar\nu})$, defined as:

\begin{equation}
\label{eq:conical-alternate}
\bar\mu={\mu-p^2 \over 1-q^2}, \qquad 
\bar\nu={\nu-p^2 \over 1-q^2}, 
\end{equation}
\noindent
so that $\bar\mu \geq0$ and $\bar\nu \leq 0$. Substituting these
coordinates 
in Equation~\ref{eq:ppsi-alternate} and taking the sum over all eight octants then leads to the simplified
expression

\begin{equation}
\label{eq:ppsi-bar-alternate}
P(\Psi)\!=\!{t(1\!+\!t^2)\over \pi} \sum\limits_i \!
          \int\limits_0^{\bar\mu^+} {\mathrm{d}\bar\mu \,(\bar\mu-\bar\nu) \over
          \sqrt{-{\bar h}(\bar\mu) {\bar h}(\bar\nu)}} 
\left| {\displaystyle \mathrm{d} t^2\over 
                    \displaystyle \mathrm{d}\bar\nu} \right|_{\bar\nu=\bar\nu_i}^{-1},
\end{equation}

\noindent
where $\bar\nu_i=\bar\nu_i(\bar\mu)$ are all solutions of $t^2(\bar\mu,\bar\nu)=0$ in
the interval $-(1-T)\leq \bar\nu\leq 0$, and

\begin{equation}
\label{eq:barh}
{\bar h}(\bar{\tau})=\bar\tau (T-\bar\tau)(\bar\tau+1-T).
\end{equation}
\noindent
We now express Equation~\ref{eq:tdtmu-sinthint-t} in terms of
$\bar\mu$ and $\bar\nu$ as given by Equation~\ref{eq:conical-alternate}, to
arrive at:

\begin{equation}
\label{eq:tdtnu-sinthint-t}
t^2 = -{\bar\mu \over \bar\nu}, \qquad 
{\mathrm{d} t^2 \over \mathrm{d} \nu} = -{t^2 \over \bar\nu},
\end{equation}

\noindent
Substituting the above expressions into
Equation~\ref{eq:ppsi-bar-alternate}, we finally obtain:
\begin{equation}
P(\Psi)\!=\!{(1\!+\!t^2)^2\over \pi t^2} 
\!\left\{
\begin{array}{ll}
 \!\! \int\limits_0^{(1\!-\!T)t^2} \!\!\!\!
    {\displaystyle\bar\mu \mathrm{d} \bar\mu \over 
      \displaystyle \sqrt{P_4(\bar\mu)}}, 
            &(0\leq t^2\leq{\displaystyle {T\over 1\!-\!T}}), \nonumber \\
  \null & \nonumber \\
 \!\! \int\limits_0^{T} \!
    {\displaystyle \bar\mu \mathrm{d} \bar\mu \over 
      \displaystyle \sqrt{P_4(\bar\mu)}}, 
                  &({\displaystyle {T\over 1\!-\!T}}\leq t^2), 
\end{array}\right.
\end{equation}
where $P_4$ is a polynomial of degree 4 in $\bar\mu$, given by
\begin{equation}
P_4(\bar\mu)=(T-\bar\mu)(\bar\mu+1-T)([1-T]t^2-\bar\mu)(\bar\mu+Tt^2). 
\end{equation}
It can be shown that: 
\begin{equation}
P(\Psi; T) = P({\textstyle {\pi \over 2}}-\Psi; 1-T), 
\end{equation}
so that we need to evaluate $P(\Psi; T)$ only for $0\leq T\leq 1/2$. 

\begin{table}
\caption{Special values of $P(\Psi)$ for the case where
  $\theta_\mathrm{int} = \theta_f$.}
\begin{center} 
\begin{tabular}{|l|l|}
\hline 
\noalign{\vskip -8pt}
\hline
&\\[-6pt]
$\Psi$  & $P(\Psi)$ \\
\hline
\noalign{\vskip 5pt}
0 &${1\over \pi} 
     \!-\! {(2T\!-\!1)\over \pi\sqrt{T(1\!-\!T)}}
      \arctan\sqrt{1\!-\!T \over T}$ \\
\noalign{\vskip 5pt}
${\textstyle {\pi \over 2}}\!-\!\theta_\mathrm{int}$ & ${1\over \pi} 
       \Bigl\{{[\sqrt{T(1\!-\!T)} -\!1] \over T(1\!-\!T)} \ln(1\!-\!2T)\!-\! 
               \ln(\sqrt{T}\!+\!\sqrt{1\!-\!T}) \Bigr\}$ \\
\noalign{\vskip 5pt}
${\textstyle {\pi \over 2}}$ & $ {1\over \pi} 
                \!-\! {(1\!-\!2T)\over \pi\sqrt{T(1-T)}}
                 \arctan\sqrt{T\over 1\!-T}$ \\
\noalign{\vskip 5pt}
\hline 
\end{tabular}
\label{tab:ppsi-specialvalues}
\end{center}
\end{table}

$P(\Psi)$ can be expressed in terms of incomplete elliptic
integrals. It diverges logarithmically for $\Psi=\theta_\mathrm{int}$, and is
elementary for $\Psi=0$, $\pi/2-\theta_\mathrm{int}$, and $\pi/2$. The expressions
for these special cases are given in
Table~\ref{tab:ppsi-specialvalues}.

The entire function $P(\Psi)$ is elementary for $T=1/2$, and is 
given by
\begin{equation}
\label{eq:ppsi-thalf}
P(\Psi; 1/2) = -{\displaystyle {2 \ln |\cos 2\Psi| \over \pi \sin^2 2\Psi}}, 
\end{equation}
which satisfies $P(\Psi)=P({\textstyle {\pi \over 2}}-\Psi)$. It follows that $P(0;
1/2)=P(\pi/2; 1/2)=1/\pi$, and $P(\Psi; 1/2)$ diverges logarithmically
at $\Psi=\pi/4$. In this case the cumulative distribution is
elementary as well. It is given by
\begin{equation}
{\bar P}(\Psi)={1\over 2} + {\ln|\cos2\Psi| \over \pi\tan 2\Psi}
                - {\arcsin(\cos2\Psi) \over \pi}. 
\end{equation}
This equals $1/2$ when $\Psi=\pi/4$, in accord with the symmetric
nature of $P(\Psi)$.

\section{Kinematic misalignment as sole shape tracer}

As derived in the previous section, observed kinematic misalignment $\Psi$
depends on the intrinsic rotation misalignment
$\theta_\mathrm{int}$ and the intrinsic shape of a galaxy. The
dependence on intrinsic shape is solely given by the triaxiality $T$
of the system, and as such, we could try to infer the intrinsic shape
distribution of our galaxy sample from the observed misalignement
distribution, parametrising the intrinsic shape with $T$ only. As we
did before with the observed histogram of ellipticity, we now
approximate the observed histrogram of kinematic misalignment
(Figure~\ref{fig:hist_eps}) with a sum of Gaussians, whose standard
deviation is given by the measurement errors. We mirror the resulting
distribution around $\Psi = 90^\circ$. As a clear rotation axis is not always
easy to identify for the slow rotators, the measurement errors for
individual galaxies are rather large (up to $90^\circ$), resulting in
a rather flat distribution of kinematic misalignment.

\begin{figure*}
\begin{center} 
\begin{tabular}{|l|l|l}

\psfig{figure=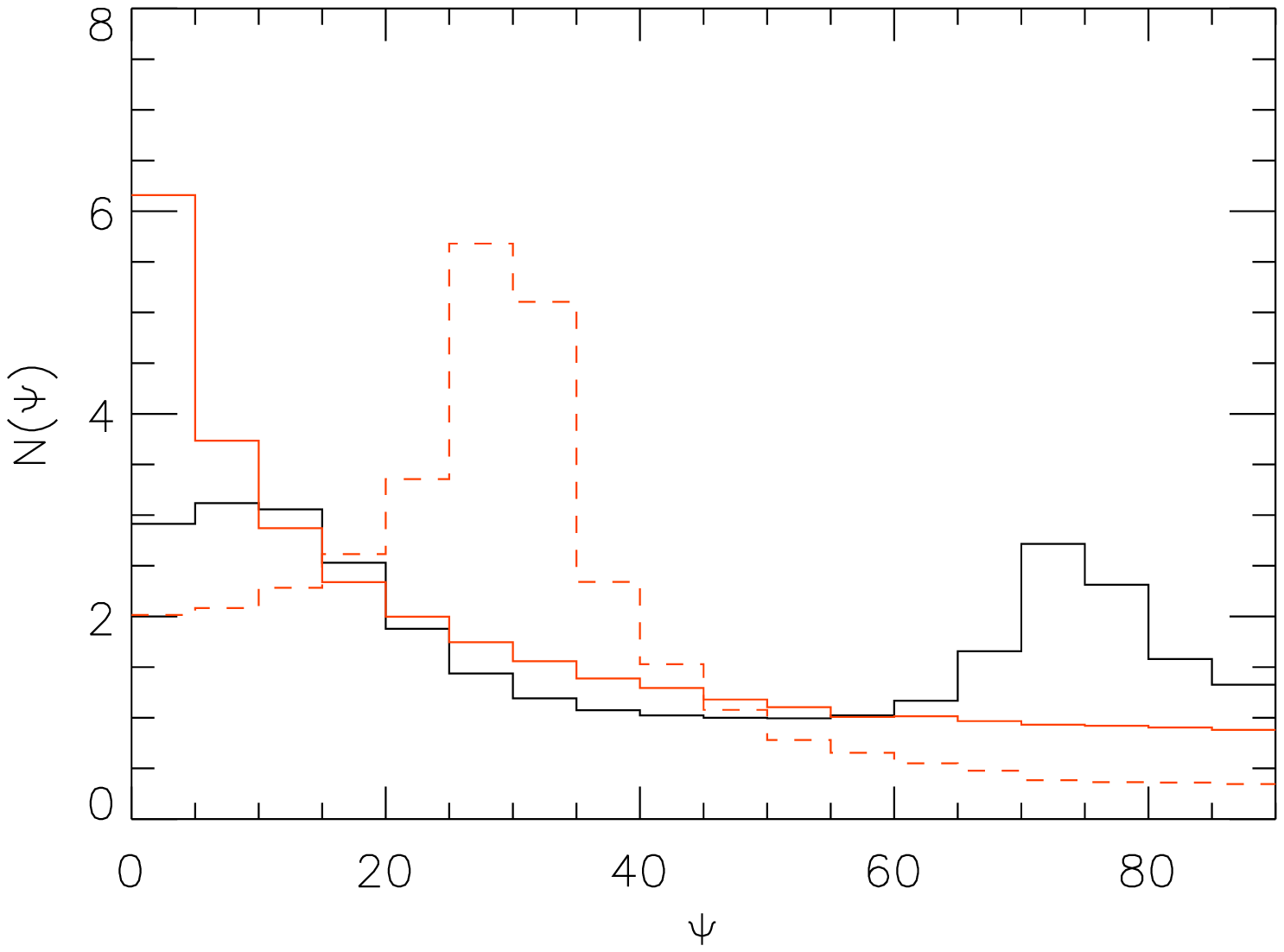,width=5.2cm} &
\psfig{figure=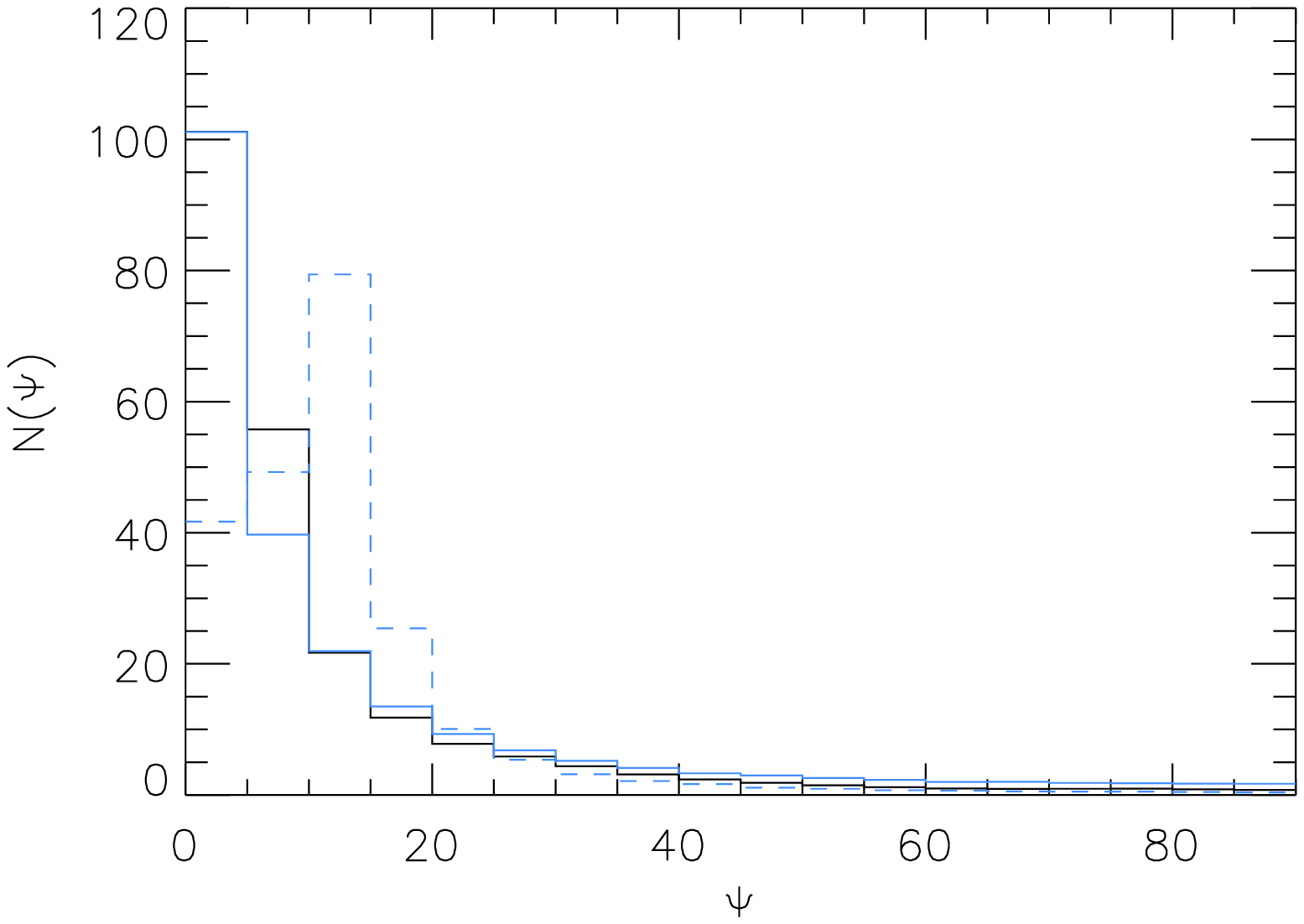,width=5.2cm} &
\psfig{figure=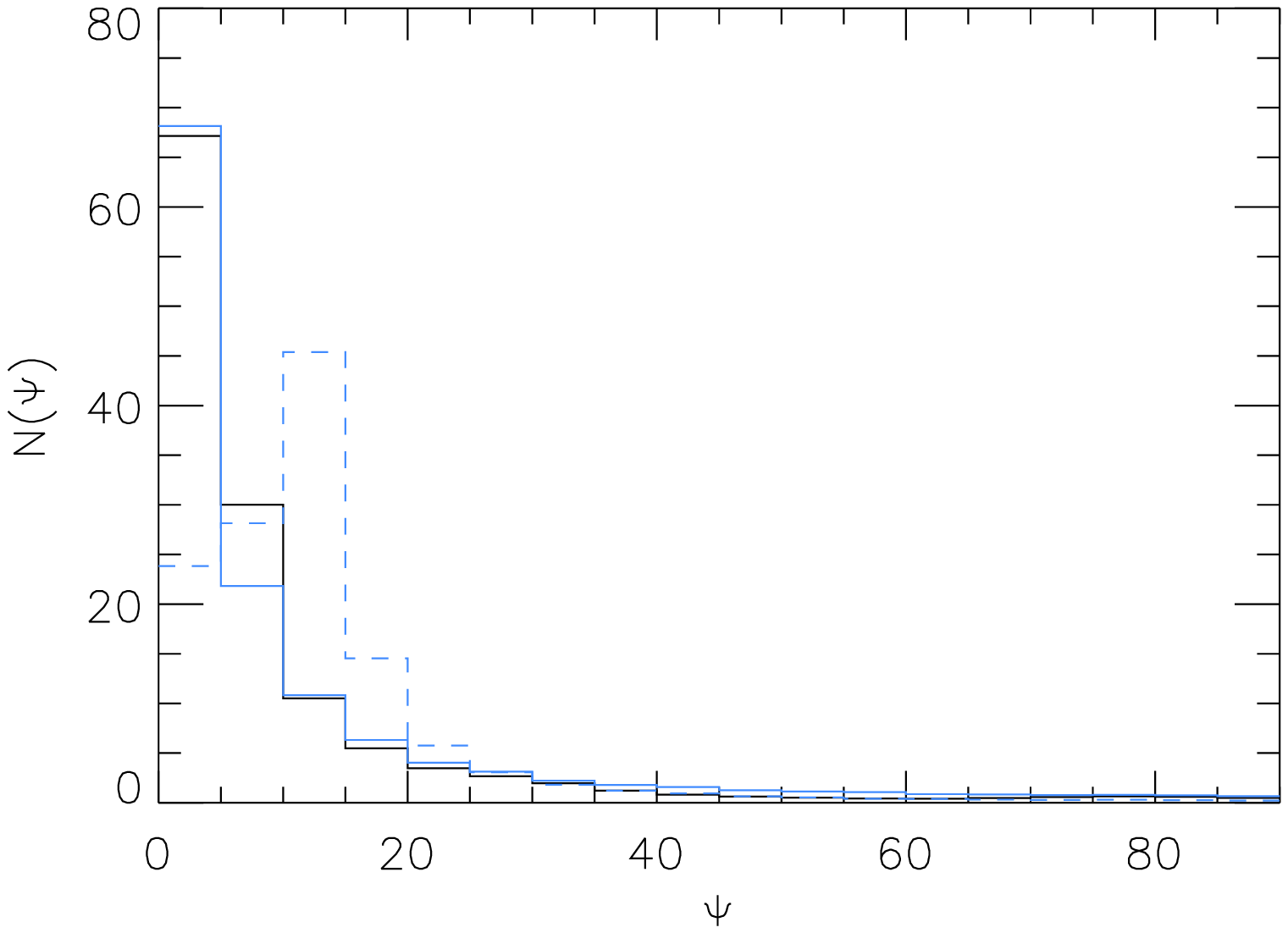,width=5.2cm} 
\end{tabular}
\end{center}
\caption{Observed distributions of kinematic misalignment (black
  histogram), constructed from Gaussians representing individual
  galaxies and their measurement errors (see text for detail). From
  left to right we show the slow rotators, fast rotators, and a
  'clean' sample
of fast rotators excluding barred and interacting
galaxies. Overplotted we show best-fit models assuming constant
triaxiality and no intrinsic misalignment (solid coloured lines) or intrinsic
misalignment scaling with triaxiality as $\theta_\mathrm{int} =
\theta_\mathrm{f}$ (dashed coloured lines). The intrinsically aligned
models give by eye a good fit to the observed kinematic misalignment
distributions, but yield unrealistically high triaxiality values ($T$
= 1.0, 0.45 and 0.35 for the slow rotators, fast rotators, and clean
sample, respectively). 
}
\label{fig:mod_kinmis}
\end{figure*}

We first assume that the intrinsic misalignment is zero
($\theta_\mathrm{int} = 0$). This is not a very realistic assumption
as especially highly triaxial models are expected to display
significant intrinsic misalignment, but does showcase the maximum
allowed triaxiality, as intrinsic misalignment does not contribute to
the observed kinematic misalignment (see
Equation~\ref{eq:def-Psi}). Generating model galaxies with random
viewing angles using Monte Carlo simulations, and binning the observed
and simulated samples in bins of $5^\circ$, we find with a simple
$\chi^2$ fit that the best-fitting intrinsic shape for the slow
rotators would have a triaxiality of $T=1$, which corresponds to a
prolate shape. As prolate galaxies are extremely rare in our sample, and previous analyses have shown that the
slow rotators in our sample are only mildy triaxial (e.g. Paper III),
we cannot take this result at face value. Similarly, the fast
rotator sample is best fitted with a shape distribution of $T=0.45$,
which is unrealistically high. This best-fitting value goes down to
$T=0.35$ if we exclude galaxies that are barred or interacting (Paper
II). We show the resulting fits in Figure~\ref{fig:mod_kinmis}, with
solid coloured lines.

A more realistic model would be to allow the intrinsic misalignment to
increase with increasing triaxiality, by assuming as before that
$\theta_\mathrm{int} = \theta_\mathrm{f}$ (see
Equation~\ref{eq:round}). We then obtain a best-fitting distribution
$T=0.25$ for the slow rotator sample, and $T=0.05$ for the fast
rotator sample. This last value does not change when excluding barred
and interacting systems. We show these fits too in
Figure~\ref{fig:mod_kinmis}, with dashed lines. Although these values
seem more realistic, the fits are worse than for the model with no
intrinsic misalignment. The results of this analysis show that
  kinematic misalignment alone is not a good tracer of intrinsic
  shape. The probability distribution $P(\Psi)$ strongly depends on
  the intrinsic misalignment in the model: it has a
  singularity for $\Psi = \theta_{\mathrm{int}}$, as shown in
  e.g. figure 9a of Franx et
  al. (1991)\nocite{1991ApJ...383..112F}. The triaxiality $T$ in
  contrast has a much milder influence on $P(\Psi)$ (e.g. figure 5a of
Franx et al. 1991).\nocite{1991ApJ...383..112F} We therefore warn against over-interpreting
this simple analysis, as valuable information on the shapes of galaxies
(their ellipticities) has not been used: indeed, this
exercise shows the importance of including both shape and misalignment
information when recovering intrinsic shape distributions. We 
refer the reader to the results presented in the main body of this
paper as more reliable representations of the intrinsic shapes.

\label{lastpage}
\end{document}